\documentclass[
12pt]{iopart}

\usepackage{graphicx}
\usepackage{float}
\usepackage[latin1, utf8]{inputenc}
\usepackage{hyperref}
\usepackage{dsfont}
\usepackage{caption}
\usepackage{subcaption}
\usepackage{color}
\usepackage{multirow}  
\usepackage{algorithm}
\usepackage[noend]{algpseudocode}

\newcommand{\sltc}{{$SL(2,{C})$}}
\newcommand{\SUT}{{$SU(2)$}}

\newcommand{\slnext}{\texttt{sl2cfoam-next}}

\newcommand{\eqref}[1]{[\ref{#1}]}

\newcommand{\nn}{\nonumber}

\newcommand{\AG}{A_{\Gamma}}
\newcommand{\Ai}{A \left(   j , \{ i_n \} \right)}
\newcommand{\AiMC}{A \left(   j , [ i_n ] \right)}
\newcommand{\Aisquared}{A^2 \left(   j , \{ i_n \} \right)}
\newcommand{\Aip}{\AG (  j , \{ i_n' \} )}
\newcommand{\ia}{\{ i_n \}}
\newcommand{\iaMC}{\left[ i_n \right]}
\newcommand{\iap}{\{ i_n' \}}

\newcommand{\Nf}{\frac{1}{Z}}

\newcommand{\Wsix}[6]{\left \{ \begin{array}{ccc} #1 & #2 & #3 \\ #4 & #5 & #6 \end{array}\right \} }

\usepackage[colorinlistoftodos,prependcaption,textsize=footnotesize]{todonotes}

\begin{document}
\title[
Markov Chain Monte Carlo methods for graph refinement in Spinfoam Cosmology]{
Markov Chain Monte Carlo methods\\ for graph refinement in Spinfoam Cosmology}

\author{Pietropaolo Frisoni}
\ead{pfrisoni@uwo.ca}
\address{Dept.\,of Physics \& Astronomy, Western University, London, ON N6A\,3K7, Canada}

\author{Francesco Gozzini}
\ead{gozzini@thphys.uni-heidelberg.de}
\address{
Institut f\"ur Theoretische Physik, Philosophenweg 16, 69120 Heidelberg, Germany}

\author{Francesca Vidotto}
\ead{fvidotto@uwo.ca}
\address{Dept.\,of Physics \& Astronomy, Dept.\,of Philosophy, and Rotman Institute, \\ Western University, London, ON N6A\,3K7, Canada}


\vskip2em
\begin{abstract}
\noindent We study the behaviour of the Lorentzian Engle-Pereira-Rovelli-Livine spinfoam amplitude with homogeneous boundary data, under a graph refinement going from five to twenty boundary tetrahedra. This can be interpreted as a wave function of the universe, for which 
we compute boundary geometrical operators, correlation functions and entanglement entropy.  The numerical calculation is made possible by adapting the Metropolis-Hastings  algorithm, along with recently developed computational methods appropriate for the deep quantum regime. We confirm that the transition amplitudes are stable against such refinement.  We find that the average boundary geometry does not change, but the new degrees of freedom correct the quantum fluctuations of the boundary and the correlations between spatial patches. The expectation values are compatible with their geometrical interpretation and the correlations between neighbouring patches decay when computed across different spinfoam vertices. 
\end{abstract}

\maketitle 

\section{Introduction}    
\noindent The last years have seen a lively development of numerical methods in the covariant, or \emph{spinfoam},  formulation of Loop Quantum Gravity \cite{book:Rovelli_Vidotto_CLQG}.  A key step in this direction has been the introduction of Markov chain Monte Carlo (MCMC) methods and their application to the study of the EPRL propagator \cite{Spinfoam_on_a_Lefschetz_thimble}.  In \cite{Spinfoam_on_a_Lefschetz_thimble} the authors exploited the known properties of the single-vertex semiclassical limit to perform a Monte Carlo sampling over particular subspaces of the complexified parameter space, using semiclassical boundary states \cite{Bianchi_coherent_spin_networks}. Exploiting the stationary phase approximation \cite{Bianchi_propagator}, the authors found a good agreement with the semiclassical results
obtained via analytical methods. 

The evaluation of spinfoam amplitudes in the Engle-Pereira-Rovelli-Livine model (EPRL) \cite{flatness_ex_marsigliesi, Dona2019, self_energy_paper, review_numerical_LQG}, the introduction of effective models \cite{Effective_spinfoam_1, Effective_spinfoam_2} and the numerical study of cuboid renormalization \cite{Steinh_cuboidal_renorm} have shed considerable light on several aspects of the theory, such as the role of the Immirzi parameter, the accidental flatness constraints and the refinement limit.
 
This paper introduces a technique that can be applied to some calculations in the regime where the number of degrees of freedom is large, but the relevant spin quantum numbers are small. We combine the Metropolis-Hastings algorithm \cite{MH_original_paper} with some recently developed high performance computing techniques in covariant LQG \cite{Francesco_draft_new_code}. Not being based on analytical approximations, this method requires minimal knowledge of the spinfoam geometry. On the other hand, the algorithm becomes source-demanding as the complexity of the spinfoam increases. We test the algorithm in the case of single 4-simplex, and then we use it to study a spinfoam with six vertices in the bulk and twenty nodes on the boundary. This corresponds to the cellular decomposition obtained from one elementary 4-simplex by splitting each of the five boundary tetrahedra into four tetrahedra. The resulting spinfoam is a refinement of the 4-simplex vertex which does not add any internal (dynamical) face to the spinfoam two-complex. In the following, we refer to it as the ``star" spinfoam.  We restrict the calculation to the homogeneous sector where the spins of all boundary links have the same value. In this sector, the spinfoam degrees of freedom are given by the boundary intertwiners, which encode the shapes of the boundary tetrahedra. We compute the amplitude as a function of these variables numerically, and use the Monte Carlo sampling to study expectation values of different boundary operators and their correlations. For the sake of completeness, we investigate both the BF and EPRL models. 

Spinfoam amplitudes with on the boundary a regular graph and homogeneous data can be interpreted as cosmological states \cite{Bianchi:2010zs}
In particular, when a single boundary states is considered, the amplitude can be seen as a transition from nothing into a 3-dimensional geometry, compatible with the 4-dimensional Lorentzian bulk \cite{Vidotto:2011qa,Vidotto:2015bza}. More precisely, the computed amplitude is a truncation of the spinfoam vertex expansion of the \emph{nothing-to-geometry} transition amplitude. This provides a spinfoam Lorentzian version of the Hartle-Hawking wave function of the universe \cite{Hartle:1983ai}. We refer to the literature in spinfoam cosmology for the physical interpretation of these states \cite{Vidotto:2010kw,Roken:2010vp,Bianchi:2011ym,Hellmann:2011jn,Kisielowski:2011vu,Livine:2011up,Kisielowski:2012yv,Rennert:2013pfa,Rennert:2013qsa,Gielen:2013kla,Gielen:2013naa,Gielen:2016dss,Vilensky:2016tnw,Gielen:2014uga,Bahr:2017eyi}.   

These states are rich enough to describe a boundary geometry that is regular on average but allows quantum fluctuations.  
The recent advance in numerical methods applied to the computation of spinfoam amplitudes have opened the possibility to compute concrete observables with spinfoam cosmological states. 
The first was computation of this kind was introduced in \cite{Gozzini_primordial} using a single 4-simplex. In this paper we would like to focus on the numerical methods. The cosmological spinfoam states provide an interesting framework to investigate novel techniques.

A key open question in the spinfoam approach to quantum gravity is the convergence of the amplitudes under refinement of the two complex of the spinfoam. In this regard, we are able to confirm numerically that the refinement studied is stable for the boundary observable, in the sense that there is excellent agreement between expectation values computed on the single vertex graph and on the refined graph. The correlations turn out to be different, as well as the quantum information entropy between different boundary nodes, reflecting the finer scale at which they become accessible.   The results provides a quantitative estimate of the quantum correlations between different spatial patches in the manifold boundary whose truncation is represented by the boundary spin network. 

The paper is organized as follows. In Section \ref{sec:boundary_state} we define the boundary state that we use in our analysis. In Section \ref{sec:MCMC} we discuss the Markov chain Monte Carlo method applied to the spin sums over the boundary degrees of freedom. In Section \ref{sec:The_4_simplex} we test the Monte Carlo sampler to the case of single 4-simplex, where calculations can be performed using deterministic approaches. In Section \ref{sec:The_star} we study the star spinfoam, investigating the numerical results of geometrical operators (boundary angles and volumes) and the related correlations. We also discuss the entanglement entropy between boundary nodes, considering different partitions into subsystems. 

The code used for all the computations described in this paper is available on GitHub \cite{Frisus_star_model_repo}\footnote{The code works on any operating system with an updated version of the Julia programming language (the code was tested with Julia 1.6.2).}
All the computations were performed on Compute Canada's Cedar, Graham, Beluga and Narval clusters (\href{https://www.computecanada.ca/}{www.computecanada.ca}). The computational resources employed for this paper can be quantified as approximately $8 \cdot 10^4$ CPU hours.
\section{The boundary state}
\label{sec:boundary_state}
We study the boundary state $| \psi_0 \rangle$ introduced in \cite{Gozzini_primordial}, which studied the simplest triangulation of a 3-sphere, emerging from a single 4-simplex. In general, let $\Gamma$ be a graph with $L$  links and $N$ nodes. The LQG Hilbert space for the graph is:
\begin{equation}
\label{eq:Hilbert_space_LQG}
\mathcal{H}_{\Gamma} = L_2 \left[ SU(2)^L / SU(2)^N \right] \ .
\end{equation}
The spin network basis in $\mathcal{H}_{\Gamma}$ is made by the states $|\{ j_l \}, \{ i_{n} \} \rangle$
(from now on, we omit the $\Gamma$ subscript), where $\{ j_l \} $ is a set of half-integer spins and $ \{ i_n \} $ an intertwiner set, $n = 1 \dots N$,  $l = 1 \dots L$. An intertwiner $i_n$ is a basis element of the invariant subspace of the tensor product of 4 $SU(2)$ representations at the node $n$. In the following we fix all the spins to be equal, namely $j_l = j$. We denote a boundary spin network state of this reduced space as: 
\begin{equation}
\label{eq:ia_def}
|j, \{ i_n \} \rangle \equiv |j, i_1 \dots i_N \rangle = |j, i_1 \rangle \otimes \dots \otimes |j, i_N \rangle  \ ,
\end{equation}
suppressing the curly brackets for the spin label $j$, as there is one common spin attached to all the links. We define the state $| \psi_0 \rangle$ in the Hilbert space \eqref{eq:Hilbert_space_LQG} by 
\begin{equation}
\label{eq:amplitude_spin_int_basis_bra_ket}
\langle j, \ia | \psi_0  \rangle \equiv \Ai  \ ,
\end{equation}
where $\Ai$ is the LQG amplitude of the state in the spin network basis. The amplitude \eqref{eq:amplitude_spin_int_basis_bra_ket} can be interpreted as the amplitude associated to the transition nothing-to-$|j, \{ i_n \} \rangle$. Hence $| \psi_0 \rangle$ gives the natural state that is projected out of the empty state by the LQG dynamics. The amplitude function depend on the common spin $j$ on the links and on all the $N$ intertwiner indices. The state $| \psi_0 \rangle$ is therefore defined as:
\begin{equation}
\label{eq:cosm_state}
| \psi_0 \rangle = \sum_{\ia} \Ai  | j, \ia \rangle \ .
\end{equation}
The sum is over all possible values of all the intertwiners in the set $\{ i_n \}$, compatible with triangular inequalities. If $j_l = j$ then every intertwiner $i_n$ can assume integer values between $0$ and $2j$. This gives a total of $(2j + 1)^N$ boundary basis elements that enter the sum \eqref{eq:cosm_state}. Following the geometrical interpretation of the covariant LQG phase space in terms of twisted geometries \cite{Freidel:2010aq} we might interpret the constraint $j_l = j$ as imposing strongly at the quantum level that all the areas of the faces of the boundary tetrahedra must be equal. The intertwiner degrees of freedom model the ``shape" of the boundary tetrahedra, and these are relational observables at given value $j$.  They are directly linked to the boundary 3d dihedral angles, as discussed in Section \ref{subsec:angle_operator}. The definition \eqref{eq:cosm_state} doesn't depend on the details of the triangulation. In fact, the triangulation determines how the amplitude $\Ai$ must be computed.
\subsection{Expectation values}
\label{subsec:Expectation_values}
We consider local geometrical operators acting on single boundary nodes of $\mathcal{H}$. For each operator, we specify the matrix elements in the basis states \eqref{eq:ia_def}. We start defining the normalized expectation value on the boundary state \eqref{eq:cosm_state} of an operator $O_k$, acting on the Hilbert space associated to the $k$-th node, as:
\begin{equation}
\langle O_k \rangle \equiv \Nf \langle \psi_0 | O_k | \psi_0 \rangle \ .
\end{equation}
The normalization factor is computed as:
\begin{equation}
\label{eq:normalization_factor}
Z \equiv \langle \psi_0 | \psi_0 \rangle  = \sum_{\ia} \Ai^2 \ . 
\end{equation}
From \eqref{eq:cosm_state} we write: 
\begin{equation}
    \langle \psi_0 | O_k | \psi_0 \rangle = \Nf \sum_{\ia}  \sum_{\iap} \Ai \Aip   \langle j, \iap | O_k | j, \ia \rangle \ .
\end{equation}
By using the orthogonality of the spin-network states \eqref{eq:ia_def} we find:
\begin{equation}
\label{eq:iap_o_ia}
  \langle j, \iap | O_k| j, \ia \rangle   = \delta_{i_1', i_1} \dots \langle j, i_k' | O_k | j, i_k \rangle \dots \delta_{i_N', i_N} \ ,
\end{equation}
therefore we conclude:
\begin{equation}
\label{eq:<On>}
\langle O_k \rangle = \Nf   \sum_{\ia} \sum_{i_k' = 0}^{2j} \Ai A \left(  j , \ia , i_k' \right) \langle j,  i_k' | O_k | j, i_k \rangle \ ,
\end{equation}
where $A \left(  j , \ia , i_k' \right)$ is defined as:
\begin{equation}
\label{eq:A_synthetic_exp}
 A \left( j , \ia , i_k' \right) \equiv A \left(  j , i_1 \dots i_k' \dots i_N \right) \ ,
\end{equation}
namely, the amplitude computed with $i_k'$ in place of $i_k$. Since $i_k \in \{ i_n \}$, the sum over $i_k$ is contained in the sum over the set $\{ i_n \}$. It is now straightforward to compute $\langle O_k O_m \rangle $, which turns out to be:
\begin{equation}
\label{eq:<OnOm>}
\hskip-25mm
\langle O_k O_m \rangle = \Nf \sum_{\ia} \sum_{i_k' = 0}^{2j} \sum_{i_m' = 0}^{2j} \Ai \AG \left( \{ j \}, \ia , i_k', i_m' \right) \langle j, i_k' | O_k | j, i_k \rangle  \langle j, i_m' | O_m | j, i_m \rangle \ ,
\end{equation}
where the meaning of $\AG \left( j, \ia , i_k', i_m' \right)$ is transparent by looking at \eqref{eq:A_synthetic_exp}. That is, we refer to the amplitude with $i_k'$ instead of $i_k$ and $i_m'$ in place of $i_m$. In the case of diagonal operators $D_k$ in the spin-network basis, equations \eqref{eq:<On>} and \eqref{eq:<OnOm>} become respectively:
\begin{equation}
\label{eq:<Dn>}
\langle D_k \rangle = \Nf \sum_{\ia} A^2 \left( j , \ia \right) \langle j, i_k | D_k | j, i_k \rangle \ ,
\end{equation}
\begin{equation}
\label{eq:<DnDm>}
\langle D_k D_m \rangle = \Nf \sum_{\ia} A^2 \left( j , \ia \right) \langle j, i_k | D_k | j, i_k \rangle  \langle j, i_m | D_m | j, i_m \rangle \ .
\end{equation}
Normalized correlations are defined as:
\begin{equation}
C(O_k, O_m) = \frac{\langle O_k O_m \rangle-\langle O_k \rangle \langle O_m \rangle}{(\Delta O_k) \ (\Delta O_m)} \ ,
\label{eq:correlations}
\end{equation}
where the quantum spread is:
\begin{equation}
\Delta O_k = \sqrt{\langle O_k^2 \rangle-\langle O_k \rangle^2} \ .
\label{eq:spread}
\end{equation}
The fact that the connected correlation function \eqref{eq:correlations} between the nodes $k$ and $m$ is non-vanishing turns out to be a necessary condition in order to have correlated fluctuations between the shapes of the tetrahedra dual to nodes $k$ and $m$ \cite{Bianchi:2006uf,Livine:2006it,Alesci:2008ff,Bianchi:2009ri,Bianchi:2011hp}.
\section{Monte Carlo over intertwiner space}
\label{sec:MCMC}
As discussed in Section \ref{subsec:Expectation_values}, the expectation value of an operator requires to sum over all possible eigenstates of the quantum system. Numerically, this rapidly becomes intractable as the number of degrees of freedom increases. In the present context, for a graph with $N$ boundary intertwiners there are $(2j+1)^N$ values to compute and to sum. Suppose that the amplitude function $\Ai$ can be computed in $10^{-6}$ seconds on a reference hardware (the real time is typically orders of magnitude larger). For $20$ boundary tetrahedra, which is the case of the star spinfoam discussed in Section \ref{sec:The_star}, a spin $j = 2$ computation would take 3 years. Obviously, we cannot use blind summation if we want to approach this problem numerically. Clearly even parallelizing the computation on multiple machines cannot solve this issue in the case of many boundary degrees of freedom. \\ 
A solution is Monte Carlo summation. This is a technique that it is used to compute expectation values of random variables. We obtained the best results by adapting the Metropolis-Hastings algorithm \cite{MH_original_paper,MH_review} to the discrete sums over the boundary intertwiners. For the sake of completeness, we briefly describe the Metropolis-Hastings algorithm\footnote{We refer to the original paper \cite{MH_original_paper} or to the numerous texts available for a deeper description of the algorithm.} in its general form.

\subsection{Metropolis-Hastings algorithm}
\label{subsec:M-H_review}
Let's consider a quantity $O$ which can be computed as:
\begin{equation}
\label{eq:Metropolis_O_general_form}
O = \sum\limits_{x \in \chi} \tilde{f}_{\chi} \left( x \right) o \left( x \right) \ ,    
\end{equation}
where $x$ is a (possibly multidimensional) discrete variable on a state space $\chi$ which must be summed over, while $\tilde{f}_{\chi}$ is a probability distribution function on $\chi$, so that:
\begin{equation}
\sum\limits_{x \in \chi} \tilde{f}_{\chi} \left( x \right) = 1 \ .
\end{equation}
From now on, we define $\tilde{f}_{\chi}$ as the target distribution and we omit the state space label $\chi$. Since the target distribution is normalized, we write:
\begin{equation}
\label{eq:target_distr}
\tilde{f} \left( x \right) \equiv \frac{f \left( x \right)}{\sum\limits_{x} f \left( x \right)} \ ,
\end{equation}
from which:
\begin{equation}
O = \frac{ \sum\limits_{x} f \left( x \right) o \left( x \right) }{\sum\limits_{x} f \left( x \right)} \ .
\end{equation} 
If the target distribution \eqref{eq:target_distr} can be computed up to a multiplying constant, the Metropolis-Hastings algorithm allows to construct on state space $\chi$ an ergodic Markov chain with length $N_{MC}$: $x_{1} , \ x_{2} , \ x_{3} \dots , x_{n} \ \dots , x_{N_{MC}}$ such that $x_{n}$ is converging (in distribution) to $\tilde{f}$, exploring the space $\chi$ progressively. If we define:
\begin{equation}
O_{N_{MC}} = \frac{1}{N_{MC}} \sum\limits_{n = 1}^{N_{MC}} o \left(x_n \right) \ ,
\end{equation}
then, since the chain can be considered as a statistical sample, the law of large numbers ensures that:
\begin{equation}
\lim_{N_{MC} \to \infty} O_{N_{MC}}  = O \ .
\end{equation}
The computation is stochastic in nature and the correct result is found only in the limit of an infinite number of samples. This allows us to write:
\begin{equation}
O_{N_{MC}} \approx O \hspace{3mm} \textrm{for $N_{MC} \gg 1$} \ .
\end{equation}
That is, we obtain an estimate of the original sum \eqref{eq:Metropolis_O_general_form}. The soundness of the procedure comes from known theorems on Monte Carlo summation and we can estimate the error done by comparing many different runs. Since the simulation is Markovian and the chain itself can be considered as a statistical sample, the latter usually depends on the starting value. In the following, we choose randomly the starting point of the Markov chain. The initial steps, while the chain is in the thermalization phase, are typically removed as burn-in iterations. In order to transit from the chain state $x_{n}$ to $x_{n + 1}$, we require a proposal distribution $q$ defined on space $\chi$. If $q$ is positive everywhere, then the Metropolis-Hastings algorithm preserves $\tilde{f}$ as the stationary distribution to which the chain is progressively converging. In the random walk Metropolis-Hastings, the proposal distribution consists in a local exploration of the neighborhood of the current value $x_n$ of the Markov chain. That is, the proposed value $x_n'$ is simulated as:
\begin{equation}
x_n' = x_n + \delta x_n \ ,
\end{equation}
where $\delta x_n$ is a random perturbation with distribution $g$. That is, the proposed state $x_n'$ is sampled from a probability distribution $g \left( x_n' | x_n \right)$, which suggests a candidate given the previous sample value $x_n$. As proposal distribution, we choose a truncated normal distribution rounded to integers centered around $x_n$ with standard deviation $\sigma$:
\begin{equation}
\label{eq:proposal_explicit}
g \left( x_n' | x_n \right) =  \mathcal{N}_{d,t}(x_n, a, b; \sigma) \ ,
\end{equation}
where the definition of $\mathcal{N}_{d,t}$ is reported in \ref{app:truncated_gaussian}. The full algorithm is summarized in the flowchart \ref{numericalcode}, in which we report the steps in order to implement the random walk Metropolis-Hastings and build the Markov chain.
\begin{algorithm}[tb]
\caption{Random walk Metropolis-Hastings}\label{numericalcode}
\begin{algorithmic}[1]
\State Choose the number of iterations $N_{MC}$, the burnin parameter $b$ and the standard deviation $\sigma$ of the proposal distribution $g$.
\State Set a random initial configuration $x$ and compute $f \left( x \right)$
\State Set initial multiplicity to $1$
\For{$n = 1 \dots N_{MC}$}
\State Generate a candidate $x'$ from $x$ according to the proposal distribution $g$ 
\If{$x' = x$} 
\State Increase the multiplicity by $1$
\State \textbf{continue}
\Else
\State Compute $f \left( x' \right)$ 
\State Compute $p = \textrm{min} \Bigl\{ 1 , \frac{f \left( x' \right)}{f \left( x \right)} \frac{g \left( x | x' \right)}{g \left( x' | x \right)} \Bigl\} $
\State Generate a uniform random number $r$ between $0$ and $1$
\If{$r < p$} 
   \If{$n > b$} 
      \State Store $x$ and $f(x)$ with the corresponding multiplicity
   \EndIf 
\State Set $x = x'$, $f(x) = f(x')$   
\State Set the multiplicity to $1$
\Else
   \State Increase the multiplicity by $1$
\EndIf 
\EndIf
\EndFor
\State Dump to disk the $x$'s, $f(x)$'s and the corresponding multiplicities.
\end{algorithmic}
\end{algorithm}
The multiplicity factors and the storage of $f(x)$'s have been introduced just as a matter of efficiency. In fact, this considerably speeds up the algorithm and the consequent computation of operators. Technically, the Markov chain obtained at the end of the algorithm \ref{numericalcode} has a length less than $N_{MC}$ (as this depends on the acceptation ratio). Since it is sufficient to take into account the multiplicity of each single chain state in order to restore the original length, in the following we refer to the Markov chain obtained at the end of the algorithm \ref{numericalcode} as having length $N_{MC}$ without losing any generality. \\ 
Since $x_n$ depends on the previous element along the Markov chain, this induces a non-zero correlation between $x_n$ and $x_{n+d}$. The correlation between $x_n$ and $x_{n+d}$ is defined as the autocorrelation at lag $d$. For a Markov chain that converges to a stationary distribution, the autocorrelation should indeed decrease as the lag increases. Although the most common approach is to evaluate the autocorrelation of operators, a measure of the degree of autocorrelation of the Markov chain is represented by the autocorrelation of the sequence of the amplitudes of the states. If we define the average of the states amplitude as:
\begin{equation}
\bar{f}(x)  = \frac{1}{N_{MC}} \sum\limits_{n=1}^{N_{MC}} f(x_n) \ ,
\end{equation}
the definition of the autocorrelation function at lag $d$ associated with the sequence $f(x_1) \dots f(x_n)$ can be written as:
\begin{equation}
\label{eq:autocorrelation}
\textrm{ACF} \left( x_1 \dots x_{N_{MC}} ; d \right) = \frac{\sum\limits_{n=1}^{N_{MC} - d} \left( f(x_n) - \bar{f}(x) \right) \left(  f(x_{n+d}) - \bar{f}(x) \right)}{\sum\limits_{n=1}^{N_{MC}} \left( f(x_n) - \bar{f}(x) \right)^2 } \ .
\end{equation}
In order to obtain an unbiased estimate the statistical fluctuations due to the Monte Carlo sampling, we can compare the results of operator's expectation values over different runs. That is, we store multiple Markov chains according to algorithm \ref{numericalcode}, computing operators for each one of them. This is extremely useful in determining the convergence of the Markov Chain to the stationary distribution and the corresponding unbiased statistical dispersion of the operators. This is discussed in Section \ref{subsec:Exp_values_with_MC}.
\subsection{Expectation values with Monte Carlo}
\label{subsec:Exp_values_with_MC}
In order to apply the Metropolis-Hastings algorithm discussed in Section \ref{subsec:M-H_review} to the computation of spinfoam observables of Section \ref{subsec:Expectation_values}, a direct comparison between equations \eqref{eq:<Dn>} and \eqref{eq:Metropolis_O_general_form} is enlightening. Namely, if we associate to the state space $\chi$ the intertwiners' boundary space \eqref{eq:ia_def}, so that $x = \{ i_n \}$, then the target distribution becomes:
\begin{equation}
\label{eq:squared_prop_amp}
\tilde{f} \left( x \right) = \frac{ A^2 \left( j , \ia \right)}{\sum\limits_{\ia} A^2 \left( j , \ia \right)} \ .
\end{equation}
Namely, we run a stochastic sampling routine that extract draws of intertwiners $[i_{1} \dots i_{N}]$ from their whole configuration space, according to the Markov chain. The proposal distribution corresponds to a discrete multivariate truncated normal distribution and each intertwiner is proposed sampling from an independent one-dimensional distribution. The $a$, $b$ parameters in \eqref{eq:proposal_explicit} for each intertwiner are $0$ and $2j$, respectively. After storing the intertwiner draws we can use them to compute expectation values of operators. We can summarize the introduction of the Monte Carlo with the following substitution in the formulas of the expectation values of the operators:
\begin{equation}
\label{eq:MC_sum}
\sum_{\ia} \Aisquared o \left( \ia \right) \approx  \sum_{\iaMC}  o \left( \iaMC \right) \ .
\end{equation}
In the right side of equation \eqref{eq:MC_sum}, the sum over the intertwiners is intended as the sum over the stored draws $[i_n] \equiv [i_1 \dots i_N]$ in which the intertwiners have a fixed value compatible with triangular inequalities. That is, we are no longer considering all the independent summations over the intertwiners. This hugely reduces the computational cost, making the computation feasible in the case of many boundary degrees of freedom. With \eqref{eq:MC_sum}, the normalization factor \eqref{eq:normalization_factor} becomes:
\begin{equation}
\label{eq:normalization_factor_MC}
Z \approx \sum_{\iaMC} = {\rm Number\ of\ MC\ iterations} \equiv N_{MC} \ .
\end{equation}
We can easily find the expression for the expectation values of non-diagonal operators by multiplying and dividing for $\Ai$ and then using \eqref{eq:MC_sum}, remembering \eqref{eq:normalization_factor_MC}. In fact, \eqref{eq:<On>} becomes:
\begin{equation}
\label{eq:<On>_MC}
\langle O_k \rangle \approx \frac{1}{ N_{MC}}   \sum_{\iaMC} \sum_{i_k' = 0}^{2j}  \frac{ A \left( j ,  \iaMC , i_k' \right) }{\AiMC } \langle j, i_k' | O_k | j, i_k \rangle \ .
\end{equation}
Equation \eqref{eq:<OnOm>} becomes:
\begin{equation}
\label{eq:<OnOm>_MC}
\langle O_k O_m \rangle \approx \frac{1}{ N_{MC}} \sum_{\iaMC} \sum_{i_k' = 0}^{2j} \sum_{i_m' = 0}^{2j}  \frac{ A \left(  j , \iaMC , i_k', i_m' \right) }{ \AiMC } \langle j, i_k' | O_k | j, i_k \rangle  \langle j,  i_m' | O_m | j, i_m \rangle \ .
\end{equation}
In case of diagonal operators, from \eqref{eq:<Dn>} and \eqref{eq:<DnDm>} we obtain:
\begin{equation}
\label{eq:<Dn>_MC}
 \langle D_k \rangle \approx \frac{1}{ N_{MC}} \sum_{\iaMC} \langle j , i_k | D_k | j, i_k \rangle \ ,
\end{equation}
\begin{equation}
\label{eq:<DnDm>_MC}
\langle D_k D_m \rangle \approx \frac{1}{ N_{MC}} \sum_{\iaMC} \langle j, i_k | D_k | j,  i_k \rangle \ . \langle j, i_m | D_m | j, i_m \rangle  \ .
\end{equation}
Notice that in the case of diagonal operator, it is not necessary to compute any amplitude except those necessary for sampling the draws of intertwiners. This makes the computation of diagonal operators several orders of magnitude faster than non-diagonal ones. We consider the numerical analysis up to the value $j = 6$ for the spins associated with the boundary links. In fact, this numerical approach is intended to be applied in the full quantum regime, i.e. when the spin quantum numbers are small and the semiclassical approximation is not valid. In addition, the computational complexity represented by increasing $j$ strongly depends on the type of the considered operator, as well as on the Metropolis-Hastings parameters. The value $j = 6$ allows to compute all the operators and correlation functions that we consider with a stable precision up to 3 significant digits. \\
Crucially, notice that this approach is not based on analytical approximations and requires a minimal knowledge of spinfoam geometry, which typically becomes quite complicated for models beyond the single 4-simplex. On the other hand, the price to pay is the calculation of the spinfoam amplitude at each iteration of the algorithm \ref{numericalcode}. Much effort has been devoted in recent times to the efficient computation of spinfoam amplitudes \cite{review_numerical_LQG, Francesco_draft_new_code, Dona2018}. At present, the best numerical framework to compute BF and EPRL vertex amplitudes, which can be seen as the elementary building blocks of more general triangulations, is the \slnext \ library \cite{Francesco_draft_new_code}. The calculation of the spinfoam amplitude with high performance computing techniques is the fundamental ingredient which allows to apply the Metropolis-Hastings algorithm discussed in \ref{subsec:M-H_review}. The most recent developments have made it possible to compute potentially divergent spinfoam amplitudes with many internal faces \cite{Dona_Frisoni_Ed_infrared, self_energy_paper}. This makes the algorithm presented in this paper a good candidate to be used in the case of spinfoams with a non-trivial dynamic structure. \\
After we have stored a number C of Markov chains, each with the same length $N_{MC}$ and Metropolis-Hastings parameters, we can compute the expectation value \eqref{eq:<On>_MC} of an operator $O_k$ for each chain $\langle O_k \rangle_1 \dots \langle O_k \rangle_C$ and then consider the corresponding average and standard deviation:
\begin{equation}
\mu_{\langle O_k \rangle} = \frac{\sum_{c=1}^{C} \langle O_k \rangle_{c} }{C}  \ ,
\end{equation}
\begin{equation}
\sigma_{\langle O_k \rangle} = \sqrt{\frac{\sum_{c=1}^{C} \left( \mu_{\langle O_k \rangle} - \langle O_k \rangle_{c} \right)^2 }{C}} \ .
\end{equation}
For each considered operator, we plot the corresponding gaussian distribution:
\begin{equation}
\label{eq:gaussian}
G(x; \mu_{\langle O_k \rangle}, \sigma_{\langle O_k \rangle}) = \frac{1}{\sigma_{\langle O_k \rangle} \sqrt{2\pi}} e^{- \frac{1}{2} \left( \frac{x - \mu_{\langle O_k \rangle}}{2 \sigma_{\langle O_k \rangle}} \right)^2} \ .
\end{equation}
The number of Monte Carlo iterations $N_{MC}$ for each Markov chain, the number C of averaged chains and the relevant parameters in the Metropolis-Hastings algorithm are all listed in tables in the \ref{app:M-H_parameters}. 

\section{The 4-simplex}
\label{sec:The_4_simplex}
\begin{figure}[H]
    \centering
    \includegraphics[width=10cm]{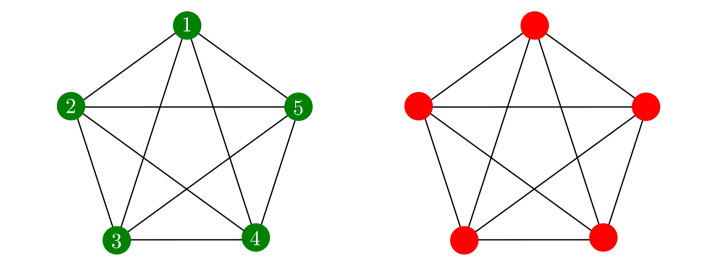}
     \caption{\label{fig:4_simplex_triangulation} Left: \textit{Geometry of the 4-simplex triangulation. Numbered circles correspond to points.} Right: \textit{The corresponding boundary spin network. Each red circle correspond to a boundary node and each line corresponds to a boundary link.}
     }
\end{figure}

The 4-simplex, or (referring to the dual triangulation) the vertex, is the simplest geometrical triangulation of the 3-sphere. It is formed by 5 tetrahedra glued on 10 faces. There are 5 boundary degrees of freedom and there are no internal faces. As show in Figure \ref{fig:4_simplex_triangulation}, there is a complete self-duality between the geometrical triangulation of the 4-simplex and the corresponding boundary graph.
The geometry in Figure \ref{fig:4_simplex_triangulation} doesn't depend on the Lorentzian character of the triangulation or not. Namely, it is the same for the $SU(2)$ topological BF or the Lorentzian EPRL 4-simplex. In order to emphasize the difference between the two models we need to explicitly represent the spinfoam associated with the amplitude. We denote such amplitudes associated with a single vertex as $V_{{\rm BF}}$ and $V_{EPRL}$. We write the corresponding expression in the general form, in which all spins have different values. As discussed in Section \ref{sec:boundary_state}, in the present context we focus on the case in which all spins are equal.
\subsection{The BF and EPRL vertex amplitudes}
\label{subsec:BF_andEPRL_vertex_amplitudes}
The vertex amplitude of the topological BF model can be written as an $SU(2)$ invariant Wigner $ \{ 15j \}$ symbol. The choice of the recoupling basis on each intertwiner determines if the symbol can be reduced to the product of lower-dimensional symbols. We choose the irreducible symmetric $\{15j\}$ symbol of \textit{first kind}, following the convention of \cite{GraphMethods}. The definition in terms of $\{ 6j\}$ symbols is:
\begin{eqnarray}
\label{eq:BF_vertex_amplitude}
\hskip-25mm
V_{{\rm BF}} \left(j_{ab}; \, i_a \right) &=&  (-1)^{\sum_{k=1}^5 j_k + i_k} \sum_s (2s + 1) 
\Wsix{i_1}{j_{25}}{s}{i_{5}}{j_{14}}{j_{15}} 
\Wsix{j_{14}}{i_{5}}{s}{j_{35}}{i_{4}}{j_{45}}
\Wsix{i_{4}}{j_{35}}{s}{i_{3}}{j_{24}}{j_{34}}
 \nn \\ 
&& \hspace{6cm }\times  
\Wsix{j_{24}}{i_{3}}{s}{j_{13}}{i_{2}}{j_{23}} 
\Wsix{i_{2}}{j_{13}}{s}{i_1}{j_{25}}{j_{12}} \nn \\ 
\hskip-25mm
&=&  \raisebox{-15mm}{ \includegraphics[width=3.5cm]{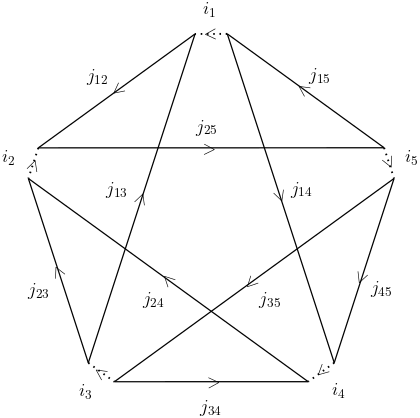}} \ ,
\end{eqnarray}
\\1em
\noindent
where $a,b = 1 \dots 5$, $a \neq b$. In order to avoid weighing down the notation, boundary dimensional factors attached to intertwiners $i_e$ and spins $j_{ab}$ have been neglected. The $ \{ 6j \}$ Wigner symbols in \eqref{eq:BF_vertex_amplitude} can be computed efficiently with libraries such as \texttt{wigxjpf} and especially \texttt{fastwigxj} \cite{Wigxjpf_library, fastwigxj_related}. \\ 
The EPRL vertex amplitude is built from the topological $SL(2,{C}))$ spinfoam vertex amplitude once that the simplicity constraints have been imposed \cite{Engle:2007wy,Engle:2007uq}. We use the form of amplitude originally derived in \cite{Speziale2016}, which results in a linear combination of $\{15j\}$ symbols weighted by one booster functions $B_4^{\gamma}$ per edge (see \ref{app:booster} for explicit formulas). We write the vertex amplitude according to the graphical notation discussed in detail in \cite{review_numerical_LQG}:
\begin{eqnarray}
\label{eq:vertex_amplitude}
\hskip-1mm
V_{EPRL}^{\gamma} \left(j_{ab} , \, i_a; \, \Delta l \right) & = \sum\limits_{l_{eq} = j_{eq}}^{ j_{eq} + \Delta l }  \sum\limits_{k_{e}} \left( \prod_{e} (2 k_e + 1) B_{4}^{\gamma} (j_{1e},l_{eq};i_{e}, k_{e}) \right)  V_{BF} \left( j_{1e}, l_{eq}; k_{e}, i_{a} \right) \nn  \hskip-5mm\\
& = \sum\limits_{ l_{eq} = j_{eq} }^{ j_{eq} + \Delta l} \raisebox{-16mm}{ \includegraphics[width=5.2cm]{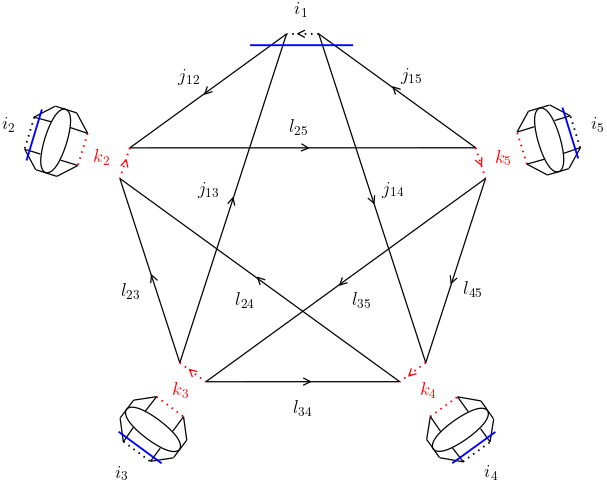}} \ ,
\end{eqnarray}
where $e,q = 2 \dots 5$, $e \neq q$. We introduced the $\Delta l$ parameter in order to truncate the summation over the auxiliary spins $l_{eq}$. The role of this truncation parameter in the context of infrared divergences has been deeply discussed in detail in  \cite{self_energy_paper, Dona_Frisoni_Ed_infrared, frisoni2021studying}. In this paper we consider $\Delta l = 20$ and $\gamma = 1.2$ as value of Barbero-Immirzi parameter. 
It is interesting to notice how the truncation parameter $\Delta l$ seems to play a minimal role in the computation of boundary observables, as already observed in \cite{self_energy_paper} in the case of infrared divergent amplitudes. In the present context, except for a slight systematic shift in the expectation values of the dihedral angles of the star model (discussed in Section \ref{subsec:angle_operator}), we found no difference\footnote{In order not to be redundant, we do not explicitly report the data in the case $\Delta l = 0$.} between the case $\Delta l=0$ and $\Delta l=20$. From a computational point of view, the advantage of using a small $\Delta l$ parameter is remarkable. The 4 spins associated with the gauge-fixed edge are $(j_{12}, j_{13}, j_{14}, j_{15})$, as the elimination of a redundant \sltc \ integration along one edge in the EPRL vertex \eqref{eq:vertex_amplitude} is necessary to ensure that the corresponding amplitude is well defined  \cite{Engle:2008ev}. As in \eqref{eq:BF_vertex_amplitude} we neglected the dimensional factors attached to the boundary intertwiners and spins.
\subsection{Testing the Monte Carlo sampler}
\label{subsec:testing_sampler}
\begin{figure}[H]
    \centering
        \begin{subfigure}[b]{75mm}
        \includegraphics[width=75mm]{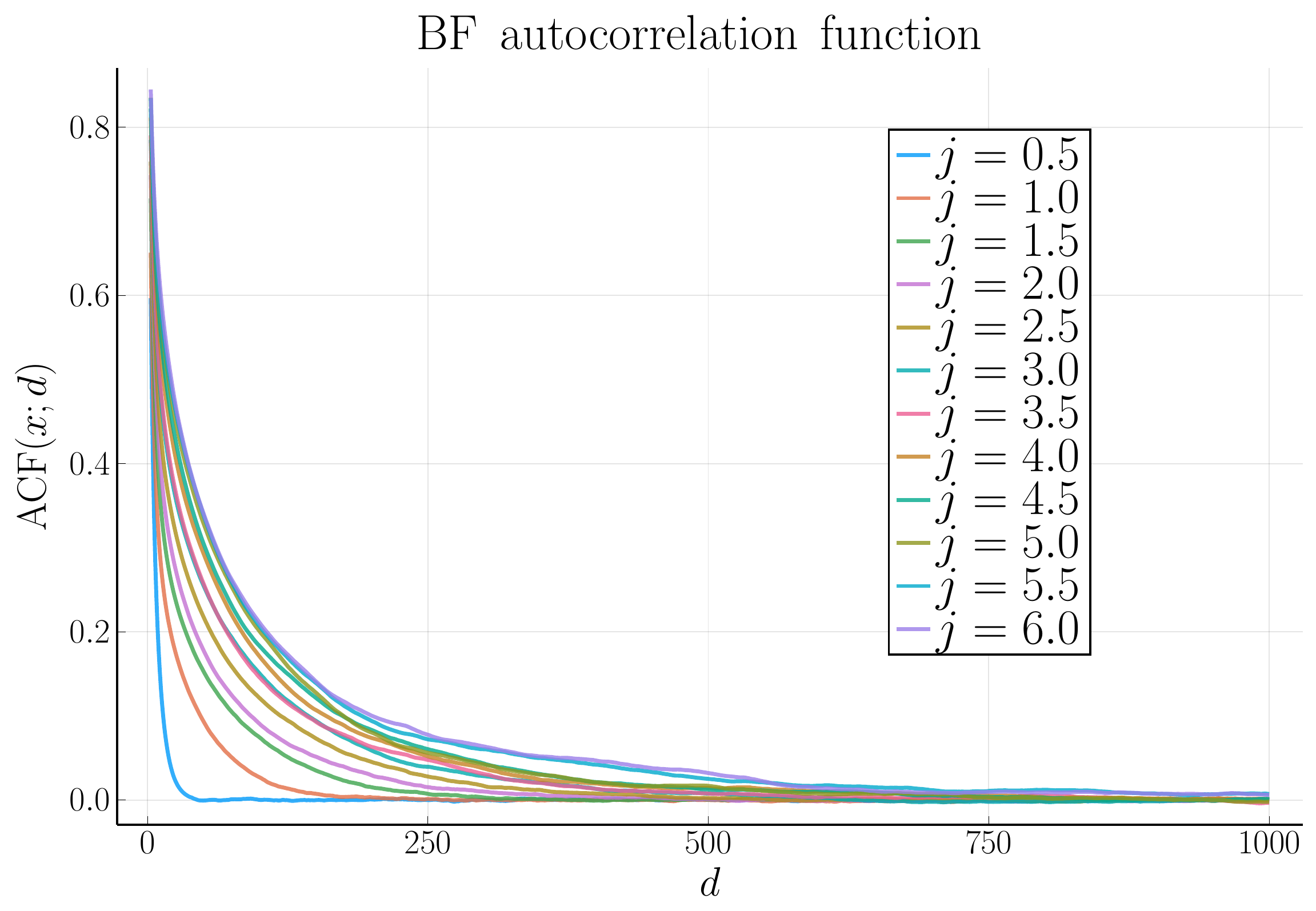}
     \end{subfigure}
    ~~~
        \begin{subfigure}[b]{75mm}
        \includegraphics[width=75mm]{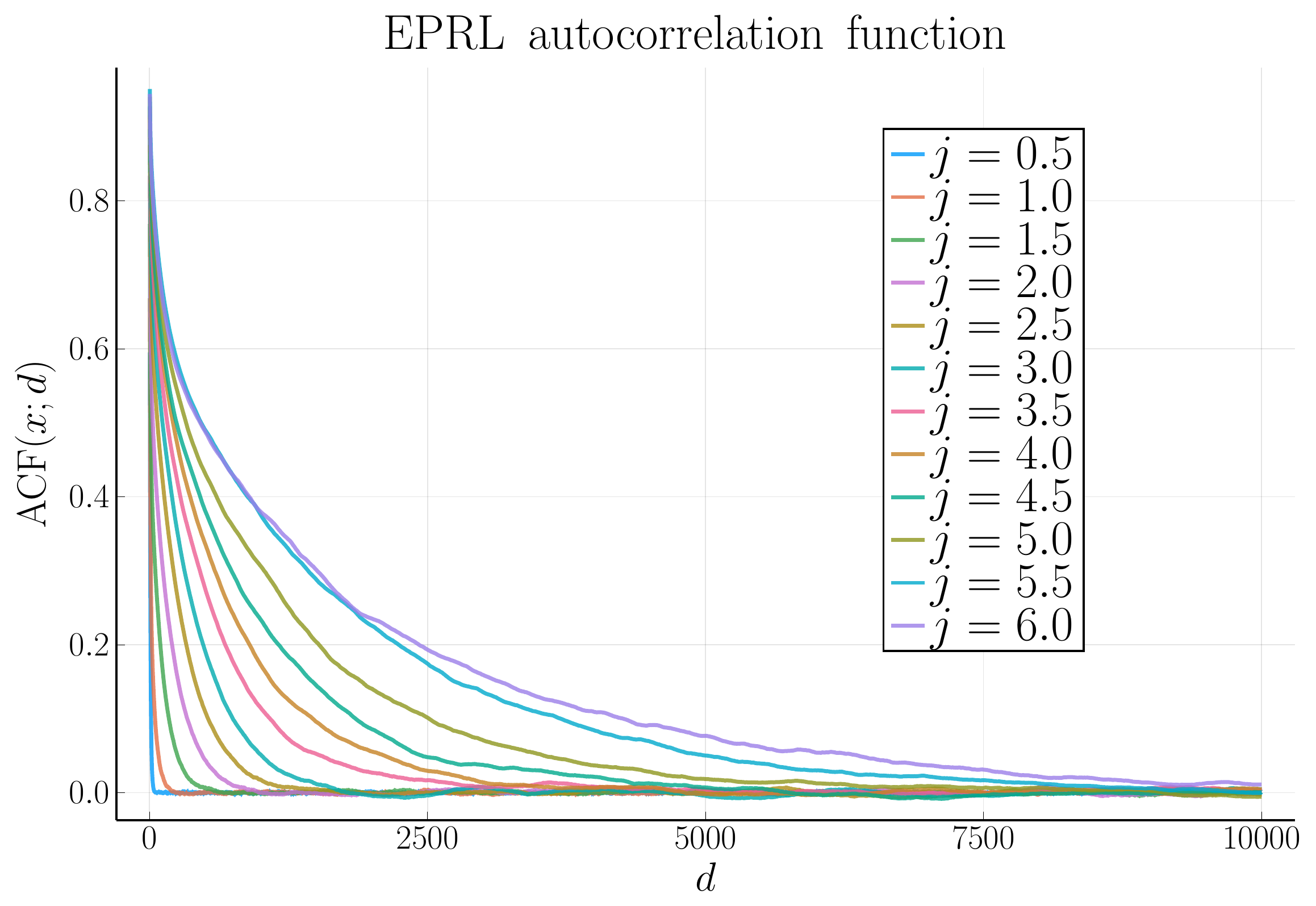}
    \end{subfigure}   
   \caption{\label{fig:single_vertex_autocorrelation} Autocorrelation function \eqref{eq:autocorrelation} of the BF and EPRL vertex amplitudes as a function of the lag $d$ for different values of $j$. As expected for a Markov chain that converges to a stationary distribution, the autocorrelation decreases as a function of the lag.}
\end{figure}

\begin{figure}[tb]
    \centering
        \begin{subfigure}[b]{75mm}
        \includegraphics[width=75mm]{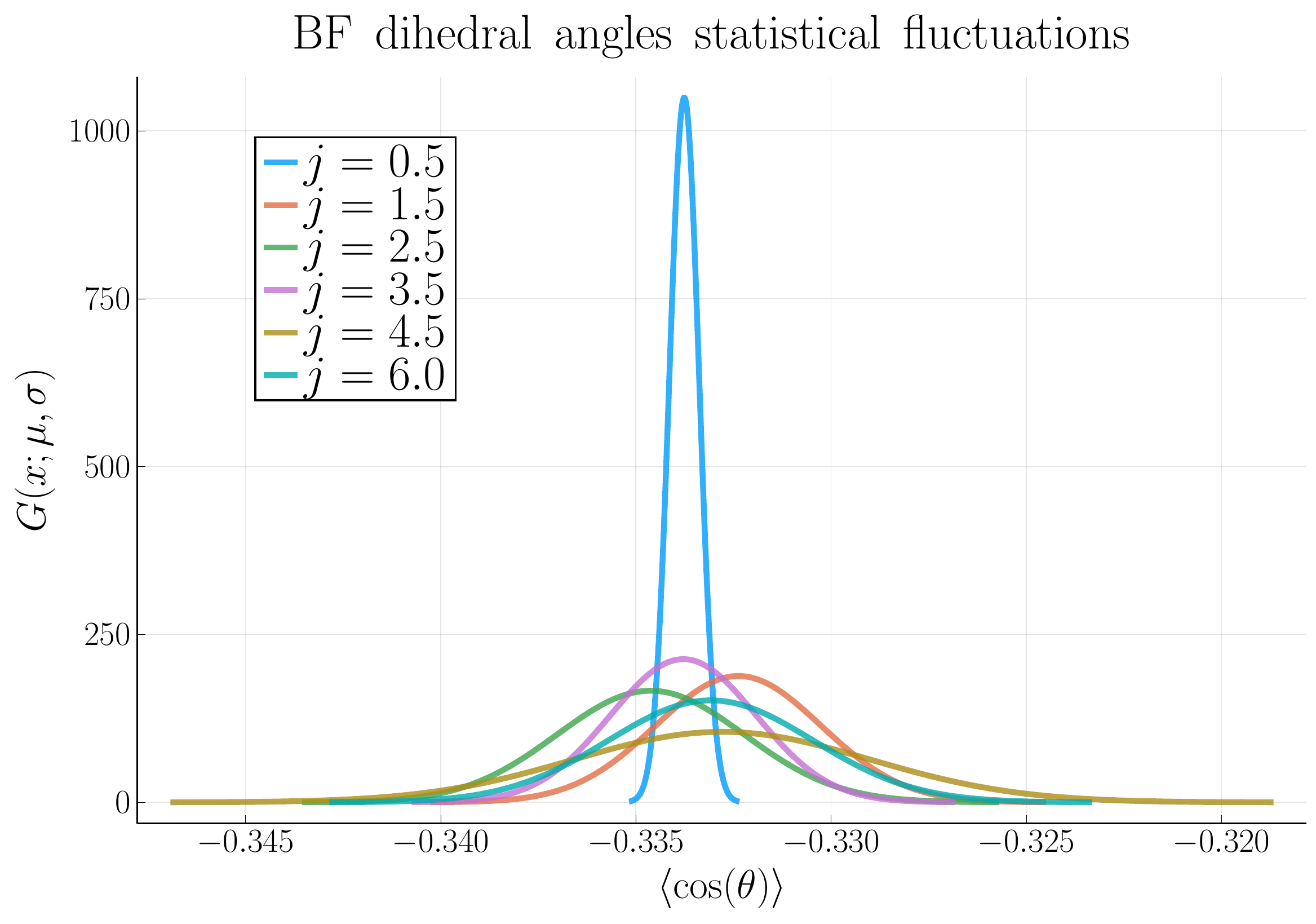}
     \end{subfigure}
    ~~~
        \begin{subfigure}[b]{75mm}
        \includegraphics[width=75mm]{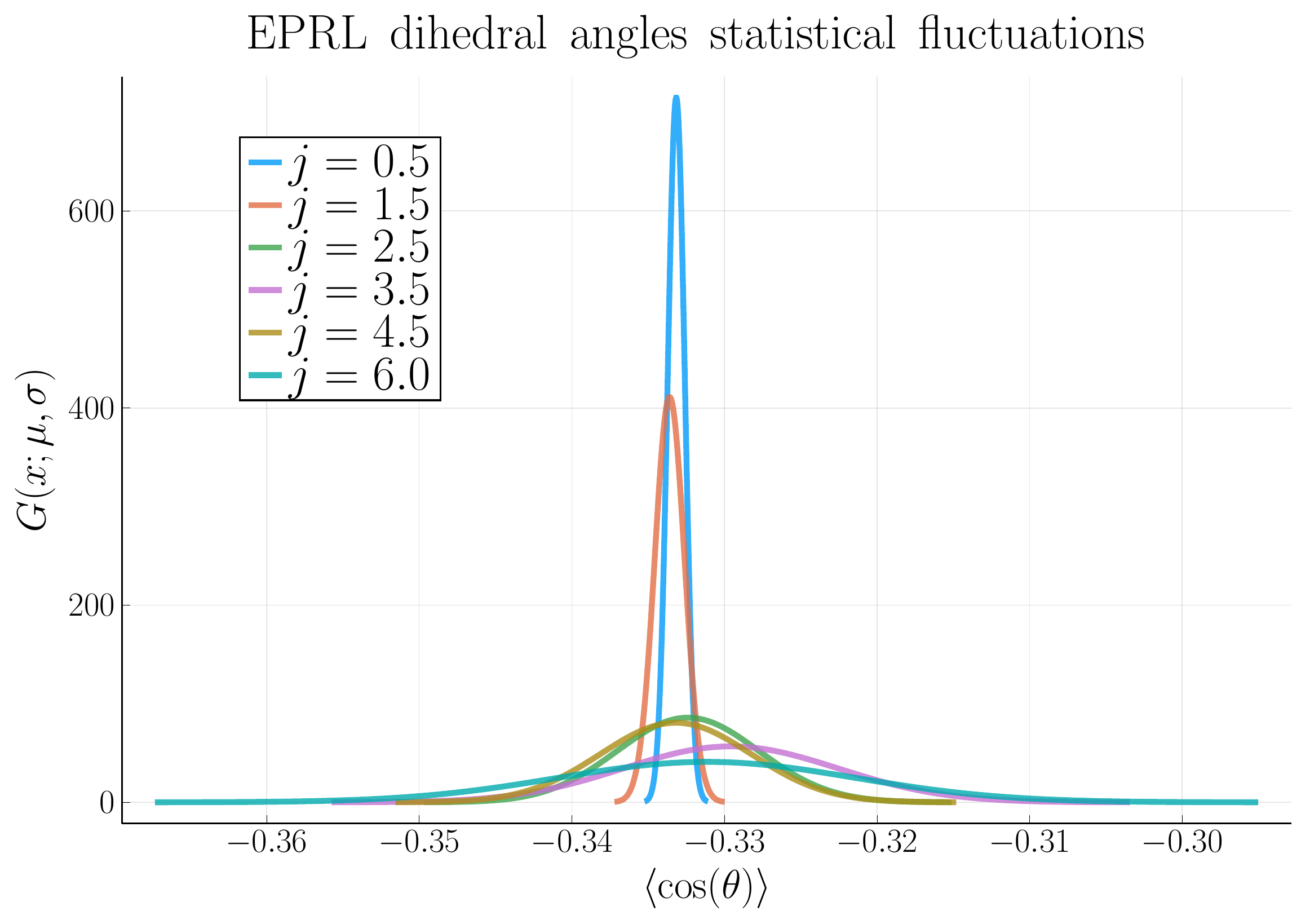}
    \end{subfigure}   
   \caption{\label{fig:single_vertex_angles_fluctuations} Gaussian distribution \eqref{eq:gaussian} of the expectation values \eqref{eq:<Dn>_MC} of the dihedral angle operator \eqref{eq:geom-angleformula} in the 4-simplex model. We averaged over several runs, computing the (average) angle $\mu_{<\cos{\theta}>}$ defined on a single node and the corresponding standard deviation $\sigma_{<\cos{\theta}>}$ for different values of $j$.}
    \end{figure}

The 4-simplex is an excellent model to test the sampling routine over the intertwiners space discussed in Section \ref{subsec:Exp_values_with_MC}. In fact, since the number $N=5$ of boundary degrees of freedom is low, it is possible to compute observables and related functions without resorting to Monte Carlo methods, allowing for a direct comparison. In order to perform the sampling algorithm it is necessary to compute the vertex amplitudes \eqref{eq:vertex_amplitude}-\eqref{eq:BF_vertex_amplitude} at fixed value of boundary spin $j$ and for all the possible $(2j+1)^5$ values of intertwiners. For the EPRL model, as mentioned in Section \ref{subsec:BF_andEPRL_vertex_amplitudes}, the computational time considerably increases with $j$, especially with a high value of the truncation parameter $\Delta l$. The computation of the EPRL vertex amplitude \eqref{eq:vertex_amplitude} for $j = 0.5 \dots 6$ has been distributed over several machines and hundreds of CPUs and it has taken about 5 days to complete. The amplitudes are available at the public repository \cite{Frisus_star_model_repo}, along with the corresponding BF counterparts \eqref{eq:BF_vertex_amplitude}. We start by looking at the autocorrelation function \eqref{eq:autocorrelation} of the vertex amplitudes sampled during the random walk. Obviously, in order to measure the autocorrelation we set $b=0$ in the flowchart \ref{numericalcode}, while all the other parameters are the same reported in \ref{app:M-H_parameters}. We report the data in Figure \ref{fig:single_vertex_autocorrelation}.
The results on the autocorrelation suggest to consider a burnin parameter $b \sim 10^2$ for the BF model and $b \sim 10^3$ for EPRL. Obviously, the most important parameter to verify the effectiveness of the sampler is the computation of observables. In this respect, a direct comparison is made easier by the fact that the external dihedral angle operator in this simple model has been studied in \cite{Gozzini_primordial}. The dihedral angle operator is the simplest operator to compute in the intertwiner basis, and it describes the cosine of the external dihedral angle $\cos(\theta_{ab})$ between two faces $a$ and $b$ of a boundary tetrahedron. Faces $a$ and $b$ depend on the recoupling basis chosen for the invariant $SU(2)$ $\{15j\}$ symbol, which appears both in the EPRL \eqref{eq:vertex_amplitude} and BF amplitude \eqref{eq:BF_vertex_amplitude}. The external dihedral angle of the tetrahedron dual to the node $n$ in the symmetry-reduced space basis states \eqref{eq:ia_def} is \cite{Gozzini_primordial}:
\begin{equation}
  \label{eq:geom-angleformula}
 \langle j, i_n  | \cos(\theta) | j, i_n \rangle =  \frac{i_n(i_n +1) - 2j(j+1) }{2 j(j+1)} \ .
\end{equation}
The dihedral angle operator \eqref{eq:geom-angleformula} is diagonal in the spin-network basis, therefore we can compute it very fast with equation \eqref{eq:<Dn>_MC}. We show the statistical fluctuations \eqref{eq:gaussian} of the expectation values \eqref{eq:<Dn>_MC} in the case of the dihedral angle operator \eqref{eq:geom-angleformula} in Figure \ref{fig:single_vertex_angles_fluctuations} for some values of $j$.
All the parameters used in the Metropolis-Hastings algorithm are reported in \ref{app:M-H_parameters}. The results show that the angle average corresponds to a regular tetrahedron, which is the result originally obtained in \cite{Gozzini_primordial} with deterministic calculations. It is clear how stochastic fluctuations in the random walk over intertwiner space tend to grow much faster for EPRL, rather than for BF, as boundary spin $j$ increases. This interesting behaviour is the main reason why we used a number of $N_{MC}$ iterations larger than one order of magnitude  in the analysis of the star model, as discussed in Section \ref{sec:The_star}. In order to have a stable precision up to the third significant digit, especially for large values of $j$, it is necessary to set a sampling number $N_{MC} \sim 10^6$, compared to $(2j + 1)^5$ sums to be carried out in the exact calculation \eqref{eq:<Dn>}. Although this indicates that the sampler works as expected, this makes the use of the random walk Metropolis-Hastings algorithm extremely inefficient in the case of single 4-simplex. However, as for other (Markov Chain) Monte Carlo methods, the advantage obtained in the case of many degrees of freedom emerges surprisingly, as we show in the case of the star spinfoam amplitude. This is due to the fact that the Metropolis-Hastings algorithm (as well as other MCMC methods) is not affected by the problem known as \textit{curse of dimensionality}. \\
As mentioned in Section \ref{sec:MCMC}, the statistical fluctuations in Figure \ref{fig:single_vertex_angles_fluctuations} are very useful in determining the convergence of the Markov Chain to the stationary distribution, as well as the dispersion of the operators. For the star model, which is the main element of analysis in the present context, we explicitly report the fluctuations \eqref{eq:gaussian} computed for all the considered operators and for the entanglement entropy.
\section{The star}
\label{sec:The_star}
\begin{figure}[H]
    \centering
    \includegraphics[width=12cm]{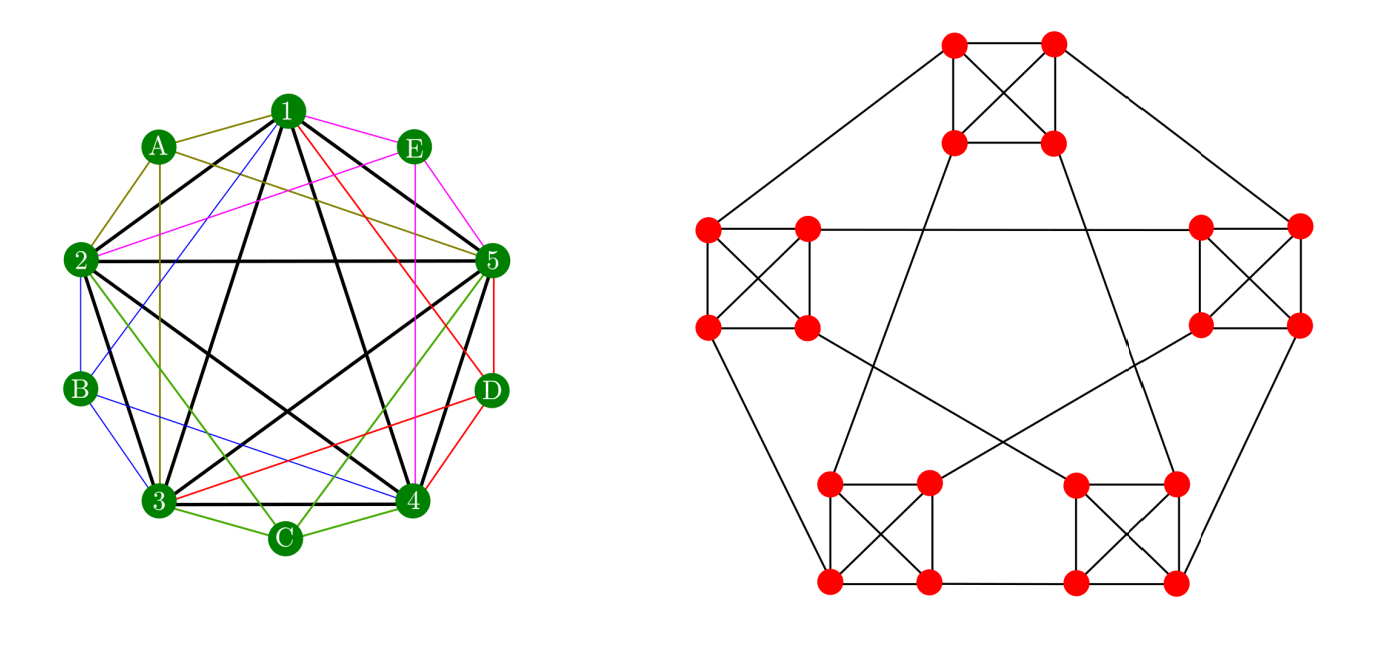}
     \caption{\label{fig:triangulation} Left: \textit{Geometry of the star triangulation. Numbered circles correspond to points and each letter is associated to a unique 4-simplex. Colored lines are shared by three tetrahedra belonging the the same 4-simplex.} Right: \textit{The corresponding boundary spin network. Each red circle correspond to a boundary node and each line corresponds to a boundary link.}
     }
\end{figure}

\begin{figure}[t]
    \centering
    \includegraphics[width=165mm]{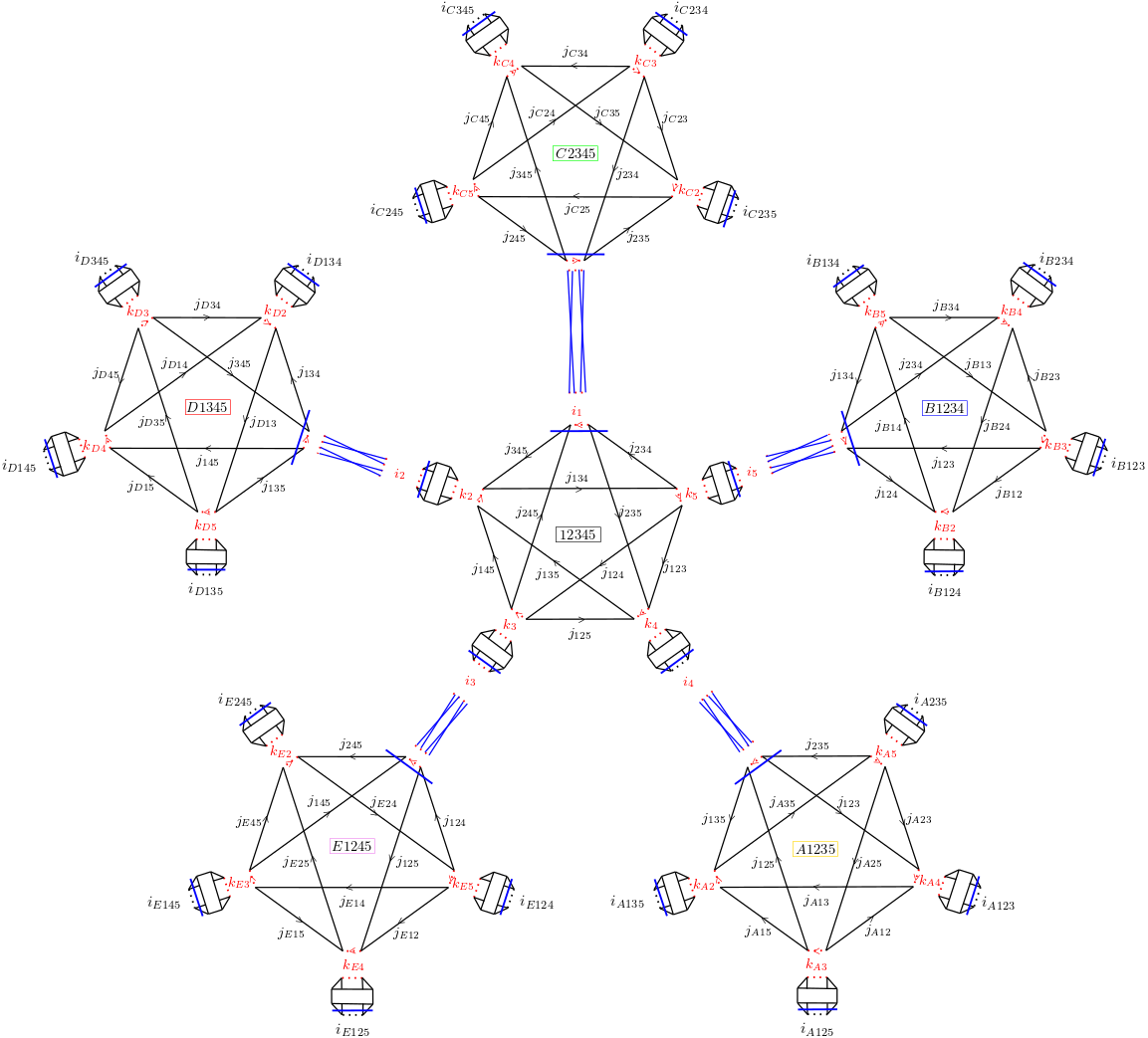}
    \caption{\emph{Graphical representation of the EPRL spinfoam amplitude corresponding to the {star triangulation geometry} described in Figure \ref{fig:triangulation}. Excluding the intertwiners connecting the booster functions with the $\{15j\}$ symbols in the same vertex, there are 5 internal intertwiners that need to be summed over.}} 
    \label{fig:star_spinfoam}
\end{figure}

After testing the algorithm discussed in Section \ref{sec:MCMC} to the 4-simplex case, we are ready to study the star spinfoam model, in which a computation without resorting to Monte Carlo methods would be impossible. The 2-complex of the star is composed by 6 vertices (one completely internal), 5 edges and has no internal faces. The boundary graph is a refinement of the 4-simplex graph obtained by splitting each of the 5 nodes into 4 nodes. The final result of this refinement process is that we obtain 20 nodes on the boundary, which correspond dually to 20 boundary tetrahedra. Therefore the full triangulated manifold is composed by five 4-simplices glued on 5 internal tetrahedra, each 4-simplex showing 4 tetrahedra on its boundary. The triangulation of the star graph is showed in Figure \ref{fig:triangulation}, along with the boundary spin network.
The triangulation of the star model is non-regular, since there are segments that are shared between 3 tetrahedra and other segments that are shared by 6 tetrahedra. Notice that the dual graph is a sort of ``magnification" of the dual 4-simplex. Iterating the same procedure we obtain a fractal structure.
The EPRL star amplitude is constructed starting from the vertex amplitudes \eqref{eq:vertex_amplitude}, shown in Figure \ref{fig:star_spinfoam}. 
The diagram for the BF is similar, with the difference that each vertex amplitude is simply given by \eqref{eq:BF_vertex_amplitude}. That is, with respect to the graphical amplitude in Figure \ref{fig:star_spinfoam}, there are no booster functions (and therefore no Y-map). In order not to be redundant, we do not report also the BF spinfoam diagram explicitly. The labels refer to the triangulation shown in Figure \ref{fig:triangulation}. The 4-simplices are labeled with one letter and four points. The boundary intertwiners are labeled by one letter and three points, which indicate the corresponding tetrahedron in the triangulation (as there is one intertwiner for each node). The links shared by three 4-simplices are labeled by three points, as they are dual to triangles. Those connecting two nodes belonging to the same 4-simplex are labeled with one letter and two points. Finally, the intertwiners connecting the booster functions with the $\{15j\}$ symbols are labeled with the position of the node in the corresponding 4-simplex. We can write the analytical expression of the amplitude associated with the star spinfoam EPRL and BF respectively as:
\begin{equation}
  \label{eq:EPRL_star_amplitude}
A_{EPRL}^{\gamma} \left( j, \, i_b , \, \Delta l \right) = \sum\limits_{i_1 \dots i_5}  V_{EPRL}^{\gamma}  \left(j, \, i_1 , \, i_2 ,\,  i_3 ,\,  i_4 , \,  i_5 ,\, \Delta l \right) \prod\limits_{a = 1}^{5} V_{EPRL}^{\gamma}  \left(j, \, i_a , \, i_b , \, \Delta l \right) \ ,
\end{equation}
\begin{equation}
  \label{eq:BF_star_amplitude}
A_{BF} \left( j, \, i_b \right) = \sum\limits_{i_1 \dots i_5} V_{BF} \left(j, \, i_1 , \, i_2 ,\,  i_3 ,\,  i_4 , \,  i_5 \right) \prod\limits_{a = 1}^{5} V_{BF} \left(j, \, i_a , \, i_b \right) \ ,
\end{equation}
where we used the expressions for the EPRL and BF vertex amplitude \eqref{eq:vertex_amplitude}-\eqref{eq:BF_vertex_amplitude}. The dependence on intertwiners for each vertex (apart from those on which it is necessary to sum over to assemble the amplitude) has been generically indicated with $i_b$ in order not to weight down the notation. The combinatorial structure of the spinfoam should be clear by looking at Figure \ref{fig:star_spinfoam}.
\subsubsection{A simple benchmark}
\begin{figure}[H]
    \centering
        \begin{subfigure}[b]{75mm}
        \includegraphics[width=75mm]{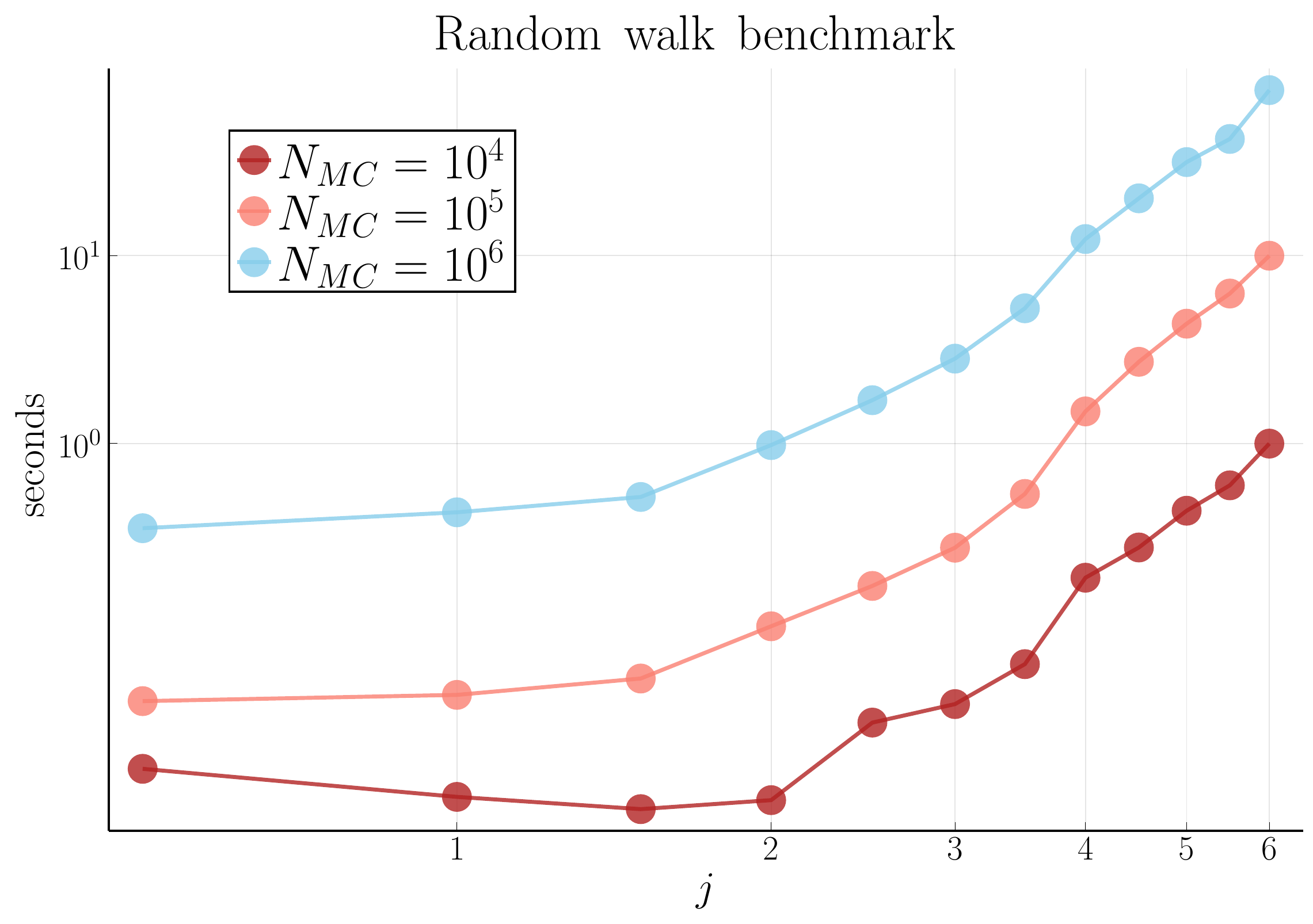}
    \end{subfigure}
   \caption{\label{fig:RW_benchmark} {\textit{Benchmark of the sampling and storage process of intertwiners draws in the random walk \ref{numericalcode} over the 20-dimensional intertwiners' space of the star spinfoam amplitude \ref{fig:star_spinfoam}. Computation time asymptotically scales as $\sim j^{4.5}$.}}}
   \end{figure}

Before discussing the expectation values, it is interesting to estimate the computational time of the Metropolis-Hastings algorithm, discussed in Section \ref{subsec:Exp_values_with_MC}, applied to the star spinfoam amplitude. As shown in equations \eqref{eq:EPRL_star_amplitude}-\eqref{eq:BF_star_amplitude} and in the flowchart \ref{numericalcode}, at each step of the Markov chain we need perform the contraction of the vertex amplitudes over 5 bulk intertwiners. This is undoubtedly the computationally most expensive part of the algorithm. For low spins, it is sufficient to perform the contraction with HPC techniques exploiting solely the CPU. We found the best performance using the \texttt{LoopVectorization} Julia package. The offloading of tensor contractions on the GPU with parallelization on the GPU cores \cite{Julia_GPU}, exploiting the recent tensor network techniques \cite{itensor}, will be implemented in future works. In fact, the best improvement is obtained for large values of the spins \cite{Francesco_draft_new_code}, making this approach more suitable for a study of the semiclassical limit of spinfoams rather than the quantum regime. We show a benchmark of the random walk sampling algorithm in Figure \ref{fig:RW_benchmark} for increasing values of the total number of iterations $N_{MC}$ in the Markov chain.
The sampling in Figure \ref{fig:RW_benchmark} has been carried out on a laptop with processor Intel(R) Core(TM) i7-10750H 2.60GHz. The acceptance rate of intertwiners draws has been set between $30\%$ and $33\%$, with a burn-in parameter $b = 10^3$. In the code available at the repository \cite{Frisus_star_model_repo} the Markov chains are automatically parallelized on the number of available CPUs, eventually distributing the computation on multiple machines. As discussed in Section \ref{subsec:Exp_values_with_MC}, building more Markov chains is useful for improving accuracy and estimating the error committed due to the statistical fluctuations of the random walk.
\subsection{Numerical results: operators}
\label{subsec:numerical_results}
We now describe the numerical values obtained for the expectation value of local geometric operators \eqref{eq:<On>} with the boundary state \eqref{eq:cosm_state} for the star spinfoam amplitude. We also compute the quantum spread \eqref{eq:spread} and correlation functions \eqref{eq:<OnOm>} between different nodes. For each geometrical operator, we discuss the results obtained both with the BF model and the EPRL model. The parameters used for the sampling of the draws employed for the computation of the operators are discussed in detail in  \ref{app:M-H_parameters}. 
\subsubsection{The dihedral angle operator}
\label{subsec:angle_operator}
\begin{figure}[H]
    \centering
        \begin{subfigure}[b]{75mm}
        \includegraphics[width=75mm]{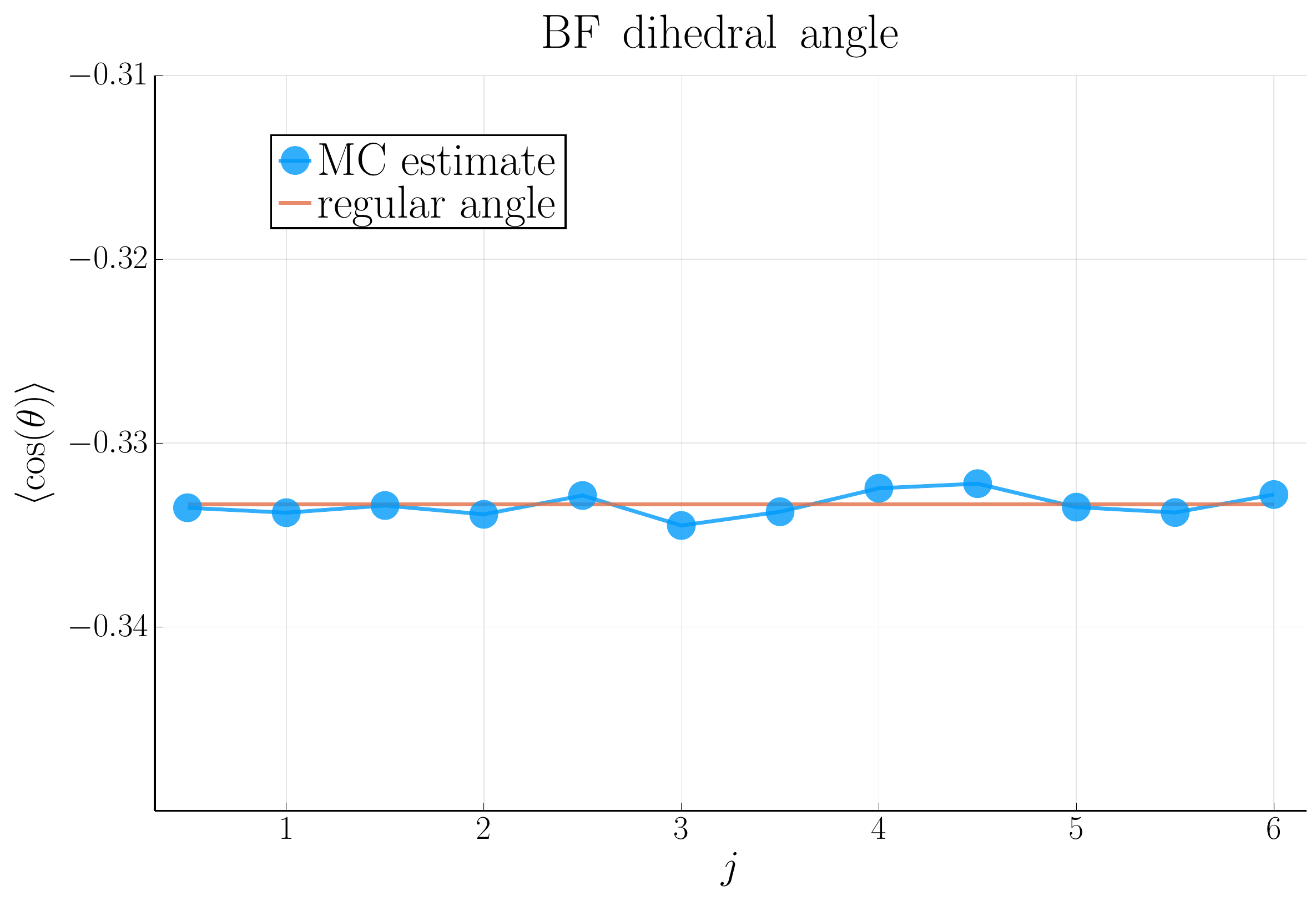}
     \end{subfigure}
    ~~~
        \begin{subfigure}[b]{75mm}
        \includegraphics[width=75mm]{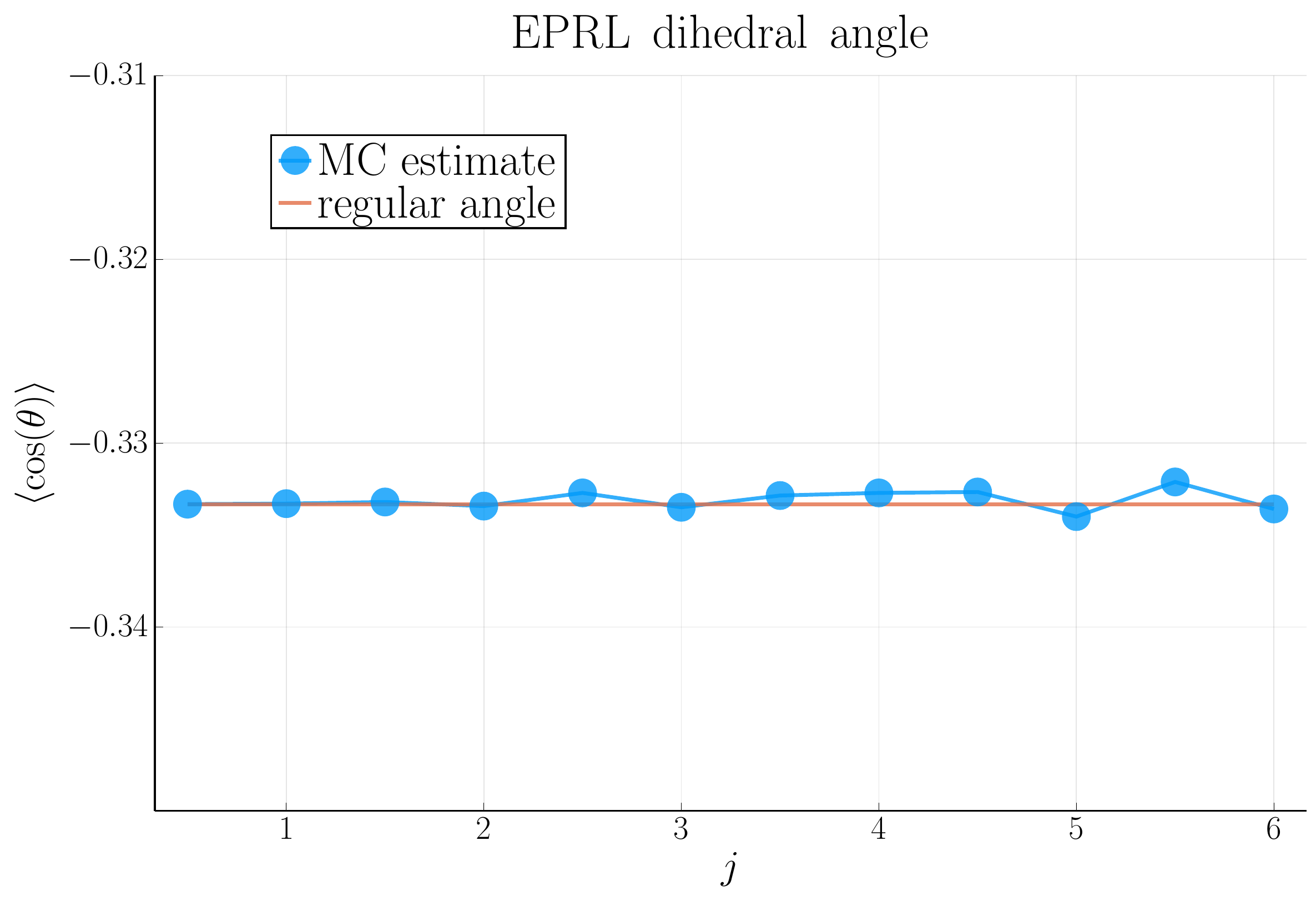}
    \end{subfigure}    
    \caption{\label{fig:angles}\textit{Expectation values \eqref{eq:<Dn>_MC} of the dihedral angle operator \eqref{eq:geom-angleformula}, averaged over all the $20$ nodes of the spinfoam. The orange line shows the value of the cosine of dihedral angle of a regular tetrahedron, which is $\cos (\theta_{{\rm regular}}) = -0.\bar{3}$.}}    
    \end{figure}

\begin{figure}[b]
     \centering
        \begin{subfigure}[b]{75mm}
        \includegraphics[width=75mm]{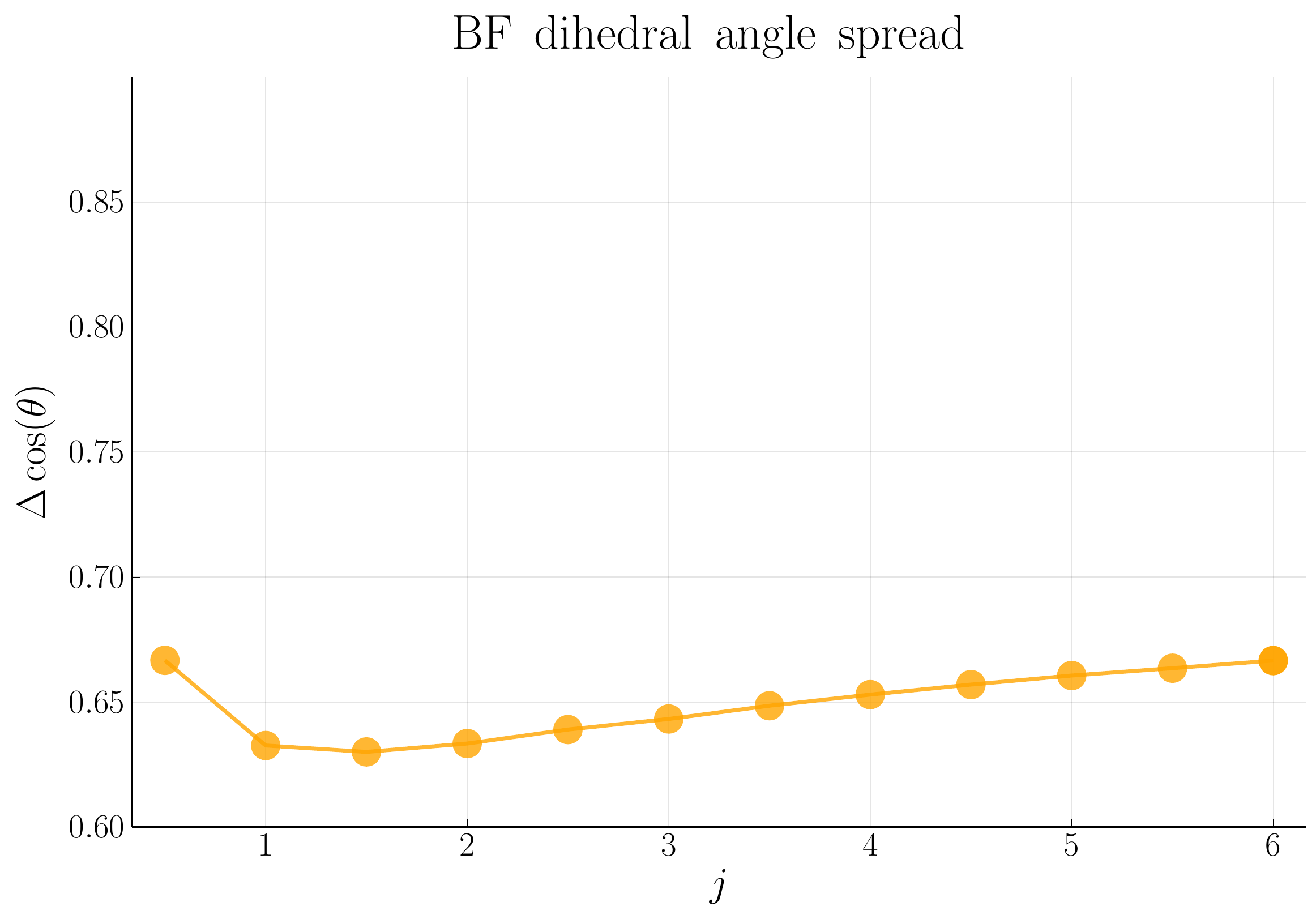}
     \end{subfigure}
    ~~~
        \begin{subfigure}[b]{75mm}
        \includegraphics[width=75mm]{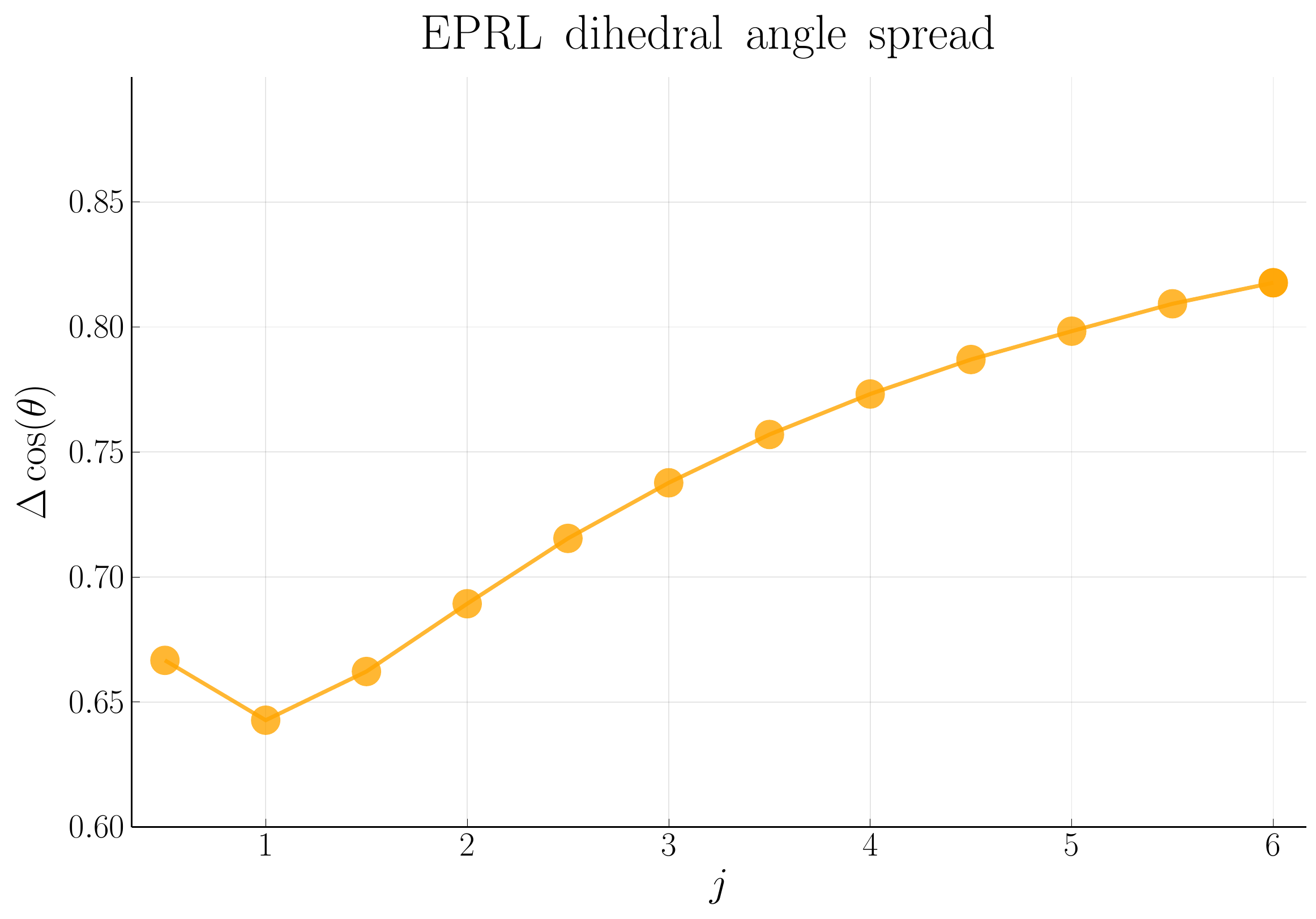}
    \end{subfigure}  
    \caption{\label{fig:angles_spread}\textit{Expectation values of the spread \eqref{eq:spread} for the dihedral angle operator \eqref{eq:geom-angleformula}, averaged over all the $20$ nodes of the spinfoam. In the EPRL model, the quantum spread increases faster as a function of the boundary spin $j$.}}
    \end{figure}

\begin{figure}[t]
    \centering
        \begin{subfigure}[b]{75mm}
        \includegraphics[width=75mm]{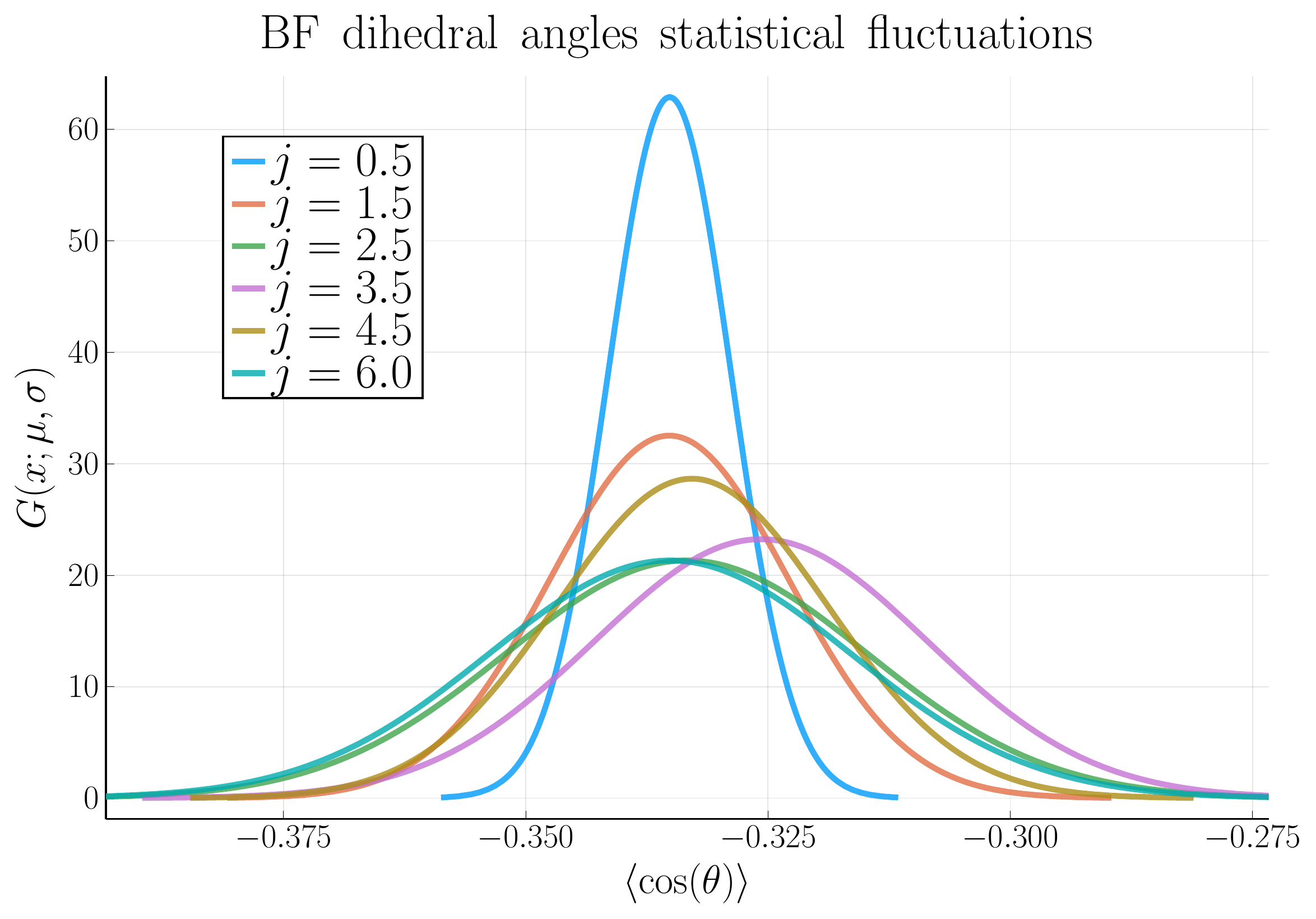}
     \end{subfigure}
    ~~~
        \begin{subfigure}[b]{75mm}
        \includegraphics[width=75mm]{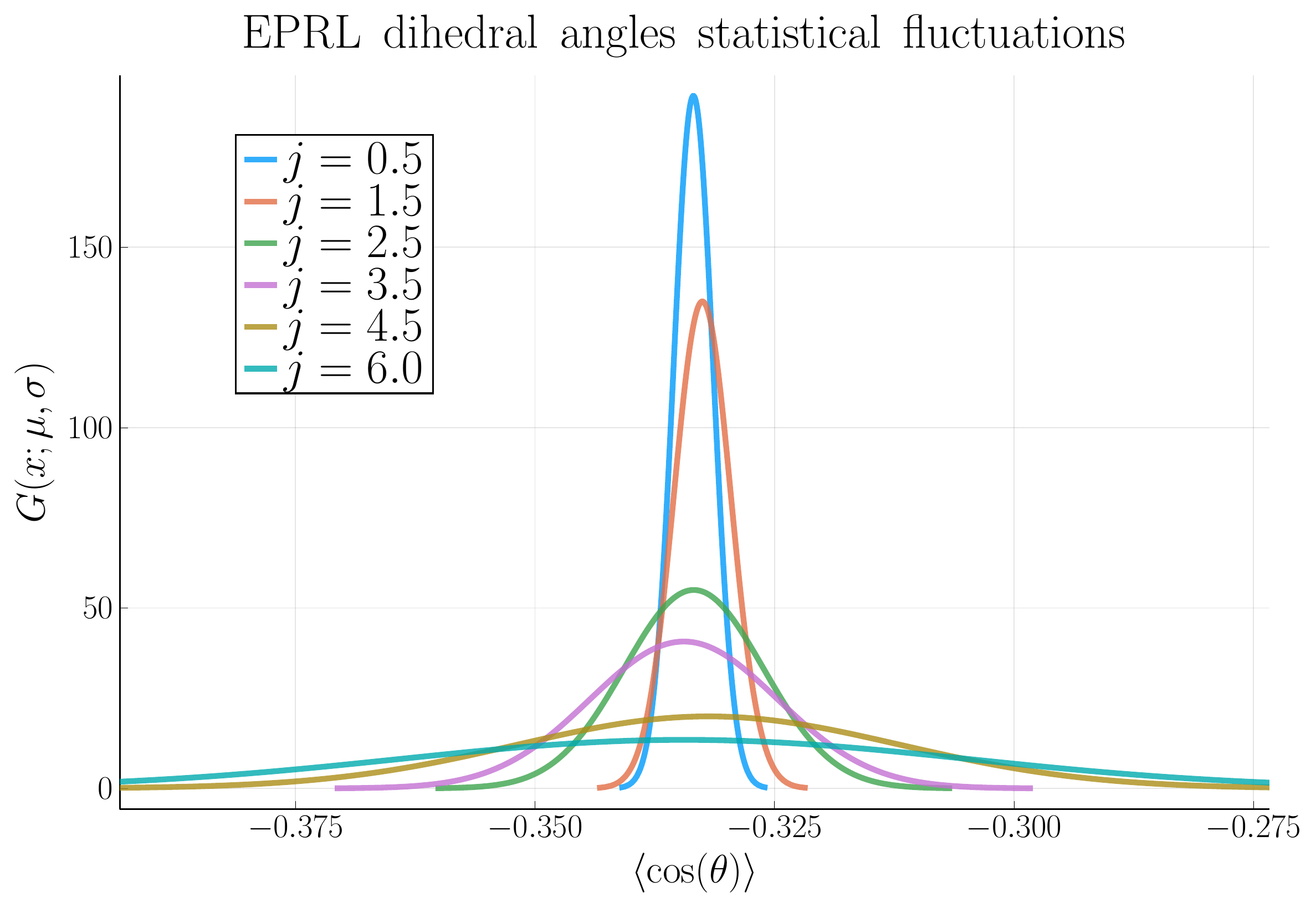}
    \end{subfigure}    
    \caption{{\label{fig:angesl_numerical_fluctuations} Gaussian distribution \eqref{eq:gaussian} of the expectation values \eqref{eq:<Dn>_MC} of the dihedral angle operator \eqref{eq:geom-angleformula}. We averaged over several runs, computing the (average) angle $\mu_{<\cos{\theta}>}$ defined on a single node and the corresponding standard deviation $\sigma_{<\cos{\theta}>}$ for each $j$. }}
\end{figure}

The dihedral angle operator has already been discussed in Section \ref{subsec:testing_sampler}, as we used it in order to test the Monte Carlo sampler. In the star  model, since all boundary tetrahedra are equal and regular, we can improve the statistic with a further average\footnote{This step is justified a posteriori once it has been verified that the expectation value of the operator over all the nodes is identical.}. Namely, we can compute the expectation value of the dihedral angle \eqref{eq:geom-angleformula} and the corresponding quantum spread \eqref{eq:spread} for all $20$ nodes of the spinfoam independently at fixed boundary spin $j$, then averaging the results, which are shown in Figure \ref{fig:angles}. The results show that the expectation value of the boundary dihedral angle \eqref{eq:geom-angleformula} is peaked to the value corresponding to an equilateral tetrahedron, which is the same result obtained in Section \ref{sec:The_4_simplex} with the simplest possible triangulation of a 3-sphere. This indicates that in the evolution from 1 to 6 vertices, the spatial metric of the boundary state still averages to that of the 3-boundary of a regular 4-simplex, i.e. to that of a 3-sphere. This is not a trivial consequence of the reduction \eqref{eq:ia_def}, 
but turns out to be a dynamical result of the global geometry. In fact, in \eqref{eq:<Dn>} we are considering the sum over all the spinfoam boundary intertwiners. Different geometries might give different (non-regular) average boundary angles, which is for example the case of the $\Delta_3$ spinfoam graph \cite{flatness_ex_marsigliesi}.
To get an idea of the gain obtained with respect to the blind summation, it is sufficient to consider that at spin $j = 6$ it is possible to compute the expectation value of an operator (for example the dihedral angle operator) stably up to the third significant digit with a number of Monte Carlo iterations $N_{MC} \sim 10^7$, as in the 4-simplex model. The exact sum would require performing $(2j+1)^{20} \sim 10^{22}$ sums. Therefore, the Metropolis-Hastings algorithm adapted to the spinfoam formalism allows to reduce the computation complexity of about 15 orders of magnitude. \\
The corresponding quantum spread $\Delta \cos (\theta)$ is shown in Figure \ref{fig:angles_spread}. It turns out to be rapidly increasing for EPRL and slightly increasing for the BF model. As originally noticed in \cite{Gozzini_primordial}, this suggests that quantum fluctuations of the metric in the Lorentzian model are wide, and are not suppressed in the asymptotic regime with few vertices. This however might be a simple consequence of the boundary state \eqref{eq:ia_def},  which fixes the areas of the boundary triangles at the quantum level, implying that the boundary angles are quantum totally spread.
The results suggest that, even if it is not a regular triangulation, the star model is suitable to discretize a closed geometry as a simplicial manifold bounded by a topological 3-sphere. In fact, in addition to the similarity with the results obtained in \cite{Gozzini_primordial}, preliminary results on the 16-cell spinfoam model, which constitutes the second regular triangulation of the 3-sphere after the 4-simplex, exhibit a striking similar behavior \cite{16_cell_model}. 
The Gaussian distributions \eqref{eq:gaussian}, measuring the statistical fluctuations in the Monte Carlo sampling, are shown in Figure \ref{fig:angesl_numerical_fluctuations}. 
    \begin{figure}[b]
    \centering
        \begin{subfigure}[b]{75mm}
        \includegraphics[width=75mm]{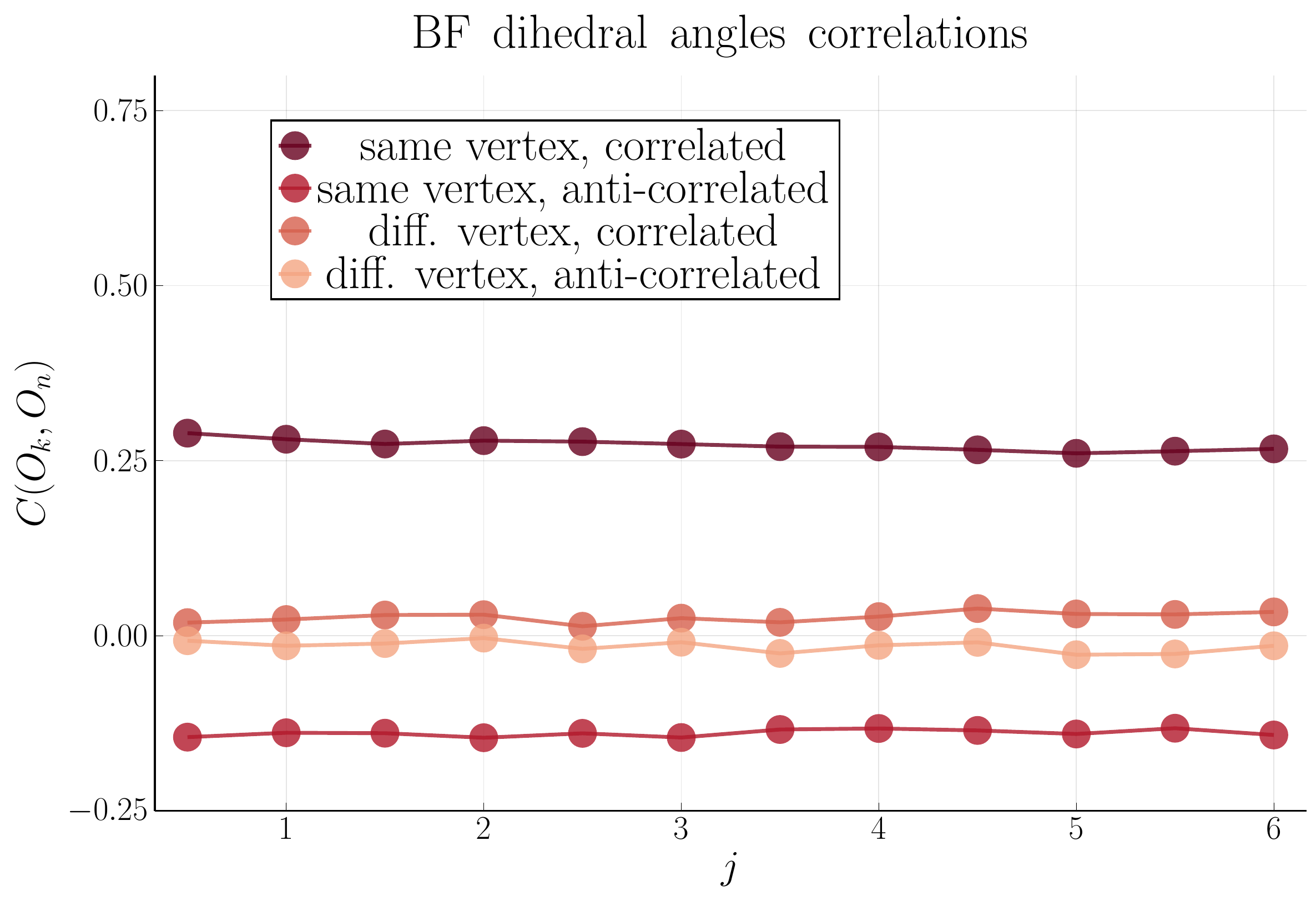}
     \end{subfigure}
    ~~~
        \begin{subfigure}[b]{75mm}
        \includegraphics[width=75mm]{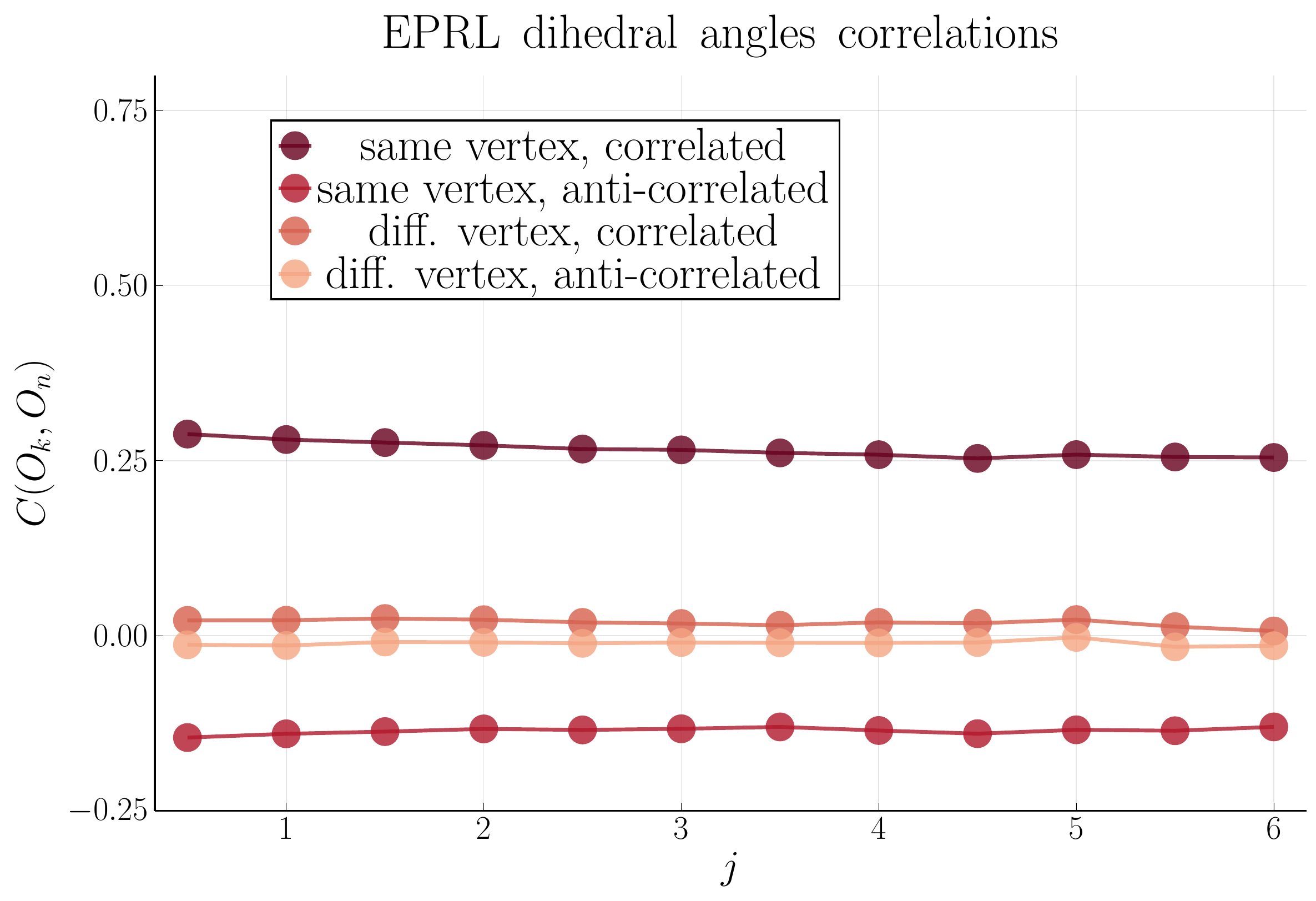}
    \end{subfigure}   
    \caption{\label{fig:angles_correlations}\emph{\textit{Expectation values \eqref{eq:<DnDm>_MC} of the correlations between dihedral angle operators \eqref{eq:geom-angleformula}. The correlations computed respectively for the EPRL and the BF models are essentially indistinguishable.}}}
    \end{figure}
For the sake of clarity, we underline that the statistical fluctuations in Figure \ref{fig:angesl_numerical_fluctuations} were computed by averaging the expectation value \eqref{eq:<Dn>_MC} for the operator \eqref{eq:geom-angleformula} on a single node over several runs, according to \eqref{eq:gaussian}, while in Figure \ref{fig:angles} we performed a further average on the 20 nodes of the spinfoam. \\
We computed the correlation functions \eqref{eq:correlations} between dihedral angles for all the $190$ independent nodes combinations of the spinfoam. The result is shown in Figure \ref{fig:angles_correlations}. For the dihedral angle operator \eqref{eq:geom-angleformula}, we found that correlations can assume two types of values, both for correlations between operators defined on nodes belonging to the same vertex and for different vertices as well, finding 4 different total possible numerical values. In the case of nodes on the same vertex, we get the same correlations originally computed in \cite{Gozzini_primordial}. In the second case, we observe that also angles between distant vertices can be (only) positively or negatively correlated, and the absolute value of the correlations is small compared to the first case. This is in agreement with the results on the entanglement entropy, as discussed in Section \ref{subsec:entanglement_entropy}. \\
The numerical results show that the EPRL and BF models give rise to essentially indistinguishable dynamic correlations in the case of the dihedral angle operator \eqref{eq:geom-angleformula}. This suggests that, at least in the approximation described in Section \ref{sec:boundary_state}, the \SUT {} topological model (typically much easier to compute) provides an excellent approximation for studying dynamical correlations.

\subsubsection{The volume operator}
\label{subsubsec:volume_operator}
\begin{figure}[H]
    \centering
        \begin{subfigure}[b]{75mm}
        \includegraphics[width=75mm]{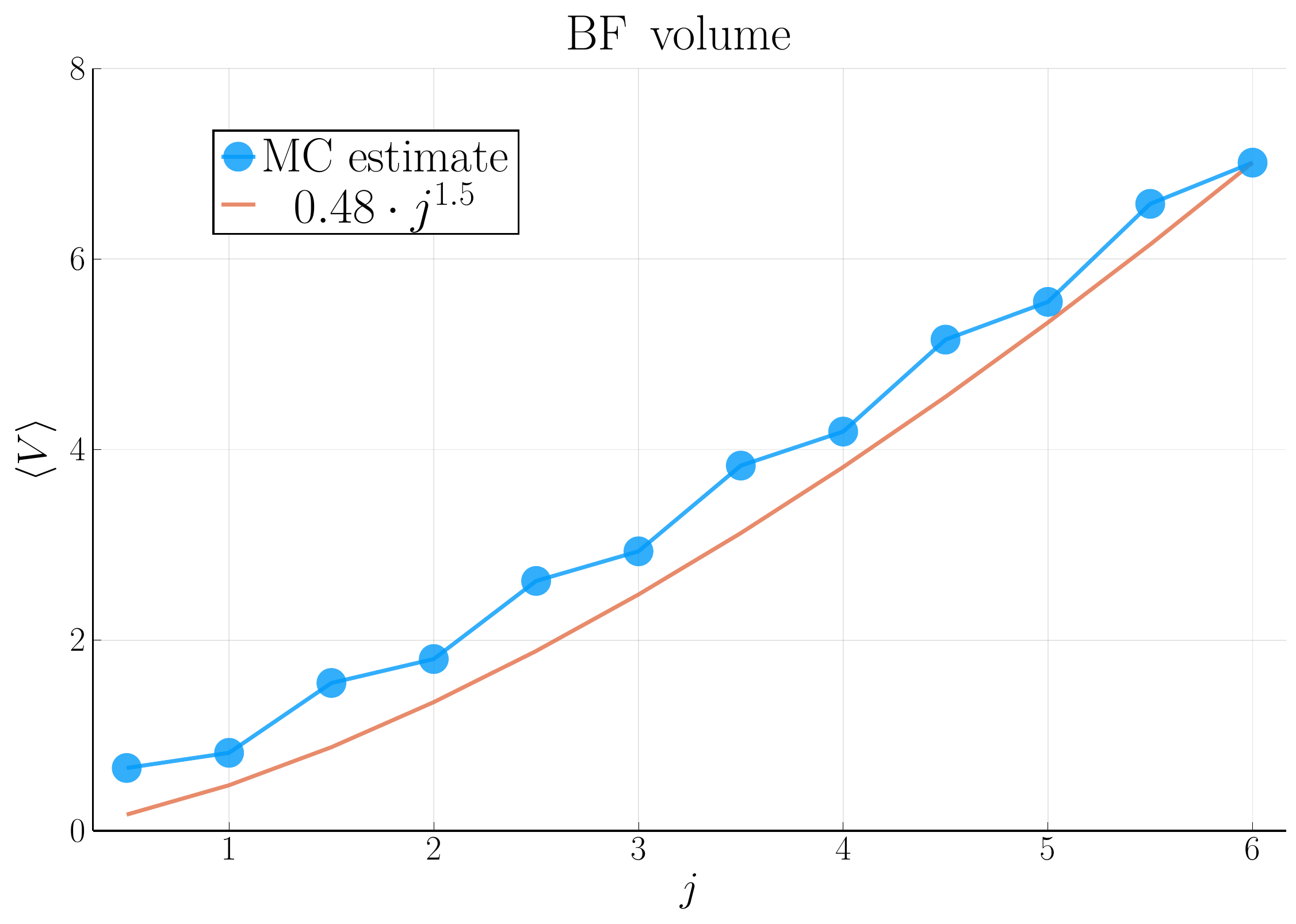}
     \end{subfigure}
    ~~~
        \begin{subfigure}[b]{75mm}
        \includegraphics[width=75mm]{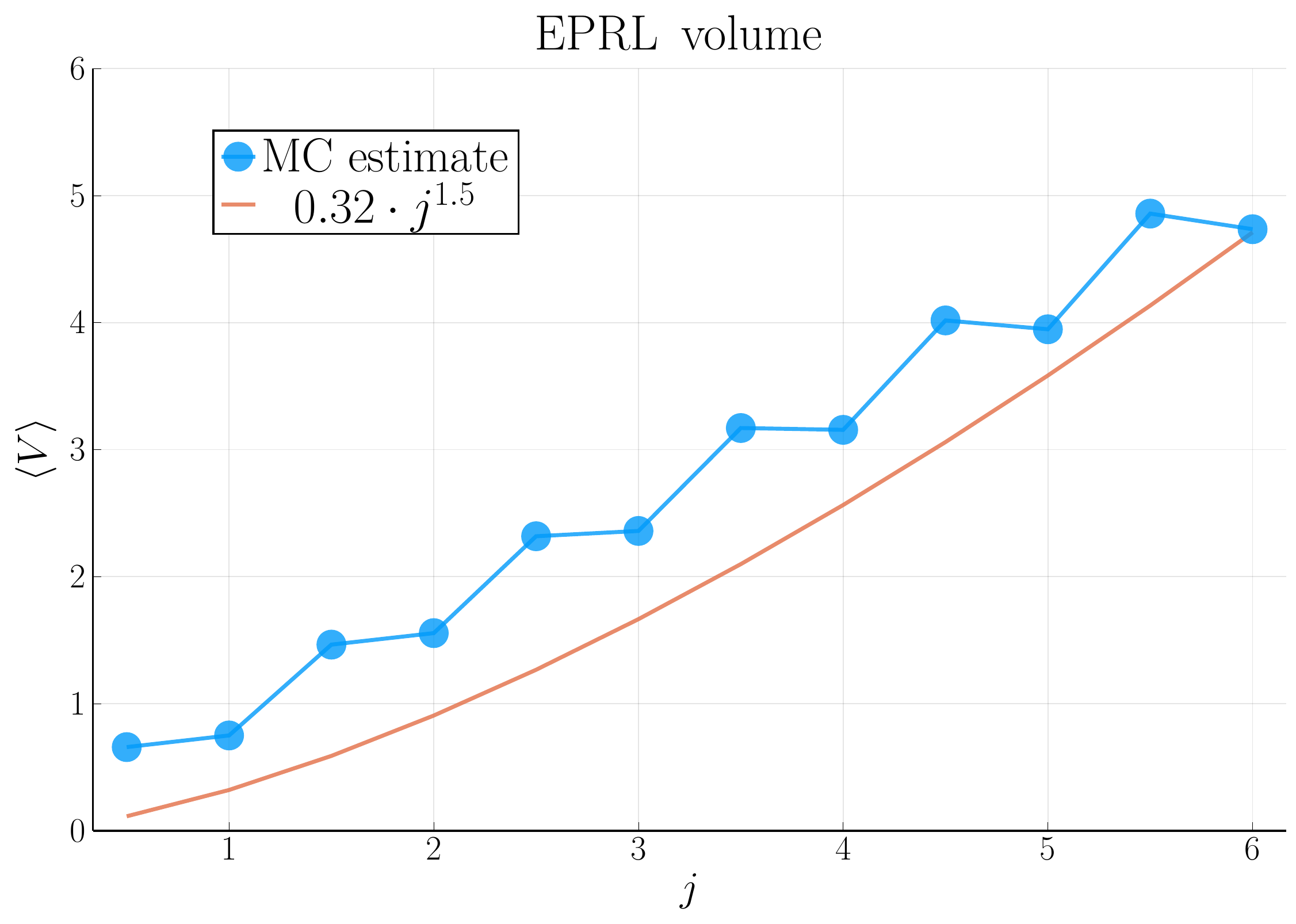}
    \end{subfigure}    
    \caption{{\label{fig:volumes} \textit{Expectation values \eqref{eq:<On>_MC} of the volume operator \eqref{eq:geom-volume_formula}. The orange line is proportional to the functional dependence between the volume of a regular tetrahedron and the area of one of its faces: $V \propto j^{3/2}$}. }}    
\end{figure}

\begin{figure}[b]
    \centering
        \begin{subfigure}[b]{75mm}
        \includegraphics[width=75mm]{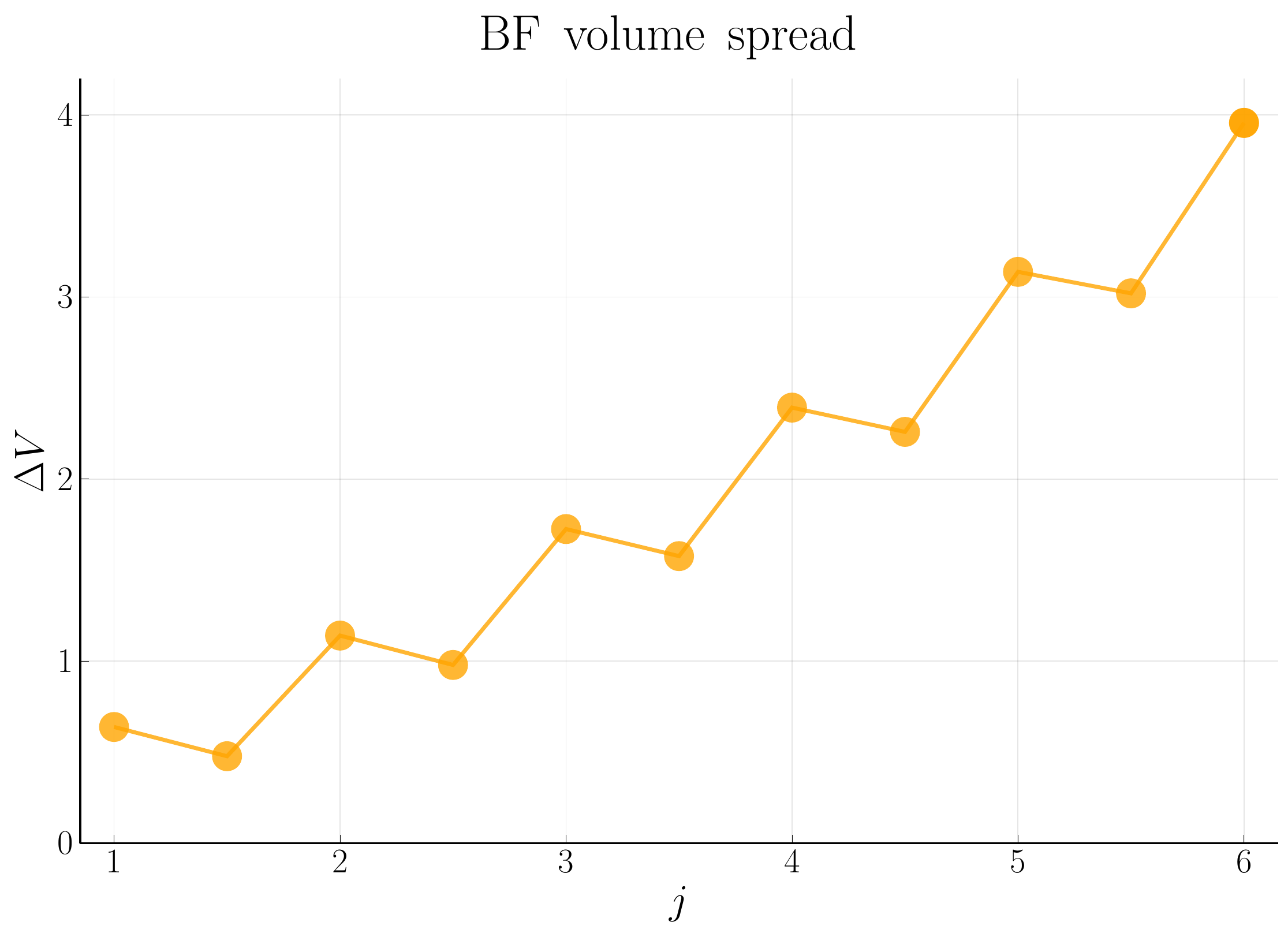}
     \end{subfigure}
    ~~~    
        \begin{subfigure}[b]{75mm}
        \includegraphics[width=75mm]{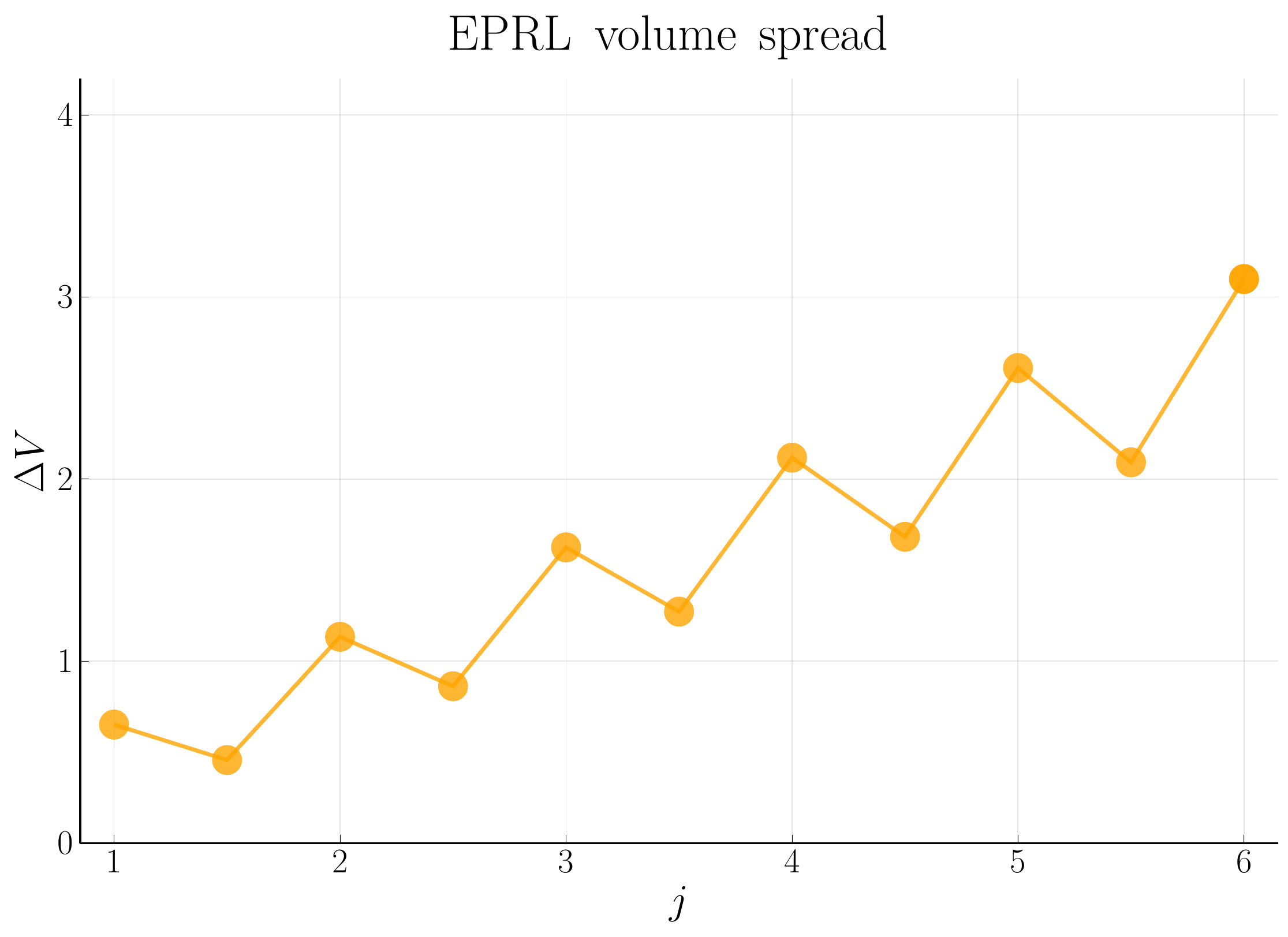}
    \end{subfigure}  
    \caption{{\label{fig:volumes_spread} \textit{Expectation values of the quantum spread \eqref{eq:spread} for the volume operator \eqref{eq:geom-volume_formula}. 
As in Figure \ref{fig:angles}, it is evident that the spectrum of the volume operator gives rise to two distinct curves for integer and half-integers spins, which turn out to be shifted with respect to each other.}}}
    \end{figure}

\begin{figure}[H]
    \centering
        \begin{subfigure}[b]{75mm}
        \includegraphics[width=75mm]{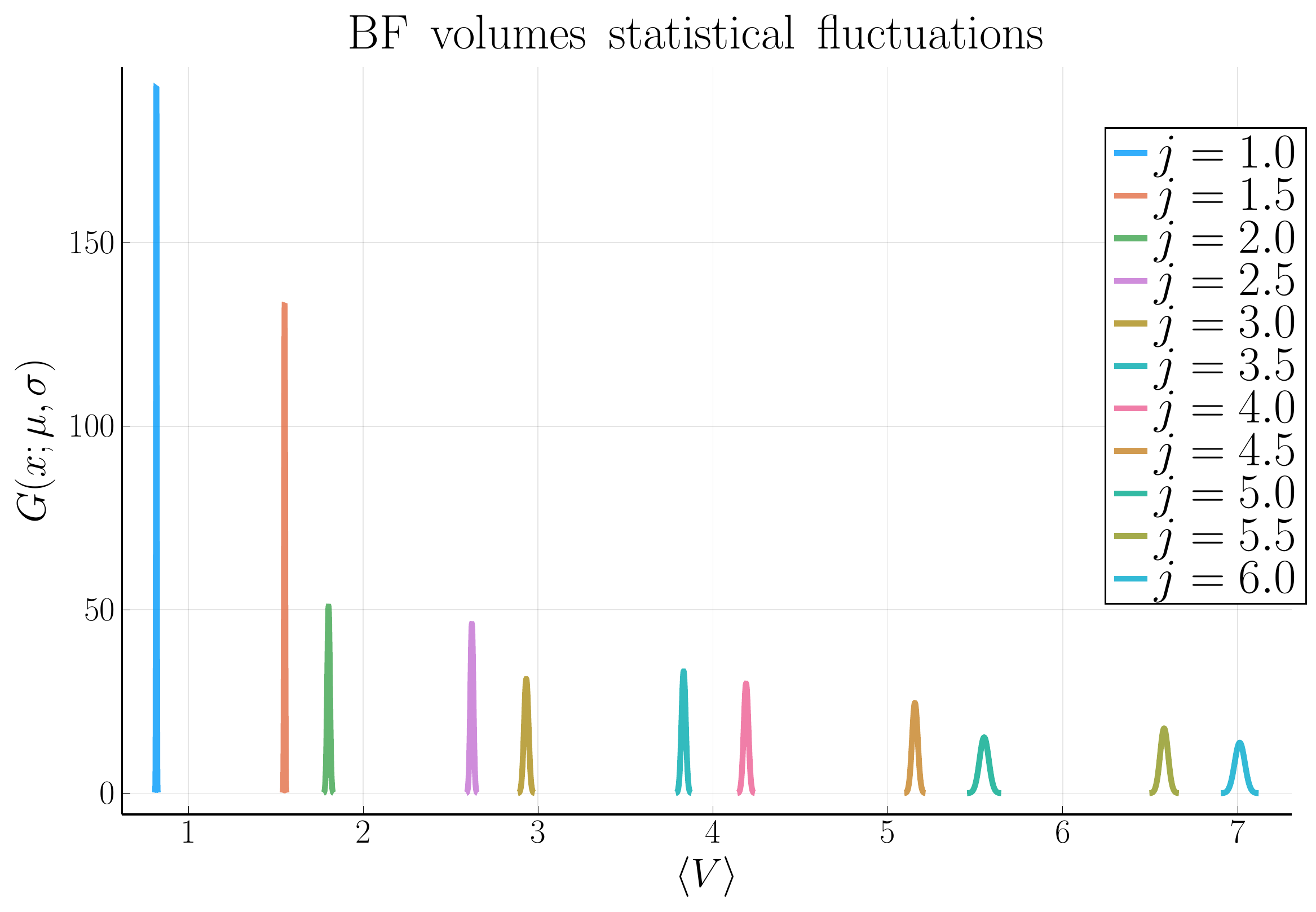}
     \end{subfigure}
    ~~~
        \begin{subfigure}[b]{75mm}
        \includegraphics[width=75mm]{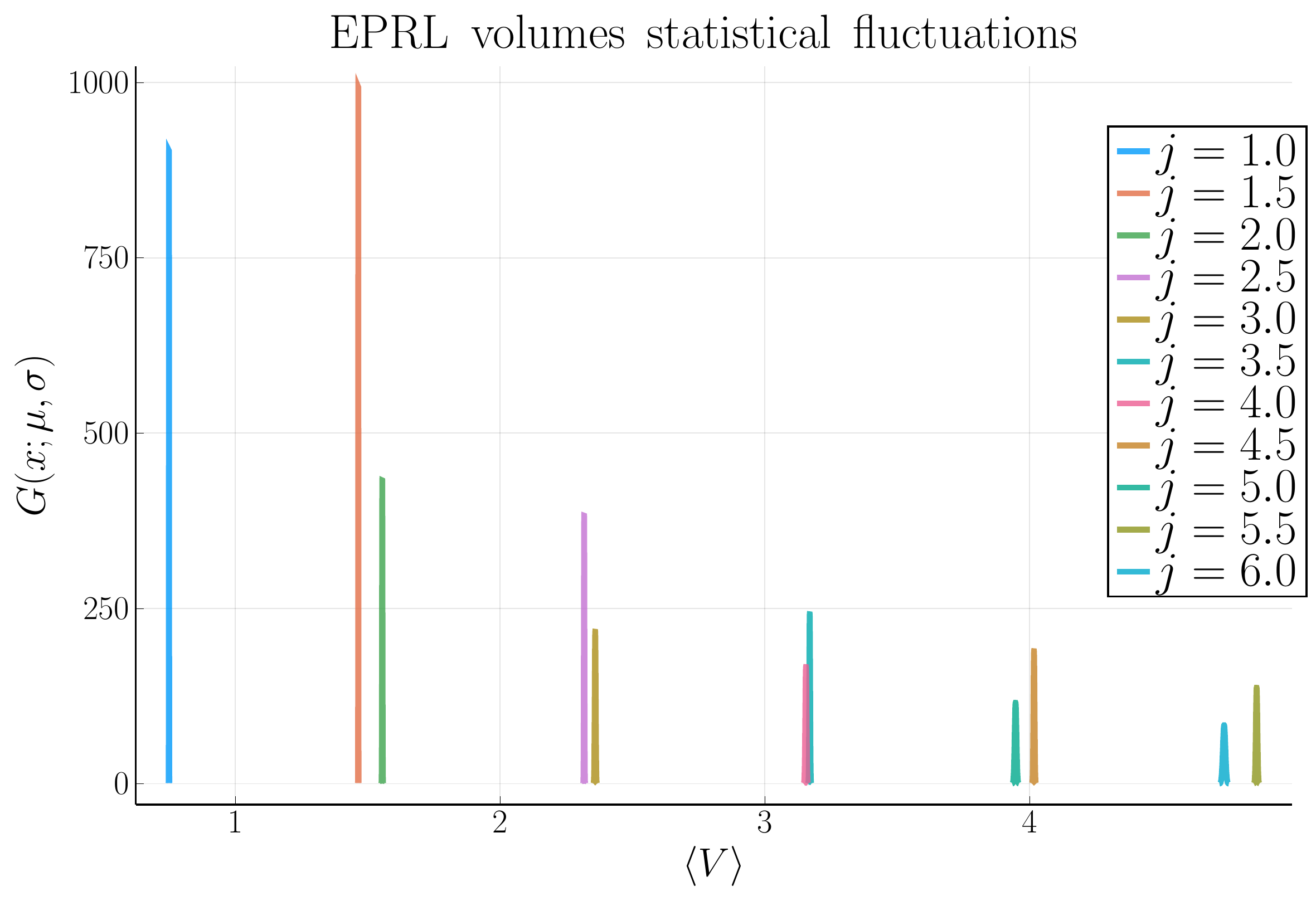}
    \end{subfigure}    
    \caption{{\label{fig:volumes_numerical_fluctuations} \textit{ Gaussian distribution \eqref{eq:gaussian} of the expectation values \eqref{eq:<On>_MC} of the volume operator \eqref{eq:geom-volume_formula}. We averaged over several runs, computing the (average) volume $\mu_{<V>}$ defined on a single node and the corresponding standard deviation $\sigma_{<V>}$ for each $j$.}}}
    \end{figure}

\begin{figure}[b]
    \centering
        \begin{subfigure}[b]{75mm}
        \includegraphics[width=75mm]{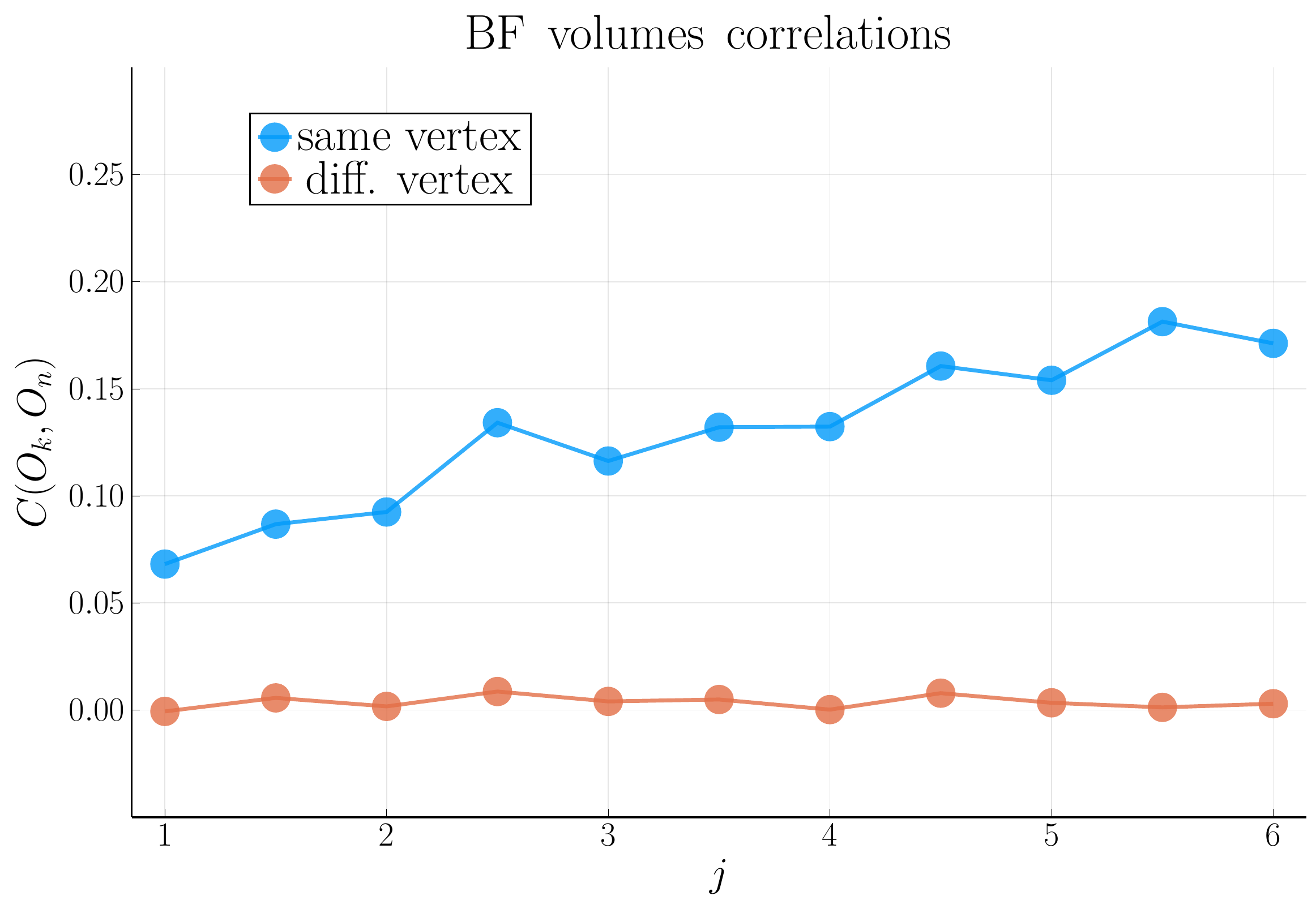}
     \end{subfigure}
    ~~~
            \begin{subfigure}[b]{75mm}
        \includegraphics[width=75mm]{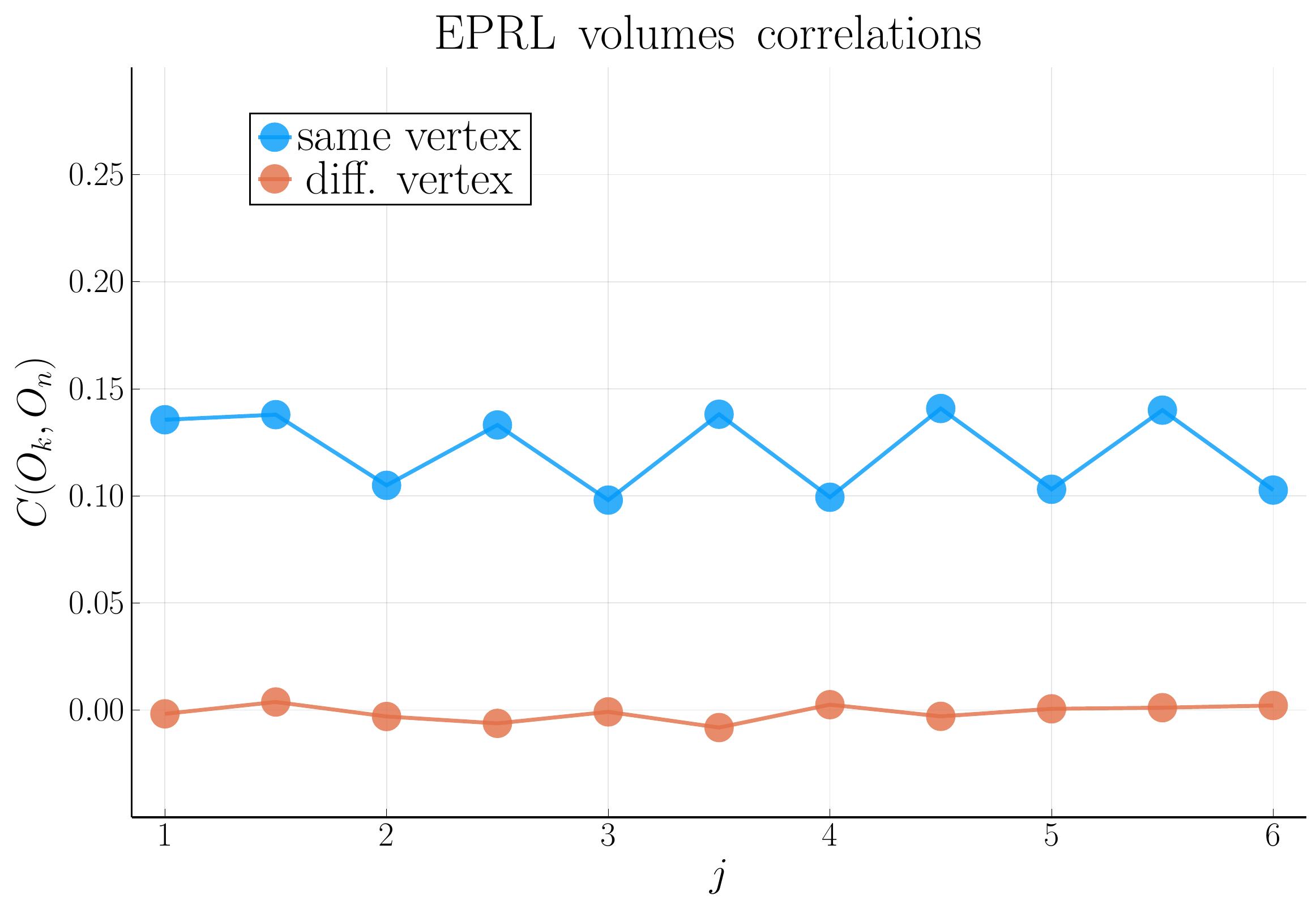}
    \end{subfigure}
    \caption{ \label{fig:volumes_correlations} \textit{Expectation values \eqref{eq:<OnOm>_MC} of the correlations between volumes. While in the BF model the correlations between volumes on the same node seem to slowly increase with the boundary spin $j$, in EPRL this does not happen. Unlike angles (see Figure \ref{fig:angles_correlations}), all nodes on the same vertex have the same correlation, while nodes on different vertices are essentially uncorrelated.}}
    \end{figure}

There are two slightly different prescriptions for the volume operator in LQG. Here we follow the Rovelli-Smolin prescription in \cite{Rovelli1994a}. Since the general expression of the volume matrix elements in the spin-network basis is not trivial \cite{book:Rovelli_Vidotto_CLQG}, here we limit ourselves to the equations in symmetric-reduced space of \eqref{eq:Hilbert_space_LQG} in which all the spins have the same value $j$ and the basis states are given by \eqref{eq:ia_def}. \\
Let $A$ be the $(2j + 1) \times (2j + 1)$ Hermitian matrix:
\begin{equation}
A = i \times \left(\begin{array}{cccccc} 
    0  & -a_{1} & 0      & 0      & \dots  & 0 \\
    a_1    &   0    & -a_2   & 0      & \dots  & 0 \\
     0     & a_{2}  & 0      & -a_3   & \dots  & 0 \\
    \vdots & \vdots & \vdots & \vdots & \ddots & \vdots \\
    \end{array}\right)
\end{equation}
where the coefficients $a_k$ are defined as:
\begin{equation}
a_k = \frac{1}{4} \frac{k^2((2j+1)^2 - k^2)}{\sqrt{4k^2 - 1}} \ ,
\end{equation}
Let $q_k$ be its real eigenvalues and $|q_k \rangle$ the corresponding eigenvectors. For each $j$ the eigenvalues come in pairs of opposite signs, plus one $0$ eigenvalue for $j$ integer. The volume operator matrix can be written as:
\begin{equation}
\label{eq:geom-volume_formula}
\langle j, i_n  | V | j, i_{n}' \rangle =   \frac{\sqrt{2}}{3} \left( 8 \pi G \hbar \gamma \right)^{\frac{3}{2}} \sum\limits_{k} \sqrt{ |q_k| } \langle  j, i_{n}'| q_k \rangle \langle  q_k | j, i_{n} \rangle. 
\end{equation}
Contrary to the dihedral angle \eqref{eq:geom-angleformula}, the volume operator \eqref{eq:geom-volume_formula} is not diagonal in the basis \eqref{eq:ia_def}. In terms of expectation values \eqref{eq:<On>_MC} and correlations \eqref{eq:<OnOm>_MC}, this involves recomputing the amplitude function for each element of the sampling, hence it is much slower then the corresponding diagonal evaluation (however it is still incomparably faster than blind summation \eqref{eq:<OnOm>} which would be required without the Monte Carlo approximation \eqref{eq:MC_sum}). This means that computing the expectation value \eqref{eq:<On>_MC} in the case of the volume operator \eqref{eq:geom-volume_formula} for more than one node of the spinfoam takes too long. The expectation values of the volumes are shown in Figure \ref{fig:volumes}, in which we neglected all the constant factors in the expression \eqref{eq:geom-volume_formula} since it simply corresponds to a homogeneous re-scaling of all points. It turns out that the scaling of the boundary volume, as a function of the boundary spin $j$, corresponds to that actually existing between the volume of a regular tetrahedron and the area of one of its faces. That is, $V \propto j^{3/2}$, as the eigenvalue of the area operator is proportional to $\sqrt{j(j+1)} \approx j$. This is what we observe in both BF and EPRL models, despite the fact that the spectrum is not the same. An interesting feature of the volume operator spectrum is that there is a systematic shift between integer spins and half-integers spins. That is, these are two slightly shifted curves.
The quantum spread of the volume operator is shown in Figure \ref{fig:volumes_spread}. The shift between the curves corresponding to integer spins and half-integers is manifest. Differently with respect to the angle operator \ref{fig:angles_spread}, for the volume operator the quantum spread increases faster for BF rather than EPRL.
The Gaussian distributions \eqref{eq:gaussian} are shown in Figure \ref{fig:volumes_numerical_fluctuations}. The average values of the volumes used as mean in the Gaussian distributions \eqref{eq:gaussian} are the same plotted in Figure \ref{fig:volumes}. 
The probability distribution for non-diagonal operators in equation \eqref{eq:<On>_MC} is not strictly positive. Interestingly, as shown in Figure \ref{fig:volumes_numerical_fluctuations}, this does not affect the convergence.
Volume correlations are shown in Figure \ref{fig:volumes_correlations}. As for the angles, the volumes correlations between nodes belonging to the same vertex are much higher than those between different vertices. For the volumes, the latter appear to be essentially zero. It is interesting to notice that, contrary to what happens with the angles, there is only one type of correlation between volumes. That is, it only exists one common value for all the correlations between volumes of tetrahedra on the same vertex, and the same is true for non-adjacent tetrahedra. 
\subsection{Numerical results: entanglement entropy}
\label{subsec:entanglement_entropy}
We now discuss the results about the computation of the entanglement entropy between different nodes, viewed as quantum subsystems of the whole graph. The entanglement entropy turns out to be the most relevant quantity to study the degree of correlation between operators, as it gives the mutual information between different subsystems \cite{nielsen_chuang_2010, Dona_entropy}. In the topological BF model, the way in which we defined the boundary state \eqref{eq:cosm_state} in Section \ref{sec:boundary_state} coincides with definition of the Bell-Network states, whose entanglement entropy is computed in \cite{Dona_entropy} for different combinations of graphs and subsystems. \\
In general, a quantum system composed of two subsystems $A$ and $\bar{A}$ has a Hilbert space given by the tensor product:
\begin{equation}
\label{eq:Hilbertspaces}
\mathcal{H} = \mathcal{H}_{A} \otimes \mathcal{H}_{\bar{A}}\,.
\end{equation}
Given the boundary state  $|\psi_0 \rangle$ in the Hilbert space \eqref{eq:Hilbert_space_LQG}, the normalized reduced density matrix of the subsystem $A$ is defined by the partial trace over its complement $\bar{A}$:
\begin{equation}
\label{eq:densitym}
\rho_A = \frac{1}{Z} \Tr_{\bar{A}} \;|\psi_0 \rangle\langle\psi_0 | \, .
\end{equation}
The entanglement entropy of the subsystem $A$ is then defined as the von Neumann entropy of the reduced density matrix  
\begin{equation}
\label{eq:EEentropy}
S_A=-\Tr\,\big(\rho_A \log\rho_A\big) \ .
\end{equation}
Using the expression \eqref{eq:cosm_state} in \eqref{eq:densitym}, after some algebraic manipulations the normalized reduced density matrix \eqref{eq:densitym} can be written as:
\begin{equation}
\label{eq:density_mat_M}
\rho_A = \frac{1}{Z} \sum_{ \{ i_a \}} \sum_{ \{ i_a' \}} M \left(  j , \{ i_a \} , \{ i_a' \} \right) \bigotimes_{a} | j, \{ i_a \} \rangle \langle j , \{ i_a' \} | \ , 
\end{equation}
where $a \in A$. The coefficients $M \left(  j, \{ i_a \} , \{ i_a' \} \right)$ are defined by tracing over the intertwiners $i_{\bar{a}}$ in the complement subsystem $\bar{A}$:
\begin{equation}
\label{eq:Matrix_coeff}
M \left( j , \{ i_a \} , \{ i_a' \} \right) = \sum_{\{ i_{\bar{a}} \}} A \left(  j , \{ i_a \}, \{ i_{\bar{a}} \} \right) A \left(  j , \{ i_a' \}, \{ i_{\bar{a}} \} \right) \ .
\end{equation}
We replaced the sum over the full set $\{ i_n \}$ with $\{ i_a \}$, namely, the intertwiners involved in the partition \eqref{eq:Hilbertspaces}. By introducing the Monte Carlo approximation \eqref{eq:MC_sum}, the expression for the density matrix becomes:
\begin{equation}
\label{eq:density_mat_MC}
\rho_A \approx \frac{1}{N_{MC}} \sum_{[i_n]} \sum_{\{ i_a' \} } \frac{ A \left(  j , \{ i_a' \}, [i_n] \right)}{ \AiMC } \bigotimes_{a} | j , [ i_a ] \rangle \langle j,  \{ i_a' \} | \ .
\end{equation}
The notation $[ i_a ]$ is a label for the set of intertwiners draws $[ i_n ]$ in which the nodes belonging to the subsystem $A$ have a value compatible with the position in the density matrix, and the meaning of the amplitude $ A \left(  j , \{ i_a' \}, [i_n] \right) $ should be clear by looking at \eqref{eq:A_synthetic_exp}. Notice that the density matrix is symmetric and $\Tr\,(\rho_A) = 1$. The entropy \eqref{eq:EEentropy} becomes:
\begin{equation}
\label{eq:Entropy_MC}
S_A \approx - \nu_i \log \nu_i \  ,
\end{equation}
where $\nu_i$ is the $i$-th eigenvalue of the density matrix \eqref{eq:density_mat_MC}. Notice that the computational time of the density matrix considerably increases as the number of nodes $N_A$ in subsystem $A$ grows, as the matrix \eqref{eq:density_mat_MC} has dimensions $(2j + 1)^{N_A} \times (2j + 1)^{N_A}$. Furthermore, at fixed number of Monte Carlo iterations $N_{MC}$, the statistical fluctuations increase along with $N_A$ since each matrix element is sampled by a set which becomes smaller and smaller. For these reasons, in the numerical computation of the density matrices we introduced a second multi-threading parallelization scheme using multiple machines. That is, each node computed the density matrix using a different Markov chain by distributing the calculation over multiple CPUs. The same hybrid parallelization scheme can be used to speed up the calculation of non-diagonal operators \eqref{eq:<On>_MC} and corresponding correlations \eqref{eq:<OnOm>_MC}.
\subsubsection{Subsystem with 1 and 2 nodes}
\begin{figure}[b]
    \centering
        \begin{subfigure}[b]{75mm}
        \includegraphics[width=75mm]{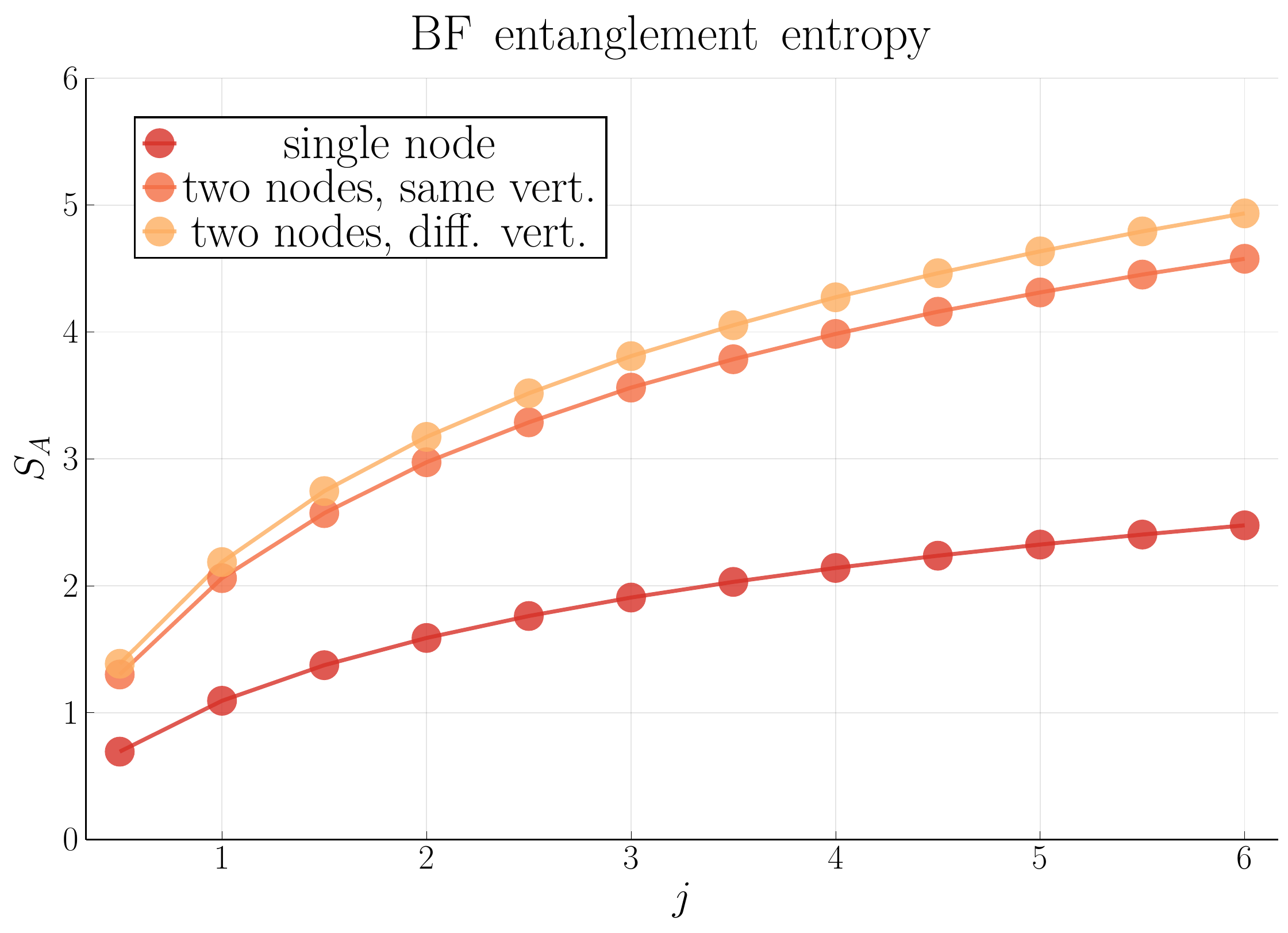}
    \end{subfigure}
    ~~~
        \begin{subfigure}[b]{75mm}
        \includegraphics[width=75mm]{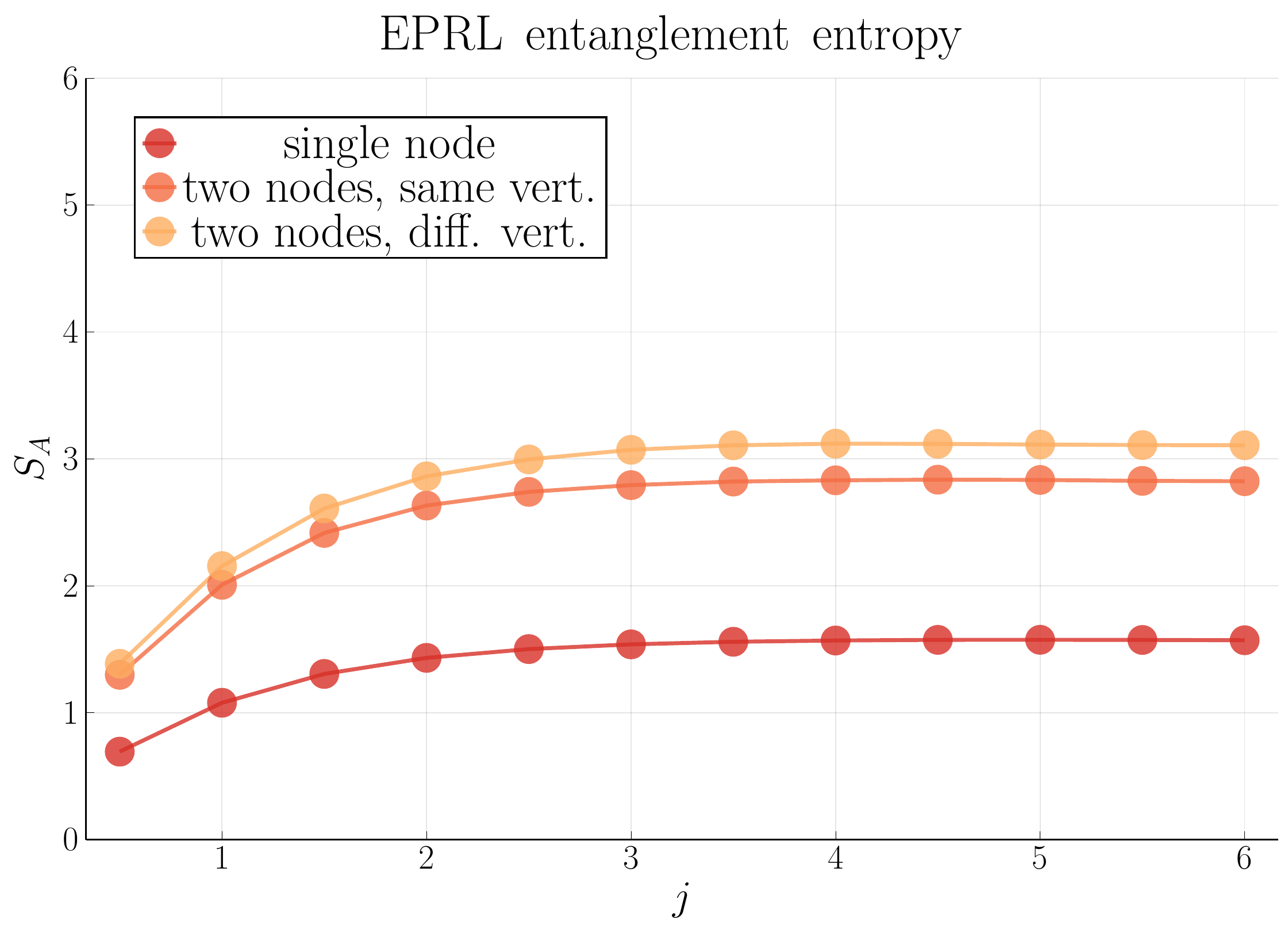}
    \end{subfigure}    
    \caption{{\label{fig:entropy} \textit{Values of the entanglement entropy \eqref{eq:Entropy_MC} for different subsystems $A$ in the partition \eqref{eq:Hilbertspaces} with $N_A = 1$ and $N_A = 2$ for the star spinfoam in Figure \ref{fig:star_spinfoam}}.}}
    \end{figure}

The values of the entanglement entropy as a function of the boundary spin $j$ are shown in Figure \ref{fig:entropy} for the subsystem $A$ in the partition \eqref{eq:Hilbertspaces} consisting in 1 and 2 nodes. We used the parameters listed in  \ref{app:M-H_parameters} for the data in Figure \ref{fig:entropy}. Notice that, by choosing a common value $j$ for all the links of the star spinfoam (see Figure \ref{fig:star_spinfoam}), we cannot distinguish correlations between nodes belonging to distinct vertices whether or not they are connected by the same link. For example, labelling the nodes according to the notation illustrated in Figure \ref{fig:star_spinfoam}, computing the entanglement entropy for the subsystems $A = \{C235, A235\}$, $A' = \{C235, D134\}$ and $A'' = \{C235, B134\}$ we obtain the same numerical value. 
The value of the EPRL entropy for the subsystem consisting of a single node is similar to the value obtained in \cite{Gozzini_primordial} for the single vertex graph. It is interesting to notice that the EPRL entropy of all considered subsystems seems to tend asymptotically to a constant value as the boundary spin $j$ increases. 
\begin{figure}[H]
    \centering
        \begin{subfigure}[b]{75mm}
        \includegraphics[width=75mm]{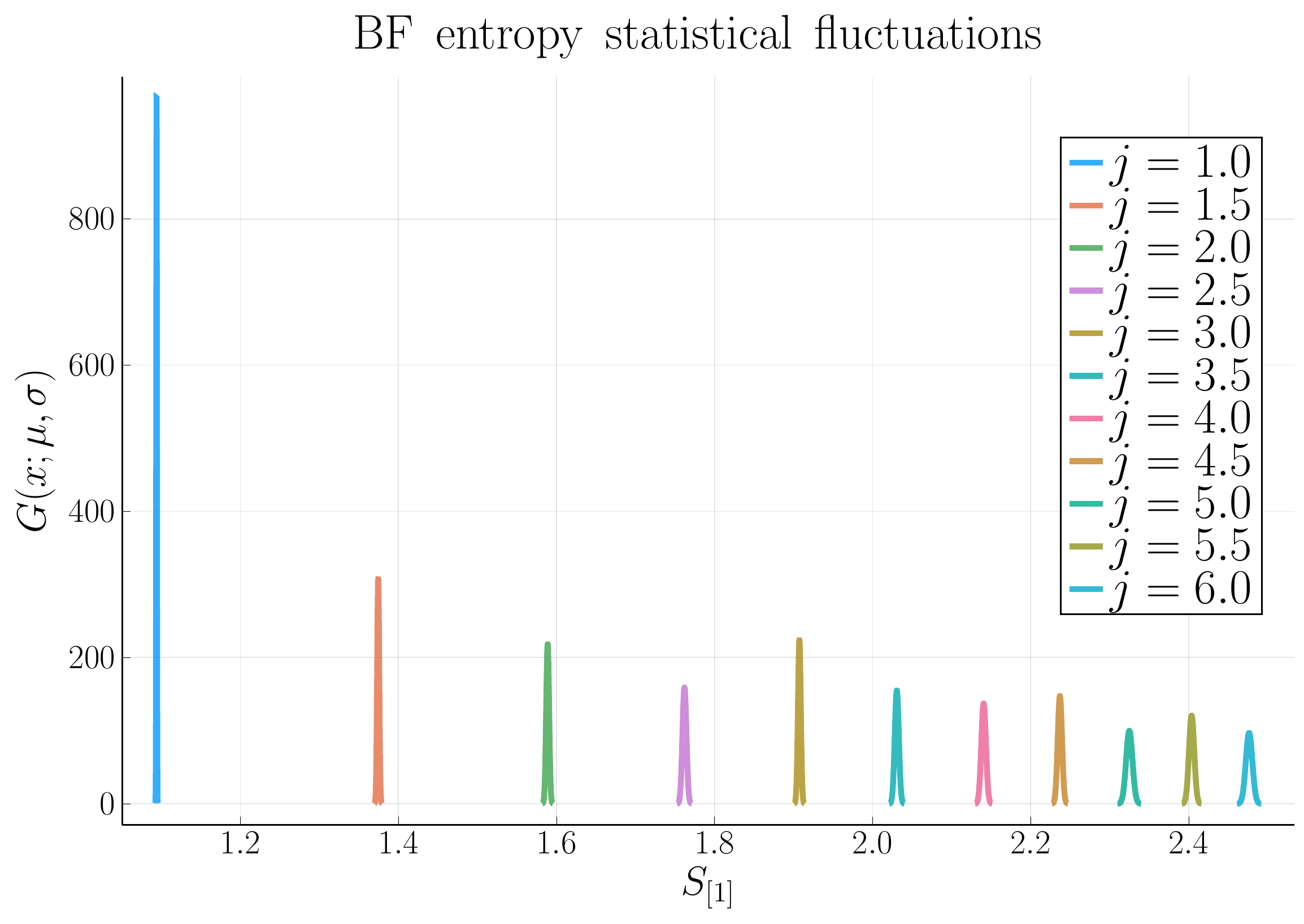}
    \end{subfigure}
    ~~~
        \begin{subfigure}[b]{75mm}
        \includegraphics[width=75mm]{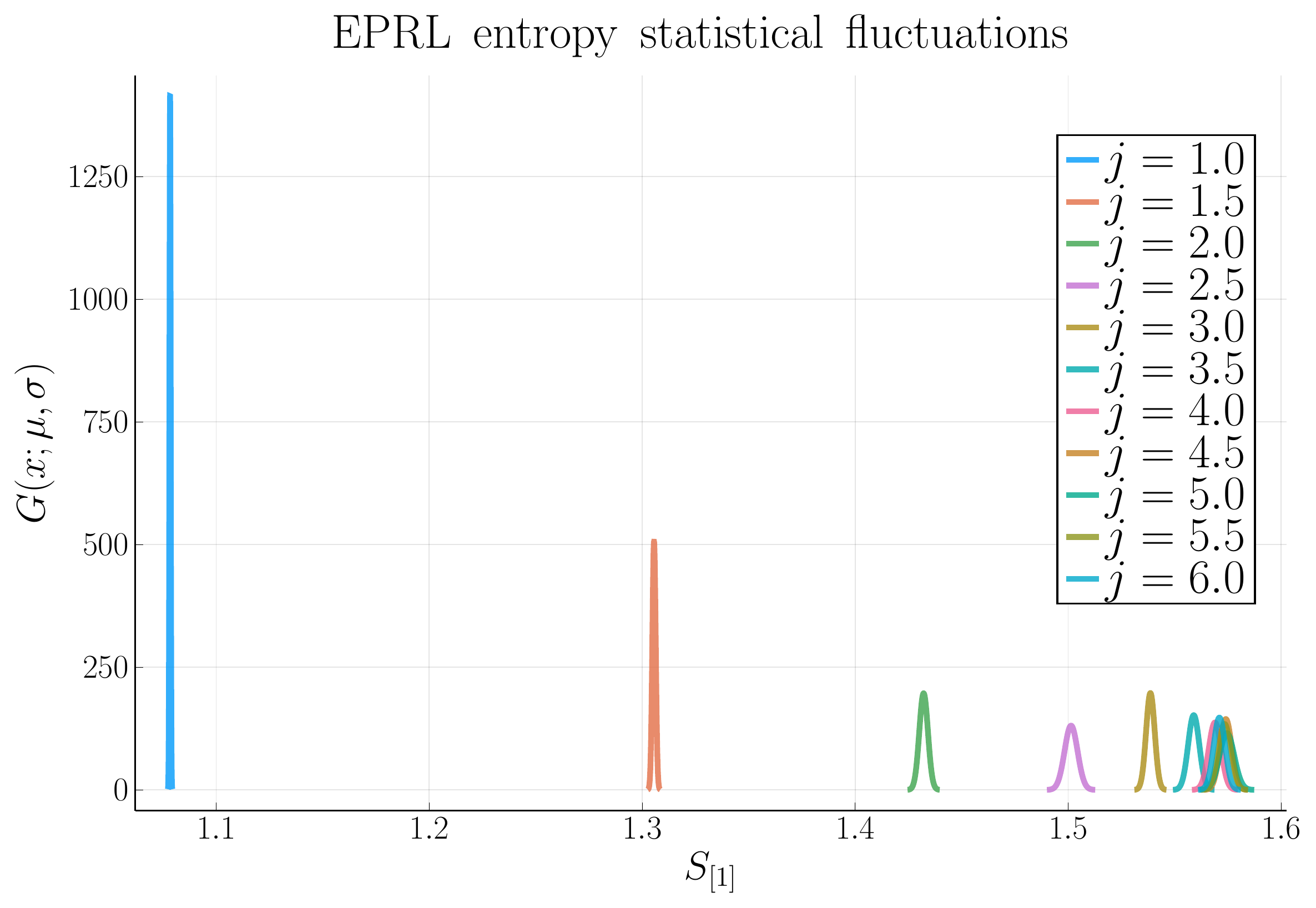}
    \end{subfigure}   
        \begin{subfigure}[b]{75mm}
        \includegraphics[width=75mm]{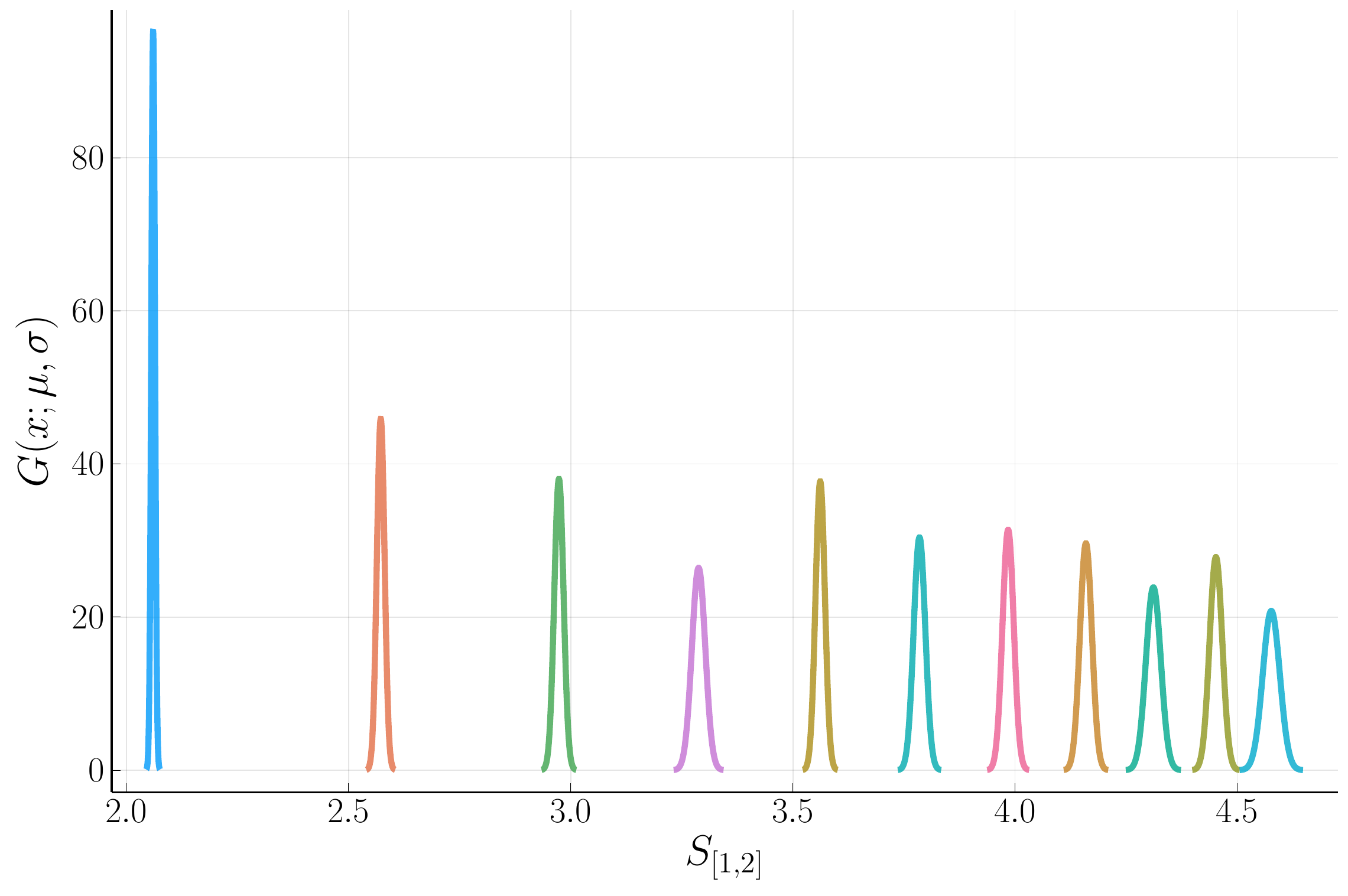}
    \end{subfigure}
    ~~~
        \begin{subfigure}[b]{75mm}
        \includegraphics[width=75mm]{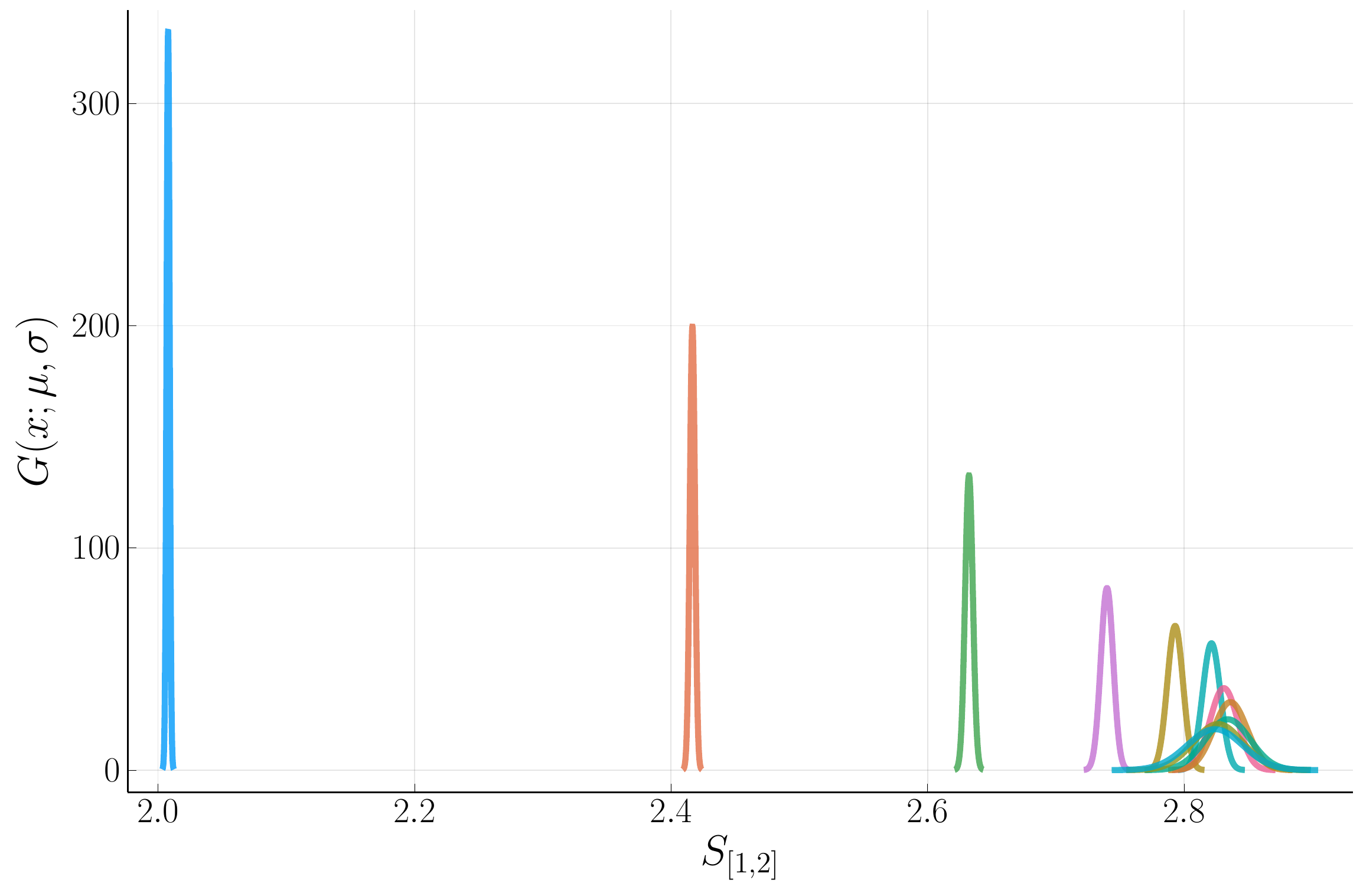}
    \end{subfigure}     
   \caption{\label{fig:entropy_numerical_fluctuations} {\textit{Gaussian distribution \eqref{eq:gaussian} of the entanglement entropy \eqref{eq:Entropy_MC} for the subsystems in Figure \ref{fig:entropy}. We proceeded as in the case of angles (Figure \ref{fig:angesl_numerical_fluctuations}) and volumes (Figure \ref{fig:volumes_numerical_fluctuations})}. Top panel: \textit{fluctuations for the subsystem with $N_A = 1$}. Bottom panel: \textit{fluctuations for the subsystem with $N_A = 2$}.}}
    \end{figure}
The statistical fluctuations of entropy are shown in Figure \ref{fig:entropy_numerical_fluctuations}. Notice that, with the same number of iterations $N_{MC}$, statistical fluctuations in Figure \ref{fig:entropy_numerical_fluctuations} are larger for $N_A=2$ for the reasons discussed above. The fact that the entropy of a subsystem composed of nodes on different 4-simplices is slightly greater than the one of the subsystem composed of nodes on the same 4-simplex, is connected to the smaller value of the correlations between nodes belonging to different 4-simplices. In order to discuss this point, we first define the mutual information $I(k,m)$ between two generic nodes $k$ and $m$ as:
\begin{equation}
\label{eq:mutual_information}
I(k,m)=S_{k}+S_{m}-S_{km} \ ,
\end{equation}
where $S_{k m}$ is the entropy of the subsystem $A$ composed by the nodes $k$ and $m$. It turns out that the mutual information \eqref{eq:mutual_information} between $k$ and $m$ actually provides a bound on correlations \cite{nielsen_chuang_2010, Dona_entropy}:
\begin{equation}
\label{eq:bound_on_corr_func}
\frac{\big(\langle O_k, O_{m} \rangle-\langle O_k \rangle \langle O_{m} \rangle \big)^2}{2\|O_k\|^2 \|O_{m}\|^2}\leq I(k, m)\,.
\end{equation} 
where $\| O_k \|$ is the norm of the local operator $O$ on the node $k$. Therefore, equations \eqref{eq:mutual_information} and \eqref{eq:bound_on_corr_func}, along with the results in Figure \ref{fig:entropy} imply that the correlation function \eqref{eq:correlations} has a more stringent upper bound for the subsystem containing nodes defined on different 4-simplices.
\subsubsection{Subsystem with 4 nodes}
\begin{figure}[H]
    \centering
        \begin{subfigure}[b]{75mm}
        \includegraphics[width=75mm]{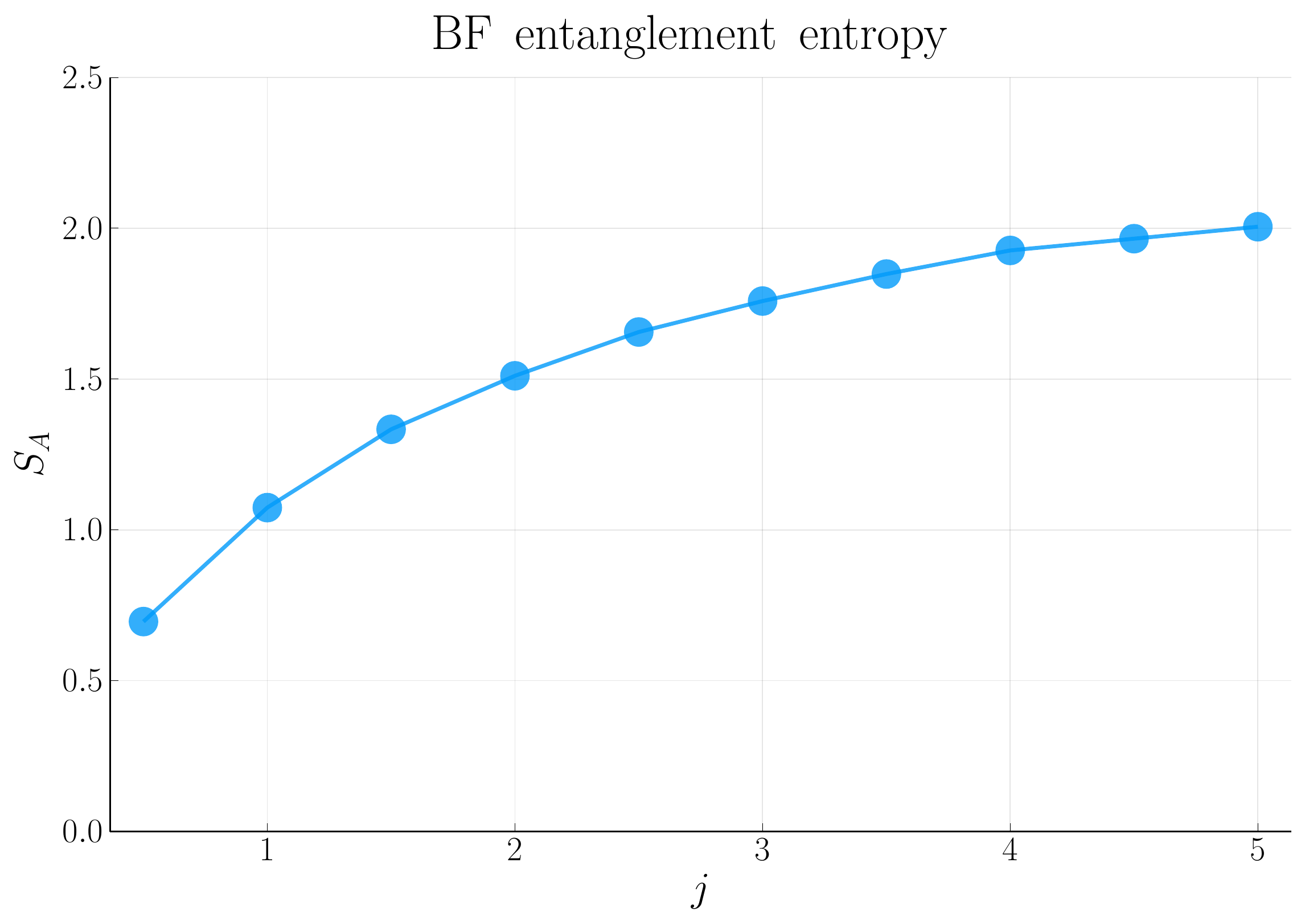}
    \end{subfigure}
    ~~~
        \begin{subfigure}[b]{75mm}
        \includegraphics[width=75mm]{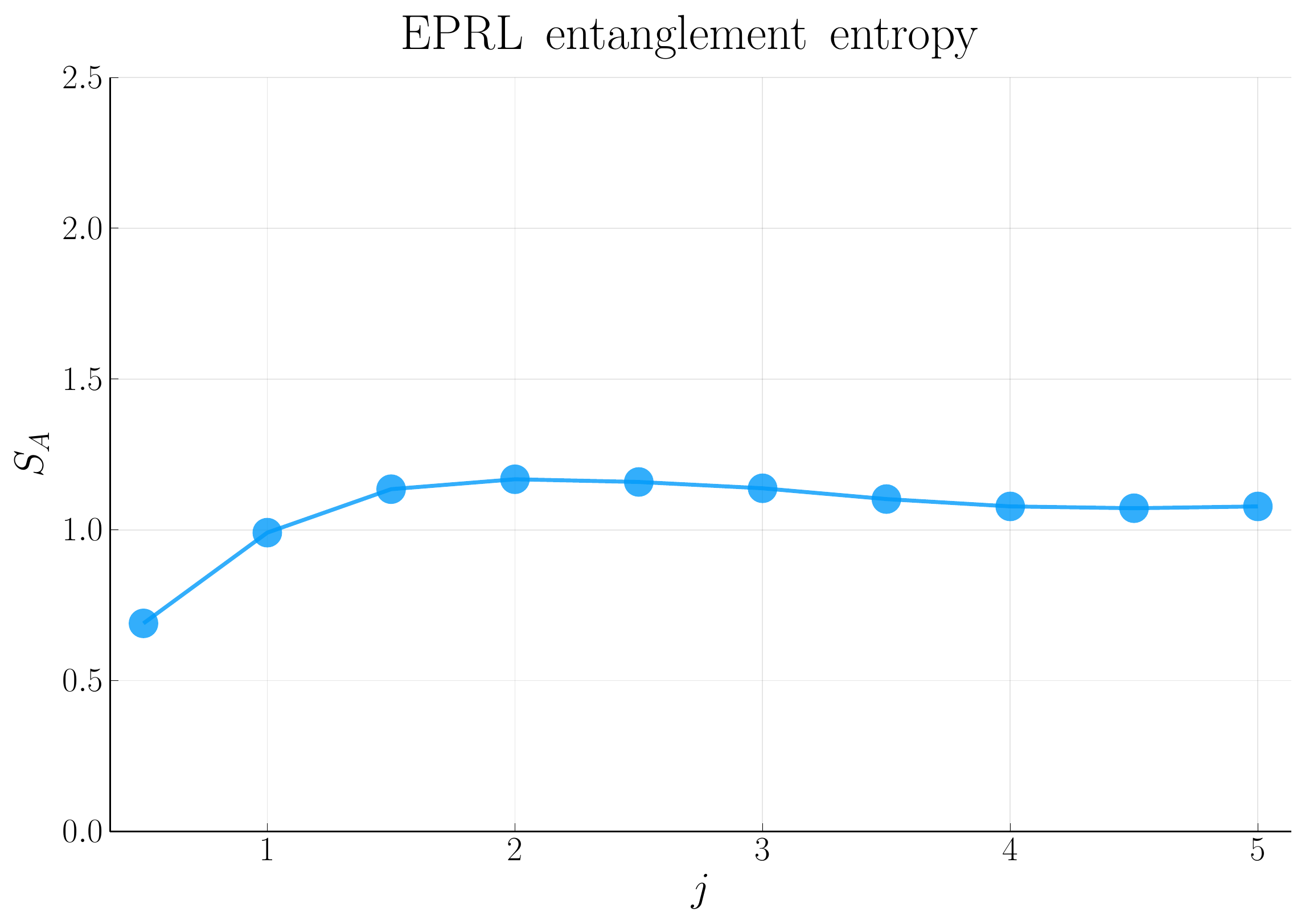}
    \end{subfigure}   
   \caption{\label{fig:entropy_four_nodes} {Values of the entanglement entropy \eqref{eq:Entropy_MC} for the subsystem $A$ in the partition \eqref{eq:Hilbertspaces} with $N_A = 4$. All the nodes in $A$ belong to the same 4-simplex.}}
    \end{figure}

For completeness, we show in Figure \ref{fig:entropy_four_nodes} the values computed for the entropy of subsystem $A$ composed of 4 adjacent nodes, that is, 4 nodes on the same 4-simplex $A = \{ i_{C235}, i_{C234}, i_{C345}, i_{C245}, \}$. For this computation, we limited the analysis to a maximum value $j = 5$ since the computational cost is significantly higher than the other calculations reported in this paper. Unlike the parameters in \ref{app:M-H_parameters}, for this specific calculation we set $N_{MC} = 10^7$ for $j=0.5 \dots 5$, averaging over 17 independent runs both for BF and EPRL. Following to the hybrid parallelization scheme discussed at the beginning of this Section, we used 17 processes, each one with 64 CPUs, for a total of 1088 CPUs. With this configuration, the total computation time for the data in Figure \ref{fig:entropy_four_nodes}, including the sampling of the intertwiners draws, took about 3 days.
\begin{figure}[H]
    \centering
        \begin{subfigure}[b]{75mm}
        \includegraphics[width=75mm]{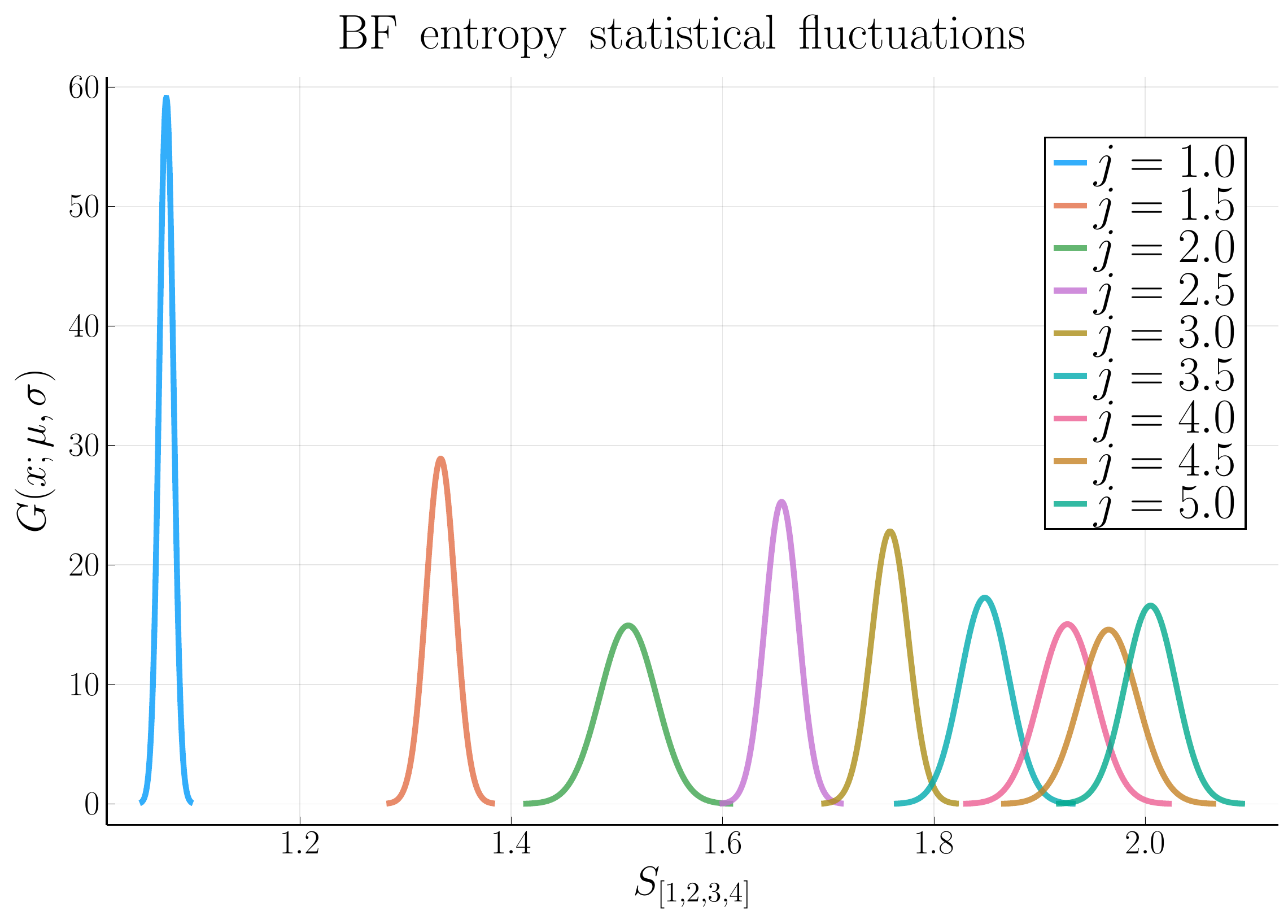}
    \end{subfigure}
    ~~~
        \begin{subfigure}[b]{75mm}
        \includegraphics[width=75mm]{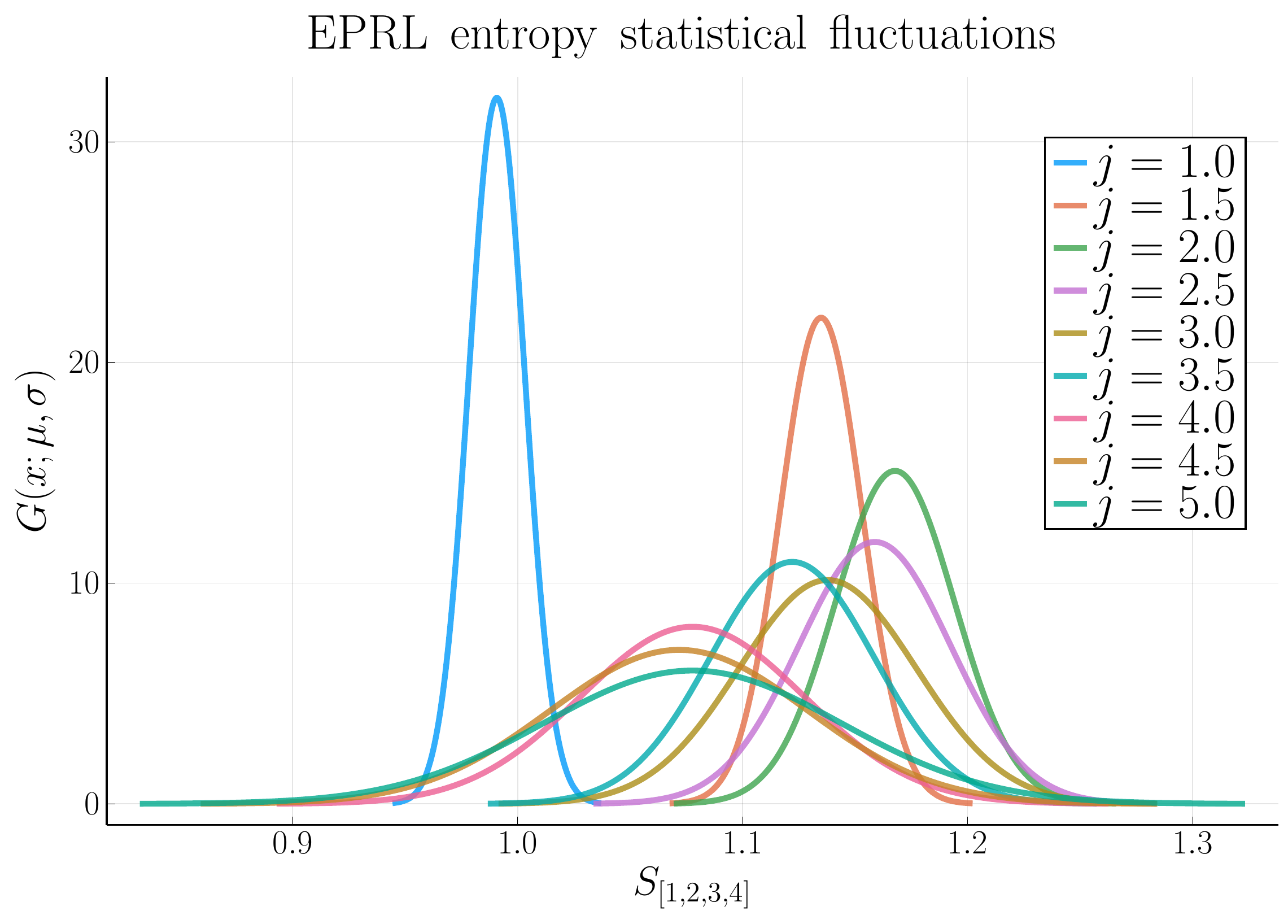}
    \end{subfigure}   
   \caption{\label{fig:entropy_four_nodes_numerical_fluctuations} {\textit{Gaussian distribution \eqref{eq:gaussian} of the entanglement entropy \eqref{eq:Entropy_MC} for the subsystem in Figure \ref{fig:entropy_four_nodes}. We used the same number of iterations $N_{MC} = 10^{7}$ for BF and EPRL, averaging over 17 runs.}}}
    \end{figure}
\pagebreak
\section*{Conclusions}    

In this paper we combined the Metropolis-Hastings algorithm \cite{MH_original_paper} with recently developed high-performance codes in LQG \cite{Francesco_draft_new_code, Dona2018} to compute the expectation value and correlation functions of operators over large spinfoam graphs in the low spins regime. After testing the method, we applied it to the computations of boundary geometrical observables, correlation functions and entanglement entropy in a spinfoam model with 20 boundary nodes, obtained as a refinement of the 4-simplex graph. We investigated both the EPRL and the BF models.
Our results show that the BF and, more significantly, the EPRL model have a well defined behavior under refinement of the boundary graph. The computed boundary geometry agrees in terms of expectation value with the geometric interpretation of the operators. We found that correlations are present in neighbouring patches but decay sharply when moving to patches that belong to different vertices, opening the way to the study of spinfoams composed of many vertices glued together. We also showed that the dynamical correlations between boundary operators in the BF and EPRL models are surprisingly similar in our case study, while the entanglement entropy shows a significant difference. \\

Our work provides important hints on the well-definiteness of spinfoam refinement. The method presented in this paper can be applied to spinfoam models with Lorentzian or Euclidean signature and to compute bulk observables. An interesting perspective would be to compare the results obtained in the spinfoam-like path integral expansion formalism \cite{Bodendorfer_2021} with the ones in the full spinfoam one. This would give interesting insights on the canonical-covariant relation. Numerical methods currently allow to perform computations using spinfoams with a much richer bulk structure than the one considered in this paper. See for example the study of infrared divergences \cite{self_energy_paper, Dona_Frisoni_Ed_infrared} or the analysis of the $\Delta_3$ and $\Delta_4$ triangulation \cite{review_numerical_LQG}. The methods described in this paper can also be applied to different choices of boundary state, although for complex coherent states one is dealing with fluctuating sampling probabilities and different Monte Carlo techniques might be more effective.
The proposed approach provides a needed complement to already existing numerical techniques in covariant LQG \cite{Spinfoam_on_a_Lefschetz_thimble}. It is effective in the regime of low spins quantum numbers with a large number of degrees of freedom, for which the other existing methods are not tailored for. The next step in the developing of this work consist in applying the algorithm presented here to study the correlations functions defined on a spinfoam model with 16 cells on the boundary: this is the next regular triangulation of the 3-sphere after the 4-simplex considered in \cite{Gozzini_primordial}. This is model is studied in \cite{16_cell_model}.  \\[3em]

\centerline{***}
\bigskip

\appendix

\newpage
\section{Discrete truncated normal distribution}
\label{app:truncated_gaussian}
We report in this appendix the definition of truncated normal distribution rounded to integers. For simplicity, we write the equations in the case of a one-dimensional variable. The probability density function of a normal distribution $\mathcal{N}(x, \sigma)$ with mean zero and standard deviation $\sigma$ is defined as:  
\begin{equation}
\mathcal{N}(x, \sigma) = \frac{1}{\sigma \sqrt{2 \pi}}e^{-\frac{x^2}{2 \sigma^2}} \ ,
\end{equation}
where $x \in {R}$. We can define the probability distribution function of a normal distribution with mean zero and standard deviation $\sigma$ rounded to integers as:
\begin{equation}
\mathcal{N}_{d}(n, \sigma) = \Phi ( n + 0.5, \sigma ) - \Phi (n - 0.5, \sigma ) \ ,
\end{equation}
where $n \in {N}$ and $\Phi ( x, \sigma ) $ is the cumulative distribution function of a normal distribution with mean zero and standard deviation $\sigma$, defined as:
\begin{equation}
\Phi (x , \sigma) = \frac{1}{\sigma \sqrt{2 \pi}} \int_{- \infty}^{x} e^{-\frac{t^2}{2 \sigma^2}} dt \ .
\end{equation}
For convenience, let's also define:
\begin{equation}
\Phi (a, b; \sigma) = \Phi (b, \sigma) - \Phi (a, \sigma) = \frac{1}{\sigma \sqrt{2 \pi}} \int_{a}^{b} e^{-\frac{t^2}{2 \sigma^2}} dt \ .
\end{equation}
The cumulative distribution function of a discrete (integer) gaussian is written as:
\begin{equation}
\Phi_{d} (n_1, n_2; \sigma) = \mathcal{N}_{d}(n_1, \sigma) + \mathcal{N}_{d}(n_1 + 1, \sigma) + \dots + \mathcal{N}_{d}(n_2, \sigma) \ .
\end{equation}
With the above definitions, we can define the probability distribution function of a truncated normal distribution rounded to integers $\mathcal{N}_{d,t}(n, n_1, n_2; \sigma)$ between $n_1$ and $n_2$ as:
\begin{equation}
\mathcal{N}_{d,t}(n, n_1, n_2; \sigma) = \frac{\mathcal{N}_d (n, \sigma)}{\Phi_{d} (n_1, n_2; \sigma)} \ .
\end{equation}
\newpage
\section{Metropolis-Hastings parameters}
\label{app:M-H_parameters}
We report in the tables below the parameters used in the Metropolis-Hastings algorithm. 
These are the parameters used for all calculations in this paper except for the data in Figures  \ref{fig:entropy_four_nodes} and \ref{fig:RW_benchmark}.
\begin{table}[H]
\label{tbl:MH_data}
\begin{center}
\begin{tabular}{|p{1cm}|p{1cm}|p{1cm}|p{1cm}|p{1cm}|p{1cm}||}
 \hline
 \multicolumn{5}{|c|}{BF 4-simplex} \\
 \hline
 $j$ & $N_{MC}$ & b & $\sigma$ & C \\
 \hline
 0.5   & $10^6$ & $10^2$ & $0.97$ & 5 \\
 1.0   & $10^6$ & $10^2$ & $0.94$ & 5 \\
 1.5   & $10^6$ & $10^2$ & $0.90$ & 5 \\
 2.0   & $10^6$ & $10^2$ & $0.90$ & 5 \\
 2.5   & $10^7$ & $10^2$ & $0.90$ & 5 \\
 3.0   & $10^7$ & $10^2$ & $0.90$ & 5 \\ 
 3.5   & $10^7$ & $10^2$ & $0.90$ & 5 \\ 
 4.0   & $10^7$ & $10^2$ & $0.90$ & 5 \\ 
 4.5   & $10^7$ & $10^3$ & $0.90$ & 5 \\ 
 5.0   & $10^7$ & $10^3$ & $0.90$ & 5 \\
 5.5   & $10^7$ & $10^3$ & $0.90$ & 5 \\
 6.0   & $10^7$ & $10^3$ & $0.90$ & 5 \\ 
 \hline
\end{tabular}
\quad
\hskip12mm 
\begin{tabular}{|p{1cm}|p{1cm}|p{1cm}|p{1cm}|p{1cm}|p{1cm}||}
 \hline
 \multicolumn{5}{|c|}{EPRL 4-simplex} \\
 \hline
 $j$ & $N_{MC}$ & b & $\sigma$ & C \\
 \hline
 0.5   & $10^7$ & $10^3$ & $0.97$ & 5 \\
 1.0   & $10^7$ & $10^3$ & $0.94$ & 5 \\
 1.5   & $10^7$ & $10^3$ & $0.90$ & 5 \\
 2.0   & $10^7$ & $10^3$ & $0.90$ & 5 \\
 2.5   & $10^7$ & $10^3$ & $0.90$ & 5 \\
 3.0   & $10^7$ & $10^3$ & $0.90$ & 5 \\ 
 3.5   & $10^7$ & $10^3$ & $0.90$ & 5 \\ 
 4.0   & $10^7$ & $10^3$ & $0.90$ & 5 \\ 
 4.5   & $10^7$ & $10^3$ & $0.90$ & 5 \\ 
 5.0   & $10^7$ & $10^3$ & $0.90$ & 5 \\
 5.5   & $10^7$ & $10^3$ & $0.90$ & 5 \\
 6.0   & $10^7$ & $10^3$ & $0.90$ & 5 \\ 
 \hline
\end{tabular}
\vskip1em
\begin{tabular}{|p{1cm}|p{1cm}|p{1cm}|p{1cm}|p{1cm}|p{1cm}||}
 \hline
 \multicolumn{5}{|c|}{BF star} \\
 \hline
 $j$ & $N_{MC}$ & b & $\sigma$ & C \\
 \hline
 0.5   & $10^6$ & $10^3$ & $0.40$ & 32 \\
 1.0   & $10^6$ & $10^3$ & $0.39$ & 32 \\
 1.5   & $10^6$ & $10^3$ & $0.37$ & 32 \\
 2.0   & $10^6$ & $10^3$ & $0.35$ & 32 \\
 2.5   & $10^6$ & $10^3$ & $0.35$ & 32 \\
 3.0   & $3 \cdot 10^6$ & $10^3$ & $0.35$ & 32 \\ 
 3.5   & $3 \cdot 10^6$ & $10^3$ & $0.35$ & 32 \\ 
 4.0   & $5 \cdot 10^6$ & $10^3$ & $0.35$ & 32 \\ 
 4.5   & $5 \cdot 10^6$ & $10^3$ & $0.35$ & 32 \\ 
 5.0   & $5 \cdot 10^6$ & $10^3$ & $0.35$ & 32 \\
 5.5   & $7 \cdot 10^6$ & $10^3$ & $0.35$ & 32 \\
 6.0   & $7 \cdot 10^6$ & $10^3$ & $0.35$ & 32 \\ 
 \hline
\end{tabular}
\quad
\hskip12mm 
\begin{tabular}{|p{1cm}|p{1cm}|p{1cm}|p{1cm}|p{1cm}|p{1cm}||}
 \hline
 \multicolumn{5}{|c|}{EPRL star} \\
 \hline
 $j$ & $N_{MC}$ & b & $\sigma$ & C \\
 \hline
 0.5   & $10^7$ & $10^3$ & $0.40$ & 32 \\
 1.0   & $3 \cdot 10^7$ & $10^3$ & $0.39$ & 32 \\
 1.5   & $3 \cdot 10^7$ & $10^3$ & $0.35$ & 32 \\
 2.0   & $3 \cdot 10^7$ & $10^3$ & $0.35$ & 32 \\
 2.5   & $3 \cdot 10^7$ & $10^3$ & $0.35$ & 32 \\
 3.0   & $5 \cdot 10^7$ & $10^3$ & $0.37$ & 32 \\ 
 3.5   & $5 \cdot 10^7$ & $10^3$ & $0.37$ & 32 \\ 
 4.0   & $5 \cdot 10^7$ & $10^3$ & $0.38$ & 32 \\ 
 4.5   & $5 \cdot 10^7$ & $10^3$ & $0.39$ & 32 \\ 
 5.0   & $5 \cdot 10^7$ & $10^3$ & $0.40$ & 32 \\
 5.5   & $8 \cdot 10^7$ & $10^3$ & $0.40$ & 32 \\
 6.0   & $8 \cdot 10^7$ & $10^3$ & $0.40$ & 32 \\ 
 \hline
\end{tabular}
\end{center}
\caption*{TABLES:~~~\textit{Parameters used in the Metropolis-Hastings algorithm. From left to right: $j$ is the spin attached to the links of the star spinfoam, $N_{MC}$ is the number of Monte Carlo iterations, $b$ is the number of burn-in iterations, $\sigma$ is the standard deviation of the truncated normal proposal distribution and $C$ corresponds to the number of Markov chains that we averaged to improve the statistic and measure the statistical fluctuations.}}
\end{table}
A general difference that we observed between the BF and EPRL model is a greater statistical fluctuation in the expectation values of operators for EPRL as $j$ increases, despite the dimension of the intertwiners' space being the same. In order to reduce the statistical fluctuations in EPRL, we tried both to increase the number of Markov chains to be averaged by an order of magnitude (in the code \cite{Frisus_star_model_repo} the latter are automatically parallelized on the available cores) and to increase the number of Monte Carlo iterations $N_{MC}$. We found good precision in both cases and for this paper we decided to use the data obtained with the second approach, as shown in the tables. 

Notice the role that the Metropolis-Hastings parameters play in the sampling process. While increasing the number of chains to be averaged has the effect of improving the accuracy in the determination of the operator's mean value (and the corresponding standard deviation), increasing the number of Monte Carlo iterations $N_{MC}$ implies reducing the standard deviation of the statistical sampling. A satisfying statistical precision is therefore achieved when these two parameters are sufficiently high and balanced. While we did not find relevant differences by modifying the number of burn-in iterations, we set the optimal standard deviation of the Gaussian proposal distribution $\sigma$ by requiring an acceptance rate of intertwiners draws $[i_n]$ around $30\%$ in the sampling algorithm.
\section{Booster functions} 
\label{app:booster}
The booster functions \cite{Dona2018}, \cite{article:Dona_etal_2019_numerical_study_lorentzian_EPRL_spinfoam_amplitude}, also known as B4 functions \cite{Speziale2016}, are the non compact residuals of the $SL(2, {C})$ integrals. These functions turn out to encode all the details of the EPRL model, such as the $Y -$map. We define them as\footnote{In this Appendix we don't indicate the dependence on multiple variables with the curly brackets in order not to weigh down the notation}:
\begin{equation}
\label{eq:boosterdef}
\hskip-25mm
B_4 \left( j_f, l_f ;  i ,  k \right) \equiv \frac{1}{4\pi} \sum_{\{ p_f \}} \left(\begin{array}{c} j_f \\ p_f \end{array}\right)^{(i)} \left(\int_0^\infty \mathrm{d} r \sinh^2r \, 
\prod_{f=1}^4 d^{(\gamma j_f,j_f)}_{j_f l_f p_f}(r) \right)
\left(\begin{array}{c} l_f \\ p_f \end{array}\right)^{(k)}\  = \raisebox{-12.5mm}{ \includegraphics[width=2.2cm]{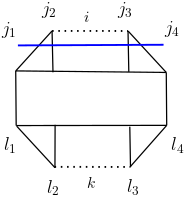}}, \hskip9mm
\end{equation}
where $d^{(\rho,k)}(r)$ are the boost matrix elements for $\gamma$-simple irreducible representations of $SL(2, {C})$ in the principal series and $\gamma$ is the Immirzi parameter. In their most general formulation, the booster functions turn out to be the $SL(2,  {C})$ analogues of the usual Clebsch-Gordan coefficients for the rotation group $SU(2)$. The semi-classical limit of booster functions was discussed in \cite{Dona2020yao}. The general explicit form of the boost matrix elements can be found in the literature \cite{Ruhl:1970fk, Speziale2016}. In the case of simple irreducible representations these turn out to be \cite{Speziale2016}:
\begin{eqnarray}
\label{eq:dsmall}
\hskip-25mm
d^{(\gamma j,j)}_{jlp}(r) =&  
(-1)^{\frac{j-l}{2}} \frac{\Gamma\left( j + i \gamma j +1\right)}{\left|\Gamma\left(  j + i \gamma j +1\right)\right|} \frac{\Gamma\left( l - i \gamma j +1\right)}{\left|\Gamma\left(  l - i \gamma j +1\right)\right|} \frac{\sqrt{2j+1}\sqrt{2l+1}}{(j+l+1)!}  
\left[(2j)!(l+j)!(l-j)!\frac{(l+p)!(l-p)!}{(j+p)!(j-p)!}\right]^{1/2} 
\hskip-1cm\nn \\
\hskip-25mm
&\hskip-5mm \times e^{-(j-i\gamma j +p+1)r}
\sum_{s} \frac{(-1)^{s} \, e^{- 2 s r} }{s!(l-j-s)!} \, {}_2F_1[l+1-i\gamma j,j+p+1+s,j+l+2,1-e^{-2r}] \ .
\end{eqnarray}
where $2F_1[a, b, c, z]$ is the hypergeometric function. \\[3em]

\begin{center}{***}
\end{center}
\bigskip
\bigskip

\paragraph*{\bf Acknowledgments}
We thank to Carlo Rovelli for many discussions on this project.   
We acknowledge the Shared Hierarchical Academic Research Computing Network (SHARCNET) for granting access to their high-performance computing resources. We thank in particular the Compute/Calcul Canada staff for the constant support provided with the Cedar and Graham clusters.
This work was supported by the Natural Science and Engineering Council of Canada (NSERC) through the Discovery Grant "Loop Quantum Gravity: from Computation to Phenomenology". We acknowledge support also from the QISS JFT grant 61466.  FV's research is supported by the Canada Research Chairs Program.
We acknowledge the Anishinaabek, Haudenosaunee, L\=unaap\`eewak and Attawandaron peoples, on whose traditional lands Western University is located.%
\\[3em]




\providecommand{\href}[2]{#2}


\end{document}